\documentclass[12pt]{article}
\usepackage[letterpaper,left=1in, right=1in, bottom=1.5in, top=1in]{geometry} 
\usepackage{amssymb}
\usepackage{amsfonts}
\usepackage{amsbsy}
\usepackage[all,v2,cmtip,2cell]{xy}
\usepackage{amsmath}
\usepackage{dsfont}
\usepackage{bbm}
\usepackage{upgreek}
\usepackage{amssymb,amscd}
\usepackage{mathrsfs}
\usepackage{amsmath,amsthm}
\usepackage{slashed}
\usepackage{dsfont}
\usepackage[utf8]{inputenc}
\usepackage{cite}
\usepackage{setspace}
\usepackage[colorlinks]{hyperref}
\hypersetup{
    citecolor = {blue}
}
\usepackage{tikz}
\usepackage{tikz-cd}
\usetikzlibrary{arrows,shapes,snakes,automata,backgrounds,petri}
\usetikzlibrary{calc,arrows,cd,decorations.markings,snakes}
\tikzset{
->-/.style args={#1rotate#2}{decoration={markings, mark=at position #1 with {\arrow[scale=1.5,rotate = #2 ]{stealth}}}, postaction={decorate}}
}
\usetikzlibrary{shapes.geometric}
\tikzset{snake it/.style={decorate, decoration=snake}}

\usetikzlibrary{decorations.markings}

\tikzset{line/.style={line width=0.25mm},
curve/.style={line,smooth,tension=1},
->-/.style={decoration={
  markings,
  mark=at position #1 with {\arrow[>=stealth]{>}}},postaction={decorate}},
-<-/.style={decoration={
  markings,
  mark=at position #1 with {\arrow[>=stealth]{<}}},postaction={decorate}},
}

\tikzset{
    partial ellipse/.style args={#1:#2:#3}{
        insert path={+ (#1:#3) arc (#1:#2:#3)}
    }
}
\tikzset{bg/.style={opacity=.5}}

\usepackage[framemethod=TikZ]{mdframed} % fancy boxes
\usepackage{tikz-cd} %for complicated commutative diagrams
\usetikzlibrary {shapes.geometric} 

\usetikzlibrary{arrows,snakes,shapes.arrows,decorations.markings}
     \tikzset{>=triangle 90}
     \tikzstyle{bbc}=[draw,circle,fill=black,scale=.75]
     \tikzstyle{rc}=[circle,fill=red,scale=.6]
     \tikzstyle{wc}=[draw,circle,scale=.75]
     
\usetikzlibrary{decorations.pathreplacing,calligraphy}

\tikzset{snake it/.style={decorate, decoration=snake}}

\tikzset{
	on each segment/.style={
		decorate,
		decoration={
			show path construction,
			moveto code={},
			lineto code={
				\path [#1]
				(\tikzinputsegmentfirst) -- (\tikzinputsegmentlast);
			},
			curveto code={
				\path [#1] (\tikzinputsegmentfirst)
				.. controls
				(\tikzinputsegmentsupporta) and (\tikzinputsegmentsupportb)
				..
				(\tikzinputsegmentlast);
			},
			closepath code={
				\path [#1]
				(\tikzinputsegmentfirst) -- (\tikzinputsegmentlast);
			},
		},
	},
	mid arrow/.style={postaction={decorate,decoration={
				markings,
				mark=at position .5 with {\arrow[#1]{stealth}}
	}}},
}

\usepackage{colortbl}
\definecolor{dgreen}{rgb}{0, 0.55, 0}
\definecolor{llightyellow}{rgb}{1.0, 0.95, 0.7}
\definecolor{llightblue}{rgb}{0.7, 0.9, 1.0}
\definecolor{llightpink}{rgb}{1.0, 0.85, 0.95}
\definecolor{llightgreen}{rgb}{0.7, 1.0, 0.4}
\colorlet{lightyellow}{llightyellow!50!white}
\colorlet{lightblue}{llightblue!50!white}
\colorlet{lightgreen}{llightgreen!50!white}
\colorlet{lightpink}{llightpink!50!white}
\definecolor{azure}{rgb}{0.0, 0.5, 1.0}
\definecolor{darkblue}{rgb}{0.15,0.35,0.7}
\definecolor{reddish}{rgb}{0.65, 0.2, 0.2}
\definecolor{brandeisblue}{rgb}{0.0, 0.44, 1.0}
\definecolor{ceruleanblue}{rgb}{0.16, 0.32, 0.75}
\definecolor{indigo(dye)}{rgb}{0.0, 0.25, 0.42}
\definecolor{grey}{rgb}{0.9,0.9,0.9}
\definecolor{dgrey}{rgb}{0.3,0.3,0.3}
\usetikzlibrary{shadings}

\usetikzlibrary {decorations.markings}

\usepackage{simpler-wick}

\usepackage{subcaption}

\makeatletter
\DeclareFontFamily{OMX}{MnSymbolE}{}
\DeclareSymbolFont{MnLargeSymbols}{OMX}{MnSymbolE}{m}{n}
\SetSymbolFont{MnLargeSymbols}{bold}{OMX}{MnSymbolE}{b}{n}
\DeclareFontShape{OMX}{MnSymbolE}{m}{n}{
    <-6>  MnSymbolE5
   <6-7>  MnSymbolE6
   <7-8>  MnSymbolE7
   <8-9>  MnSymbolE8
   <9-10> MnSymbolE9
  <10-12> MnSymbolE10
  <12->   MnSymbolE12
}{}
\DeclareFontShape{OMX}{MnSymbolE}{b}{n}{
    <-6>  MnSymbolE-Bold5
   <6-7>  MnSymbolE-Bold6
   <7-8>  MnSymbolE-Bold7
   <8-9>  MnSymbolE-Bold8
   <9-10> MnSymbolE-Bold9
  <10-12> MnSymbolE-Bold10
  <12->   MnSymbolE-Bold12
}{}

\let\llangle\@undefined
\let\rrangle\@undefined
\DeclareMathDelimiter{\llangle}{\mathopen}%
                     {MnLargeSymbols}{'164}{MnLargeSymbols}{'164}
\DeclareMathDelimiter{\rrangle}{\mathclose}%
                     {MnLargeSymbols}{'171}{MnLargeSymbols}{'171}
\makeatother

\def\be{ \begin{equation} }
\def\ee{ \end{equation}}

\newcommand{\eq}[1]{\begin{align}\begin{split}#1\end{split}\end{align}}

\def\hatU{{\widehat{U}}}
\def\hatg{{\widehat{g}}}
\def\hath{{\widehat{h}}}
\def\hatgh{{\widehat{gh}}}

\def\det{{\rm det}}

\def\dim{{\rm dim}}
\def\exp{{\rm exp}}

\def\log{{\rm log}}
\def\mod{{\rm mod}}

\def\half{\frac{1}{2}}

\def\dlangle{\langle\!\langle}
\def\drangle{\rangle\!\rangle}

\def\one{{\hbox{ 1\kern-.8mm l}}}
\def\CA{{\cal A}}

\def\CC {{\cal C}}

\def\CG {{\cal G}}
\def\CH {{\cal H}}
\def\CI {{\cal I}}

\def\CN {{\cal N}}

\def\CO {{\cal O}}
\def\CP {{\cal P}}

\def\CO {{\cal O}}

\def\CG {{\cal G}}
\def\CH {{\cal H}}
\def\CI {{{\cal I}}}

\def\CU {{\,\cal U}}

\def\IG{\mathbb{G}}

\def\ICP{\mathbb{CP}}

\def\IR{{\mathbb{R}}}

\def\IZ{{\mathbb{Z}}}

\def\rmk#1{\bigskip\noindent{\bf Remark} }
\def\cnj#1{\bigskip\noindent{\bf Conjecture:} }

\def\tildez{{\tilde{z}}}
\def\hatG{{\widehat{G}}}

\DeclareMathAlphabet{\mathpzc}{OT1}{pzc}{m}{it}

\def\Tr{ \, \textrm{Tr} \, }

\def\t{\tilde}

\def\ZZ{{\mathbb{Z}}}

\newcommand{\nc}{\newcommand}
\nc{\rnc}{\renewcommand} 

\rnc{\a}{\alpha}
\rnc{\b}{\beta}
\rnc{\d}{\delta}
\nc{\e}{\epsilon}
\nc{\z}{\zeta}
\nc{\f}{\phi}
\nc{\m}{\mu}
\nc{\n}{\nu}
\rnc{\r}{\rho}
\rnc{\k}{\kappa}
\rnc{\l}{\lambda}
\nc{\s}{\sigma}
\rnc{\t}{\tau}
\nc{\w}{\omega}
\nc{\x}{\chi}
\nc{\F}{\Phi}
\rnc{\L}{\Lambda}

\def\scrD{\mathscr{D}}
\def\scrL{\mathscr{L}}

\def\0{{(0)}}
\def\1{{(1)}}

\usepackage[margin=1cm]{caption}
\title{ 
Consequences of 
Symmetry Fractionalization \\ without 1-Form Global Symmetries 
}

\author{T.~Daniel Brennan,$^1$\footnote{ tbrennan@ucsd.edu}\ \    Theodore Jacobson,$^2$\footnote{ tjacobson@physics.ucla.edu}\ \  and Konstantinos Roumpedakis$^3$\footnote{ kroumpe1@jh.edu} \\
\\
{\it\small $^1$Department of Physics, University of California San Diego}\\
{\it\small 9500 Gilman Drive, La Jolla, CA 92093-0319, USA} \\
{\it\small $^2$Mani L. Bhaumik Institute for Theoretical Physics, }\\
{\it\small Department of Physics and Astronomy, University of California Los Angeles} \\
{\it\small 475 Portola Plaza, Los Angeles, CA 90095, USA}  \\
{\it\small $^3$William H. Miller III Department of Physics and Astronomy, Johns Hopkins University} \\
{\it\small 3400 North Charles Street, Baltimore, MD 21218, USA}
}

\date{}
\begin{document}
\maketitle
\thispagestyle{empty}

\begin{abstract}

We study the fractionalization of 0-form global symmetries on line operators in theories without 1-form global symmetries. The projective transformation properties of line operators are renormalization group invariant, and we derive constraints which are similar to the consequences of exact 1-form symmetries. For instance, symmetry fractionalization can lead to exact selection rules for line operators in twisted sectors, and in theories with 't Hooft anomalies involving the fractionalization class, these selection rules can further imply that certain twisted sectors have exact finite-volume vacuum degeneracies. Along the way, we define topological operators on open codimension-1 manifolds, which we call `disk operators', that provide a convenient way of encoding the projective action of 0-form symmetries on lines. In addition, we discuss the possible ways symmetry fractionalization can be matched along renormalization group flows. 

\end{abstract}

\newpage
\begingroup
	
	\hypersetup{linkcolor=.,linktoc=all}
	\tableofcontents

\endgroup

%
%
%
%
%
%
%
%

%
%
%%%%%%%%%%%%%%%%%%%%%%%%%%%%%%%%%%%%%%%%%%%%%%%%%%%%
\section{Introduction}
%%%%%%%%%%%%%%%%%%%%%%%%%%%%%%%%%%%%%%%%%%%%%%%%%%%%

Line operators are an indispensable tool in the study of quantum field theory (QFT). They were first introduced by Wegner~\cite{Wegner:1971app} and Wilson \cite{Wilson:1974sk} (and later expanded on by 't Hooft \cite{tHooft:1979rat}) in the context of gauge theories as a sharp probe of (de)confining dynamics. When inserted along time, line operators can be viewed as the static worldlines of probe particles, or impurities localized in space~\cite{Kondo:1964nea}. In gauge theories, such static probe charges can for instance source confining flux tubes, create long-range Coulomb fields, or induce Aharanov-Bohm phases for bulk quasiparticles.

The behavior of a line operator can be directly tied to the infrared (IR) phase structure of a bulk QFT provided the line is charged under a 1-form global symmetry. Our modern understanding of line operators has therefore grown in parallel with the development of our understanding of  
generalized~\cite{Gaiotto:2014kfa} or `categorical' symmetries (for reviews see \cite{Freed:2022qnc,McGreevy:2022oyu,Schafer-Nameki:2023jdn,Shao:2023gho,Bhardwaj:2023kri,Freed:2022iao,Brennan:2023mmt,Costa:2024wks}). 1-form global symmetries have many physical consequences: they prevent charged lines from ending on local operators, protect them from being trivialized along renormalization group (RG) flows, and provide a framework for a generalized Landau paradigm of spontaneous symmetry breaking~\cite{Gaiotto:2014kfa,lake2018,Hofman_2019,Iqbal:2021rkn,McGreevy:2022oyu,Bhardwaj:2023fca}. In particular, there is a notion of spontaneous symmetry breaking for 1-form symmetries which implies a phase boundary between any two regimes with different realizations of the symmetry (i.e. preserved vs. spontaneously broken). These phases are differentiated by whether the expectation values of charged line operators decay to zero in the limit of large loops (in any choice of scheme). In other words, the expectation value of a 1-form charged line operator serves as the order parameter in the generalized Landau paradigm.

The symmetry structure of a theory can be complicated in the case when symmetries of different form degrees coexist. For example, in certain cases naive 0-form and 1-form symmetries can mix together into a  
categorical symmetry called a 2-group (a special case of an $n$-group global symmetry)~\cite{Baez:2003yaq,Baez:2004in,Cordova:2018cvg,Benini:2018reh,Sharpe:2015mja,Gukov:2013zka,Brennan:2020ehu}. However, even in theories with basic group-like symmetries, interesting features can arise when we consider the action of 0-form global symmetries on extended operators due to a phenomenon known as symmetry fractionalization \cite{Delmastro:2022pfo,Barkeshli:2014cna,Brennan:2022tyl}. More generally, given an extended operator of dimension $p$, one must specify how it transforms under all $q$-form symmetries in the system for $q\le p$.

Symmetry fractionalization is part of the data which captures how 0-form symmetries act on line operators. First, there is an obvious action of disconnected 0-form symmetries via a permutation of the set of lines (e.g. charge conjugation symmetry in $U(1)$ gauge theory which exchanges the charge $q$ and $-q$ Wilson lines). We will not consider such symmetry actions here. Instead, we investigate the more subtle action via symmetry fractionalization, which comes from the linking between line operators and the codimension-2 \emph{junctions} of defects that generate a 0-form symmetry $G^\0$ that does not change the line type. For each line this is specified by a `fractionalization class,' generically a class in $\omega\in H^2(BG^\0,U(1))$, where the superscript `2' in the cohomology class comes from the fact that we can label a minimal trivalent junction by two $G^\0$ elements. Physically, the global symmetry fractionalization reflects the fact that the line operator describes the worldline of a particle that transforms projectively under $G^\0$ such as in spin-electromagnetism~\cite{Wang_2014,Thorngren:2014pza,Kravec:2014aza,Zou:2017ppq,Wang:2018qoy,Hsin:2019fhf,Brennan:2022tyl,Kan:2024fuu}, 
or how a ``bare" ultraviolet (UV) line operator is either screened by or traps particles that transform projectively under $G^\0$ as in the situation where monopole lines support fermion zero-modes on their worldvolumes~\cite{Jackiw:1975fn,Wang:2018qoy,Brennan:2022tyl}. 

In this paper, we will be most interested in the physical consequences of symmetry fractionalization in the absence of 1-form global symmetries. In this case, all line operators can be cut open, and the action of $G^\0$ symmetry operators on lines is constrained by the symmetry action on their endpoints. On general grounds, these endpoint operators can transform projectively under the $G^\0$ symmetry group since they are not themselves well-defined local operators. This is the basic mechanism behind $G^\0$ symmetry fractionalization on endable lines. 

The quintessential example of this phenomenon is a $G_g$ gauge theory with matter fields 
that transform under a flavor symmetry group $\hatG^\0$ but where a subgroup $\Gamma$ of the center $Z(\hatG^\0)$ is identified with a subgroup of the gauge group $\Gamma\subset G_g$. In this case, the 0-form global
symmetry group that acts faithfully on gauge-invariant local operators is $G^\0=\hatG^\0/\Gamma$. Consequently, the Wilson lines in representations that transform non-trivially under $\Gamma\subset Z(G_g)$ can terminate on charged matter fields which transform linearly under $\hatG^\0$ and (generically) projectively under $G^\0$. 

Symmetry fractionalization in this gauge theory context reflects the fact that states in the ordinary Hilbert space transform under linear representations of $G^\0$, while states which violate the Gauss law (i.e. states in certain defect Hilbert spaces) may transform projectively.   A (time-like) Wilson line insertion represents the static worldline of an infinitely heavy probe particle which sources the Gauss law. The corresponding `twisted' Hilbert space in general furnishes a projective representation of $G^\0$. Intuitively, one can imagine forming gauge-invariant `heavy-light' composite states using the infinitely heavy probe particles and the light, dynamical matter fields. Such states in general will have fractional global quantum numbers relative to those created by gauge-invariant local operators. 

In these examples, symmetry fractionalization allows the activation of discrete fractional flux in the gauge group  
at the expense of activating a correlated flux in $G^\0$. This observation has been used to learn about and leverage certain global 't Hooft anomalies~\cite{Komargodski:2017dmc,Komargodski:2017smk,Benini:2017dus,Wang:2018qoy,Cordova:2018acb,Anber:2019nze,Cordova:2019uob,Anber:2020gig,Anber:2021iip,Anber:2021lzb,Apruzzi:2021vcu,Heckman:2022suy,Lohitsiri:2022jyz,Brennan:2022tyl,Bhardwaj:2023zix,Brennan:2023tae,Anber:2023urd,Brennan:2023ynm,Brennan:2023vsa,Brennan:2023kpo} and has been extensively applied in the context of circle-~\cite{COHEN1983183,Kouno:2015sja,Iritani:2015ara,Misumi:2015hfa,Tanizaki:2017qhf,Tanizaki:2017mtm,Shimizu:2017asf,Cherman:2017tey,Komargodski:2017dmc,Yonekura:2019vyz,Kanazawa:2019tnf,Fujimori:2020zka,Nguyen:2022lie,Nardoni:2024sos} and torus-~\cite{Tanizaki:2022plm,Tanizaki:2022ngt,Hayashi:2023wwi,Hayashi:2024gxv,Hayashi:2024qkm} compactified gauge theories. While symmetry fractionalization allows us to access some of the same gauge bundles obtained by activating a 1-form symmetry, its implications for e.g. Wilson lines are weaker than the constraints from an exact 1-form symmetry. For example, symmetry fractionalization cannot prevent the flux tubes created by a Wilson line in a confining gauge theory from breaking via the production of dynamical charged particles, and the resulting perimeter-law scaling for asymptotically large contractible Wilson loops. However, symmetry fractionalization does provide information about the particles responsible for said screening --- in particular, their `fractional' quantum numbers relative to gauge-neutral composite particles.

In this paper, we will derive some physical consequences of symmetry fractionalization without 1-form global symmetries. While we often focus on gauge theories in this paper, we hasten to point out that the essential ingredients of this work only require the existence of line operators transforming in projective representations of the 0-form symmetry, i.e. symmetry fractionalization, which can occur in theories without natural gauge theory descriptions (see for instance~\cite{Sachdev:2003yk,Florens_2006,Liu_2021,Cuomo:2022xgw}). As with 1-form global symmetries, symmetry fractionalization is a statement about the action of topological operators on lines which is rigid and hence trackable along renormalization group (RG) flows.

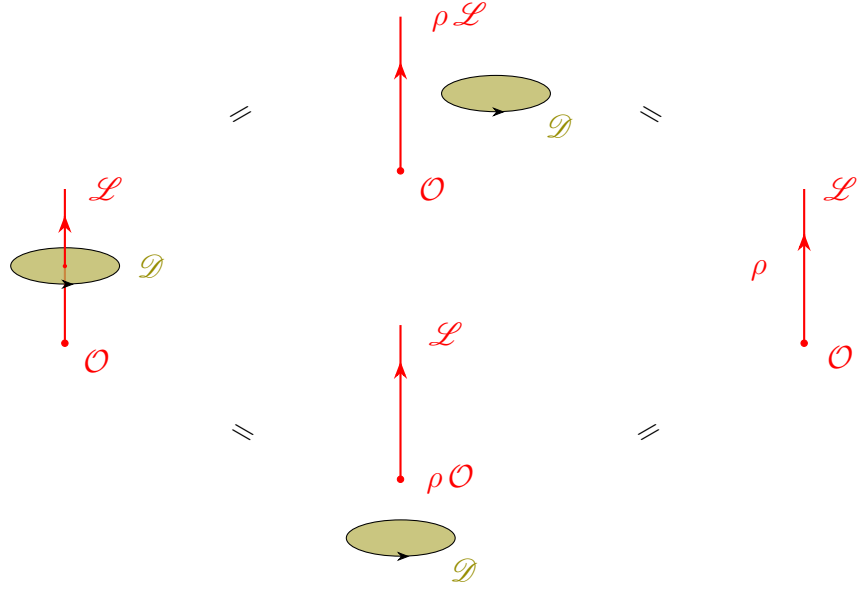
\begin{figure}[t]
    \centering
  \begin{tikzpicture}[scale=1.2]
    
      		\draw[thick, red] (-0.95,-1)  -- (-0.95,-0.35);	
            \draw[thick, red,opacity=0.5] (-0.95,-0.35)  -- (-0.95,-0.15);
            \draw[fill = olive,fill opacity=0.5] (0,0) ++ (100:2 and -.15) arc (0:360:.6 and .2);
            \draw[thick, red] (-0.95,0.7)  -- (-0.95,-0.15);
            \draw[fill, red] (-0.95,-0.15) circle (0.5pt);
            \draw[thick, red, -Stealth] (-0.95,0.4)  -- (-0.95,0.41);
            \draw[-Stealth]  (-0.95,-0.35) -- (-0.85,-0.35);
            \node[red] at (-0.5, 0.7) {$\scrL$};
            \node[red] at (-0.6,-1.2) {$\mathcal{O}$};
            \draw[fill, red] (-0.95, -1) circle (1pt); 
            \node[olive] at (0, -0.15) {$\scrD$};
        
            \draw[thick, red] (2.75,2.6)  -- (2.75,0.9);
            \draw[thick, red, -Stealth] (2.75,2)  -- (2.75,2.1);
            \draw[fill, red] (2.75, 0.9) circle (1pt); 
            \node[red] at (3.1,0.7) {$\mathcal{O}$};
            \node[red] at (3.4, 2.6) {$\rho\, \scrL$};
            \draw[fill = olive,fill opacity=0.5] (4.75,1.9) ++ (100:2 and -.15) arc (0:360:.6 and .2);            
            \node[olive] at (4.5, 1.4) {$\scrD$};
            \draw[-Stealth] (3.8,1.55) -- (3.9,1.55);

            \draw[thick, red, thick] (2.75,-0.8)  -- (2.75,-2.5);
            \draw[thick, red, -Stealth] (2.75,-1.3)  -- (2.75,-1.2);
            \draw[fill, red] (2.75, -2.5) circle (1pt); 
            \node[red] at (3.3,-2.5) {$\rho\,\mathcal{O}$};
            \node[red] at (3.25, -0.9) {$\scrL$};
            \draw[fill = olive,fill opacity=0.5] (3.7,-3) ++ (100:2 and -.15) arc (0:360:.6 and .2);            
            \node[olive] at (3.45,-3.5) {$\scrD$};
            \draw[-Stealth] (2.75,-3.35) -- (2.85,-3.35);

            \draw[thick, red, thick] (7.2,-1)  -- (7.2,0.7);
            \draw[thick, red, -Stealth] (7.2,0.2)  -- (7.2, 0.21);
            \draw[fill, red] (7.2, -1) circle (1pt); 
            \node[red] at (7.6,-1.15) {$\mathcal{O}$};
            \node[red] at (7.6, 0.7) {$\scrL$};
            \node[red] at (6.7,-.2) {$\rho$};

			\node[rotate = 30] at (1,1.5) {$=$};
			\node[rotate = -30] at (1,-2) {$=$};
			\node[rotate = -30] at (5.5,1.5) {$=$};
			\node[rotate = 30] at (5.5,-2) {$=$};
			
        \end{tikzpicture}
        \caption{An open line operator which transforms projectively under $G^\0$ is `charged' under the disk operator. Here, the two ways of unlinking the disk operator -- passing through the line (above) or through the end point (below) -- lead to a consistent action by a phase $\rho$. \label{fig:DiskLineaction} }
\end{figure}

As part of our analysis, we construct a 
topological operator which we refer to as the `disk operator' which is a useful tool for detecting symmetry fractionalization. Recall that projective representations of $G^\0$ can be viewed as ordinary linear representations of a covering group $\hatG^\0$. The disk operator can be constructed by considering a contractible junction of $G^\0$ global symmetry operators which does not lift to a junction of symmetry operators for $\hatG^\0$, and collapsing the symmetry operators as in Figure~\ref{fig:squash}. The disk operator $\scrD_c(\Sigma)$ is supported on an open codimension-1 surface $\Sigma$ and can be labeled by an element $c$ of the subgroup $\Gamma \subset \hatG^\0$ that acts trivially on genuine local operators. The disk operator realizes the symmetry fractionalization via an action on line operators as shown in Figure~\ref{fig:DiskLineaction}. Unlike a 1-form symmetry operator, the disk implements a consistent action on open lines, because the interior of the disk acts in a compensating way on non-genuine endpoint operators that transform  projectively under $G^\0$. As we will discuss in Section~\ref{sec:SymFrac}, the action of the disk operator on lines cannot be removed by local counterterms when the symmetry fractionalization class corresponds to a non-trivial element of  $H^2(BG^\0,U(1))$.\footnote{There are additional cases where the fractionalization class is valued in  $H^2(BG^\0,\IZ_N)$ which can be non-trivial even when there are no projective representations (i.e. $H^2(BG^\0,U(1))=0$).} 

In Sections~\ref{sec:selectionrulesanddegeneracies} and~\ref{sec:matching} we will discuss implications of symmetry fractionalization, some of which are similar to the constraints from exact 1-form symmetries. For instance, in Section~\ref{sec:selectionrules} we demonstrate how, even without 1-form symmetry, line operators with symmetry fractionalization are constrained to obey exact selection rules. Specifically, when a line operator $\scrL$ has a fractionalized charge associated to a projective representation in $H^2(BG^\0,U(1))$, there exist selection rules for $\scrL$ in certain $G^\0$-\emph{twisted sectors}. We show how our general formalism is able to reproduce known selection rules such as those in $SU(N_c)$ quantum chromodynamics (QCD) with $N_f$ fundamental quarks (provided $\gcd(N_c,N_f)\not=1$)~\cite{Cherman:2017tey} and the $\ICP^{N-1}$ model~\cite{Tanizaki:2017qhf} as well as new selection rules, such as for Wilson lines in gauge theories with $O(2)$ global symmetry. 

We further demonstrate in Section~\ref{sec:Degeneracy} how in certain theories with symmetry fractionalization, approximate vacuum degeneracies protected by emergent, spontaneously broken 1-form global symmetries (i.e. topological order) can be recast as \emph{exact} vacuum degeneracies in twisted sectors. These exact, finite volume degeneracies are a consequence of discrete anomalies activated by symmetry fractionalization, which can be matched in the infrared by anomalies involving an emergent 1-form symmetry. However, the degeneracies persist regardless of how badly the would-be 1-form symmetry is explicitly broken. More specifically, we consider the case where the endable, fractionalized lines in question are exactly topological and generate a discrete $(d-2)$-form global symmetry, which through symmetry fractionalization has a mixed anomaly with $G^\0$. This anomaly implies exact vacuum degeneracy in certain Hilbert spaces twisted by $G^\0$. If in the IR the topological lines cannot end and are protected by an emergent 1-form symmetry, this degeneracy is matched by a long-distance effective description with topological order, and topology-dependent vacuum degeneracy in the untwisted Hilbert space.

Finally, in Section~\ref{sec:matching} we discuss how lines with symmetry fractionalization can be tracked along RG flows. A line with symmetry fractionalization can flow to a line charged under an emergent 1-form symmetry, remain non-trivial and endable (e.g. the line could become topological, or conformal), or can be screened, i.e. flow to a worldline quantum mechanics which (aside from the coupling to $G^\0$ symmetry) is decoupled from the bulk. In the first case, the bulk of the disk operator becomes transparent and the disk flows to the codimension-2 generator of the 1-form symmetry in the IR. In general, however, the IR fate of a given line has little bearing on the IR phases of the bulk. Nevertheless, the different possibilities come with distinguishing features, and some options are ruled out in certain bulk phases, or when the lines have certain additional properties. For instance, we argue that 1) any topological line with symmetry fractionalization cannot flow to a decoupled worldline quantum mechanics, and 2) the IR phase cannot be trivially gapped if the 2-point function of fractionalized lines $\scrL$ in certain $G^\0$-twisted sectors is non-vanishing in the large separation limit.

We end our introduction with some avenues for future exploration. In the current work, we focus on the implications of symmetry fractionalization of group-like 0-form symmetries. It is natural to ask about the potential consequences of symmetry fractionalization involving more general categorical symmetries (see~\cite{Hsin:2024aqb} for a recent discussion of symmetry fractionalization for non-invertible symmetries).

Symmetry fractionalization can also occur for spacetime symmetries, like bosonic Lorentz symmetry in relativistic QFT~\cite{Thorngren:2014pza, Kravec:2014aza, Wang:2015fmi, Zou:2017ppq, Wang:2018qoy,  Hsin:2019fhf, Hsin:2019gvb, Ning:2019ffr,Kan:2024fuu,Brennan:2023kpo,Brennan:2022tyl} or fermion number~\cite{Bulmash:2021hmb,Aasen:2021vva}. For instance, in pure gauge theories on non-spin manifolds Wilson and 't Hooft lines nevertheless have a well-defined fermion number~\cite{Wang:2015fmi,Ang:2019txy}. We expect that our construction in Section~\ref{sec:diskoperator} can be used to define a disk operator that measures the spin of such lines. Similarly, line operators can carry fractional quantum numbers associated to how they transform under time-reversal~\cite{Wang:2015fmi,metlitski2015,Geiko:2022qjy,Delmastro:2019vnj}, as probed by placing the theory on non-orientable manifolds. We also expect that one can construct a disk operator that measures these quantum numbers. 

Finally, in this paper we focus on the interplay between line operators and topological symmetry operators for $G^\0$. Within this setting, in order to say anything interesting generically, the group $G^\0$ must admit projective representations. On the other hand, it is well-known that similar structures exist in QCD-like theories with vector-like $U(1)$ baryon-number symmetries~\cite{Anber:2019nze,Anber:2021iip,Tanizaki:2022ngt,Hayashi:2024gxv,Hayashi:2024qkm,Gagliano:2025gwr} despite the fact that $U(1)$ (or any subgroup thereof) does not have projective representations. The full structure involving $U(1)$ symmetries necessarily involves non-flat backgrounds, which go beyond the scope of the present work, but would be interesting to understand in the future. 

\bigskip
\noindent
\emph{Note added: during the preparation of this paper, we were made aware of the forthcoming work~\cite{transmute} which contains some overlapping ideas. }

% \pagebreak
%%%%%%%%%%%%%%%%%%%%%%%%%%%%%%%%%%%%%%%%%%%%%%%%%%%%
\section{Line Operators and Symmetry Fractionalization}
%%%%%%%%%%%%%%%%%%%%%%%%%%%%%%%%%%%%%%%%%%%%%%%%%%%%
\label{sec:SymFrac}

In this section we review the basics of symmetry fractionalization. While much of the discussion is well-known in both high energy (see e.g.~\cite{Hsin:2019gvb, Hsin:2019fhf, Yu:2020twi, Delmastro:2022pfo,Brennan:2022tyl, Bartsch:2023pzl,Bhardwaj:2023wzd}) and condensed matter communities (see e.g.~\cite{Senthil:1999czm, Barkeshli:2014cna, Chen:2014wse, Teo:2015xla, Fidkowski:2016svr, Chen:2016fxq, Barkeshli:2019vtb, Hsin:2019gvb, Hsin:2019fhf,Aasen:2021vva, Bulmash:2021hmb}), we use this section to set the stage, our notation, and highlight some subtle points which are important for the remainder of the paper.

%%%%%%%%%%%%%%%%%%%%%%%%%%%%%%%%%%%%%%%%%%%%%%%%%%%%
\subsection{Lines Neutral Under 1-Form Symmetries}
%%%%%%%%%%%%%%%%%%%%%%%%%%%%%%%%%%%%%%%%%%%%%%%%%%%%
\label{sec:no1formsymfrac}

As discussed in the introduction, our main focus is on theories without exact 1-form symmetries, but with line operators that transform projectively under a faithfully-acting 0-form symmetry which we denote by $G^\0$. In this section we spell out what we mean when we say that a 0-form symmetry `acts' on a line operator $\scrL$. To do so, it is convenient to use the language of topological symmetry operators.

The 0-form symmetry $G^\0$ is generated by topological codimension-1 operators $U_g$. In the absence of other operator insertions, these symmetry operators fuse according to the group composition law $U_g \times U_h = U_{gh}$. In the presence of a line operator, however, this composition law can be modified in a mild but controlled way --- a mere phase. Much of the paper will be concerned with leveraging this modification to derive constraints on line operator correlation functions.

%%%%%%%
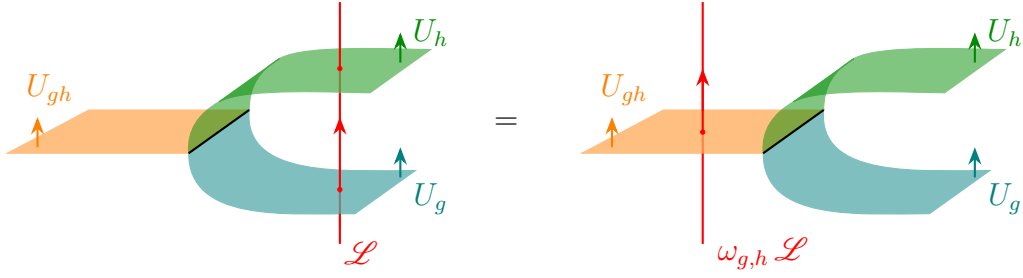
\begin{figure}[t!] %  figure placement: here, top, bottom, or page
   \centering 
    \begin{tikzpicture}[scale=0.8]

        \draw[red,thick] (2.5,-1) -- (2.5,-1.5);
        \draw[thick,red,opacity=0.5] (2.5,-0.65) -- (2.5,-1.25);
       
        \filldraw[teal,opacity=0.5] (0,0) to[out=-90, in=180, looseness=1] (2.75,-1) -- (3.75,-0.285) to[out=180, in=-95, looseness=1] (1,0.715) -- (0,0);

         \filldraw[orange,opacity=0.5] (0,0) -- (1,0.715) -- (-1.65,0.715) -- (-3,0) -- (0,0);
        
        \filldraw[red] (2.5,-0.6) circle (1pt);
         \draw[red,thick] (2.5,1) -- (2.5,-0.65);
        \draw[red,thick, -Stealth] (2.5,0.5) -- (2.5,0.6);
        \draw[thick,red,opacity=0.5] (2.5,1) -- (2.5,1.4);
        
        \filldraw[dgreen,opacity=0.5] (0,0) to[out=95, in=210, looseness=1] (0.5,0.85) -- (1.5,1.565) to[out=210,in=90, looseness=1] (1,0.715) -- (0,0);
         \filldraw[dgreen,opacity=0.5] (0.5,0.85) to[out=30, in=180, looseness=0.5] (3,1) -- (4,1.715) to[out=180, in=30, looseness=0.5] (1.5,1.565) -- (0.5,0.85);

         \filldraw[red] (2.5,1.4) circle (1pt);
         \draw[red,thick] (2.5,2.5) -- (2.5,1.4);
        
        \draw[thick, color = dgreen, -Stealth](3.5,1.5) --(3.5,2);
        \draw[thick, color = teal, -Stealth](3.5,-0.4) --(3.5, 0.1);
        \draw[thick, color = orange, -Stealth](-2.5,0.1) --(-2.5, 0.6);
       
        \draw[thick] (0,0) -- (1,0.715);
       % \draw[thick, -Stealth] (0.5,0.35) -- (0.4,0.27);
       
        \node at (-2.3,1.05) {$\color{orange}{U_{gh}}$};
        \node at (4,-0.75) {$\color{teal}{U_{g}}$};
        \node at (4,2) {$\color{dgreen}{U_{h}}$};
        \node at (2.85,-1.65) {$\color{red}{\scrL}$};

        %\draw[-Stealth] (4.5,0.5) -- (6,0.5);
        
        \node at (5.25,0.5) {$=$};

        \draw[thick,red] (8.5,0) -- (8.5,-1.5);
        \draw[thick,red,opacity=0.5] (8.5,0) -- (8.5,0.35);

        \filldraw[teal,opacity=0.5] (9.5,0) to[out=-90, in=180, looseness=1] (12.25,-1) -- (13.25,-0.285) to[out=180, in=-95, looseness=1] (10.5,0.715) -- (9.5,0);
        \filldraw[orange,opacity=0.5] (9.5,0) -- (10.5,0.715) -- (7.85,0.715) -- (6.5,0) -- (9.5,0);

        \filldraw[red] (8.5,0.35) circle (1pt);
         \draw[thick,red] (8.5,0.35) -- (8.5,2.5);
        \draw[thick,red, -Stealth] (8.5,0.7) -- (8.5,1.4);
        
        \filldraw[dgreen,opacity=0.5] (10,0.85) to[out=30, in=180, looseness=0.5] (12.5,1) -- (13.5,1.715) to[out=180, in=30, looseness=0.5] (11,1.565) -- (10,0.85);
        \filldraw[dgreen,opacity=0.5] (9.5,0) to[out=95, in=210, looseness=1] (10,0.85) -- (11,1.565) to[out=210,in=90, looseness=1] (10.5,0.715) -- (9.5,0);
        \draw[thick, color = teal, -Stealth](13,-0.4) --(13, 0.1);
        \draw[thick, color = dgreen, -Stealth](13,1.5) --(13,2);
        \draw[thick, color = orange, -Stealth](7,0.1) --(7, 0.6);

        \draw[thick] (9.5,0) -- (10.5,0.715);
       % \draw[thick, -Stealth] (10,0.35) -- (9.9,0.27);
       
        \node at (7.2,1.05) {$\color{orange}{U_{gh}}$};
        \node at (13.5,-0.75) {$\color{teal}{U_{g}}$};
        \node at (13.5,2) {$\color{dgreen}{U_{h}}$};
        \node at (9.5,-1.65) {$\color{red}{\omega_{g,h}\,\scrL}$};
        % \draw[black,opacity=0.5] (0,0) to[out=90, in=210, looseness=1] (0.5,-0.85) -- (1.5,-1.565) to[out=210,in=90, looseness=1] (1,0.715) -- (0,0);
    \end{tikzpicture}
   \caption{0-form symmetries can act projectively in the presence of a line operator. The junction of symmetry operators labeled by $g,h\in G^\0$ can act on a line $\scrL$ by a phase $\omega_{g,h}\in U(1)$. The group cohomology class $[\omega]\in H^2(BG^\0,U(1))$ measures the projective representation carried by the line, which can be interpreted as the projective transformation properties of internal (spin-defect) degrees of freedom on the line which are traced out to yield the operator $\scrL$. The arrows denote the orientations of the lines/surfaces.}
   \label{fig:projective_action}
\end{figure}
%%%%%%%

More specifically, consider a (generically non-topological) line $\scrL$ which is \emph{symmetric} under $G^\0$ in the sense of Ref.~\cite{Antinucci:2024izg} --- in other words, $G^\0$ surfaces can pass topologically through $\scrL$.\footnote{For the purpose of this discussion we assume that $G^\0$ does not have an intrinsic action on the set of lines (as can be the case with e.g. charge conjugation and time-reversal symmetries), i.e. it leaves the label of the line intact. For the more general story involving twisted cohomology, see e.g. Refs.~\cite{Barkeshli:2014cna,Delmastro:2022pfo,Bartsch:2023pzl}. } The fusion of $G^\0$ surfaces which are pierced by $\scrL$ can be modified by a phase $\omega: G^{(0)} \times G^{(0)} \to U(1)$, 
\begin{equation}
    U_g \times U_h = \omega_{g,h}\, U_{gh}\,, 
\end{equation}
so that $G^\0$ is realized projectively in the presence of the line.\footnote{The phase depends on $\scrL$, so should carry a label $\omega_{\scrL;g,h}$. We suppress the $\scrL$-dependence of the label unless we need to talk about multiple lines at a time. } Let us explain the origin of this projective phase. Rather than fusing $U_g$ and $U_h$ in one step, we first form a fusion junction $U_g\times U_h \to U_{gh}$. Since the junction is codimension-2, it has a natural linking action on line operators. Indeed, the projective phase $\omega_{g,h}$ arises when we pass the codimension-2 junction across the line $\scrL$, as shown in Figure~\ref{fig:projective_action}. Essentially, the junction of $G^\0$ symmetry operators acts on $\scrL$ in a way which is similar to a 1-form symmetry. But the fact that the junction is not a genuine codimension-2 operator, and must be attached to codimension-1 surfaces, leads to important differences. 

\begin{figure}[t!]
    \centering
        \begin{tikzpicture}[scale=.8]

      \draw[fill, red] (5.5, -1) circle (1pt);
        %\draw [thick, red] (5.5,-1) -- (5.5, 0);
        \draw [thick, red,opacity=0.5] (5.5,-1) -- (5.5, 0);
        
        \filldraw[teal,opacity=0.5] (5.5,-1) [partial ellipse=0:360:1];
  		\draw[teal,opacity=0.5] (5.5,-1) [partial ellipse=180:360:1 and 0.3];
   		\draw[teal,dotted] (5.5,-1) [partial ellipse=0:180:1 and 0.3];
	   	\draw[thick, color = teal, -Stealth](4.9,-0.25) --(4.7,0.05);
		\node[color = teal] at (4,-1) {$U_g$};
		\node[red] at (5.9, -.8) {$\mathcal O$};

        \draw [thick, red, -Stealth] (5.5,0.2) -- (5.5,0.5);       
        \draw [thick, red] (5.5,0) -- (5.5,1.6);
        \draw[fill, red] (5.5,0) circle (1pt); 
       
           \node[red] at (6, 1.2) {$\scrL$};

       % \draw [thick, -Stealth] (7.5, -0.5) -- (8.5,-0.5);

       \node at (8,-0.5) {$=$};
          \draw[fill, red] (10,-1) circle (1pt);
        \draw [thick, red, -Stealth] (10, 0.2) -- (10, 0.5);
        \draw [thick, red] (10,-1) -- (10,1.6);
        
        \node[red] at (10.5, 1.2) {$\scrL$};
        \node[red] at (11, -.8) {$R_g \cdot \mathcal O$};
       \node at (8,-3.5) {(a)};
    \end{tikzpicture}
    \hspace{2cm}
    \begin{tikzpicture}[scale=.7]

     \filldraw[orange,opacity=0.5] (0,0) -- (1,0.715) -- (-1.65,0.715) -- (-3,0) -- (0,0);
        
        \filldraw[dgreen,opacity=0.5] (0.5,0.85) to[out=30, in=180, looseness=0.5] (3,1) -- (4,1.715) to[out=180, in=30, looseness=0.5] (1.5,1.565) -- (0.5,0.85);

        \filldraw[dgreen,opacity=0.5] (0,0) to[out=95, in=210, looseness=1] (0.5,0.85) -- (1.5,1.565) to[out=210,in=90, looseness=1] (1,0.715) -- (0,0);
        \filldraw[teal,opacity=0.5] (0,0) to[out=-90, in=180, looseness=1] (2.75,-1) -- (3.75,-0.285) to[out=180, in=-95, looseness=1] (1,0.715) -- (0,0);

        \draw[thick, color = dgreen, -Stealth](3.5,1.5) --(3.5,2);
        \draw[thick, color = teal, -Stealth](3.5,-.4) --(3.5, 0.1);
        \draw[thick, color = orange, -Stealth](-2.5,0.1) --(-2.5, 0.6);

        \draw[thick] (0,0) -- (1,0.715);
       
      %  \draw[thick, -Stealth] (0.5,0.35) -- (0.4,0.27);
        \draw[red,thick] (.5,1.5) -- (.5, 4);
        \draw[red,thick, -Stealth] (.5,2.5) -- (.5, 2.6);
        \filldraw[red] (.5,1.5) circle (1pt);
        \draw[red, dashed] (0.9,0.2) arc[start angle=-100, end angle=220, radius=1.5cm];
        \draw[red,  -Stealth] (0.81, 0.25) -- (0.8,0.255);

        \node at (-2.15,1.15) {$\color{orange}{U_{gh}}$};
        \node at (4,-0.75) {$\color{teal}{U_{g}}$};
        \node at (4,2.1) {$\color{dgreen}{U_{h}}$};
        \node at (.85,4) {$\color{red}{\scrL}$};
        \node at (.85,1.5) {$\color{red}{\mathcal{O}}$};
            \node at (0.25,-2.5) {(b)};
    \end{tikzpicture}
    \caption{The $G^\0$ symmetry can act projectively on the operators $\CO$ living at endpoints of lines $\scrL$, as in (a). By moving the topological $G^\0$ junction around the endpoint of an open line as in (b), we obtain a consistency condition relating the projective phase $\omega_{g,h}$ to the representation $R$ under which the endpoint operators transform.}
    \label{fig:endpoint}
\end{figure}
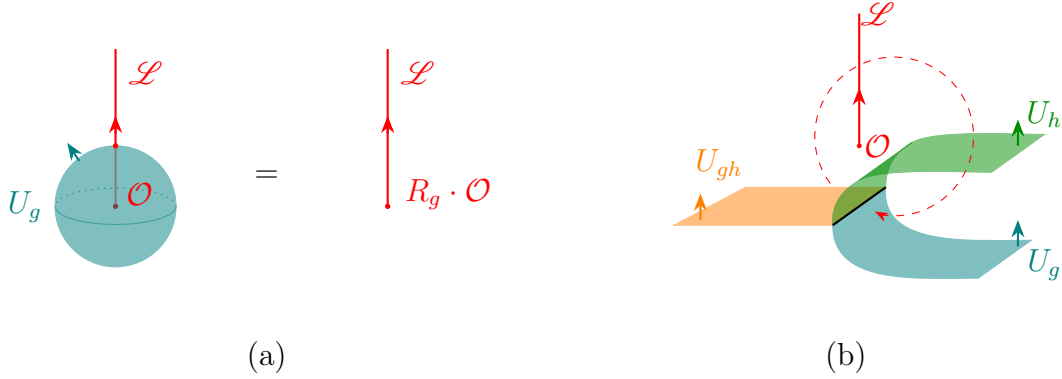

We will mostly be interested in situations where $\scrL$ is neutral under any 1-form symmetries. In this case, there is no obstruction to terminating the line.\footnote{In what follows, we assume that any line which is transparent to all codimension-2 topological operators in the theory can end. For a proof of the converse to this statement, see e.g. Ref.~\cite{Rudelius:2020orz}.  } For instance, a Wilson line in a gauge theory can terminate on a gauge-non-invariant charged matter field. More generally, it does not make sense to talk about such `endpoint operators' on their own, as they only exist at the endpoints of the line $\scrL$. Precisely the same phase $\omega_{g,h}$ appears when we consider the action of $G^\0$ on such open lines. Here we will abuse notation slightly by referring explicitly to an endpoint operator $\mathcal O$ on which $\scrL$ terminates. Such an endpoint operator is neither unique nor arbitrary --- it must transform non-trivially under $G^\0$ as $\mathcal O \to R_g \cdot \mathcal O$, with $R_g$ understood as a linear transformation on the space of endpoints (see Figure \ref{fig:endpoint} (a)). More specifically, the consistency of the topological 0-form symmetry junction in the presence of an open line requires (see Figure~\ref{fig:endpoint} (b))
\begin{equation}\label{endpointphase}
 R_{gh}^{-1}\cdot  R_g \cdot R_h \cdot \mathcal O = \omega_{g,h}\, \mathcal O\,,
\end{equation}
so we can extract the projective phase $\omega$ from the action of $G^\0$ on open lines.

%%%%%%%%%%%%%%%%%%
\subsubsection{Counterterms and Scheme-Dependence}
\label{sec:counterterms}
%%%%%%%%%%%%%%%%%%

The action of 0-form symmetry junctions described above involves contact points where the $G^\0$ operators are pierced by $\scrL$. This is a crucial difference compared to the linking action of a 1-form symmetry on a line, where no such contact points are present. The main consequence is that the phases $\omega_{g,h}$ are sensitive to local counterterms, and hence are scheme-dependent. More specifically, the precise phases one computes can be modified via a local counterterm on $\scrL$ which assigns a $U(1)$-valued phase $\alpha_g$ to the point where the line $\scrL$ intersects $U_g$. This effectively changes $R_g \to \alpha_g^{-1}\, R_g$, and accordingly 
\begin{equation}
 \omega_{g,h} \to  \omega_{g,h}\, (\delta\alpha)_{g,h}^{-1}\,, \quad (\delta\alpha)_{g,h} \equiv \frac{\alpha_g\,\alpha_h}{\alpha_{gh}}\,.  
\end{equation}
If one thinks of the operator $\scrL$ as representing the partition function of a worldline quantum mechanics coupled to the bulk, then the above change in scheme is equivalent to rephasing the unitary operators which implement the $G^\0$ symmetry in said quantum mechanics. 

If there is no choice of $\alpha_g$ that can trivialize all of the $\omega_{g,h}$, then $\omega_{g,h}$ represents a non-trivial group cohomology class $[\omega]\in H^2(BG^\0, U(1))$, which is the invariant data capturing the action of $G^\0$ on $\scrL$. In this case, we say that the line operator has a \emph{worldline ('t Hooft) anomaly} characterized by $[\omega]$. Just as with standard 't Hooft anomalies in QFTs, such worldline anomalies are preserved along symmetry-preserving RG flows and must be matched at all scales. As a result, a line $\scrL$ with a worldline anomaly cannot flow to the trivial identity line, which is incapable of reproducing the projective $G^\0$ action.\footnote{Symmetric boundary conditions are obstructed in QFTs with 't Hooft anomalous global symmetries~\cite{Jensen:2017eof,Thorngren:2020yht}. One may wonder how a worldline anomaly is consistent with the fact that the lines we are considering can end. The resolution is that the space of endpoints of the line transforms not under $G^\0$ but in a linear representation of a central extension $\hatG^\0$. %The non-trivial dimension can be interpreted as the degeneracy associated with the  `spontaneous breaking' of the symmetry at the boundary of the line. \tdb{Not sure how I feel about this last sentence.}
} This perspective on symmetry fractionalization was recently discussed in~\cite{Antinucci:2024izg}, where the authors consider the more general notion of anomalies on extended defects in QFTs. We further discuss anomaly matching constraints in Section~\ref{sec:matching}. 

As with standard 't Hooft anomalies of $d$-dimensional quantum field theories, we can cancel the worldline anomaly of $\scrL$ by placing it at the edge of a classical 2d bulk, 
\begin{equation}
    \scrL(\gamma) \, \exp\left(i \int_\sigma A^*\omega\right)\,, \quad \partial\sigma = \gamma\,,
\end{equation}
where $A$ is the background gauge field for $G^\0$ and $A^*\omega$ is the pullback of the group 2-cocycle $\omega$ to a differential 2-form (here we have switched to additive notation for $\omega$). It is often the case that one couples the Lagrangian to background fields $\widehat{A}$ for an extended group $\hatG^\0$ whose $\Gamma$ subgroup acts trivially, $\hatG^\0/\Gamma = G^\0$. In the case that a projective representation of $G^\0$ is equivalent to an ordinary representation $\mathfrak{R}$ of $\hatG^\0$, it is tempting to write a `counterterm' of the form
\begin{equation}
    \text{Tr}_{\mathfrak{R}}\, \mathcal P \, e^{i \oint_\gamma \widehat{A} }~,
\end{equation}  
to cancel a worldline anomaly. However, this is not a valid counterterm --- instead, it should be interpreted as the partition function of a quantum mechanical system with $\dim(\mathfrak{R})$ degenerate ground states forming a projective representation of $G^\0$. Such a dynamical quantum mechanical theory can cancel the worldline anomaly of the line $\scrL$, but cannot arise dynamically without enlarging the Hilbert space of the QFT (it is the analog of 't Hooft's spectator fermions in~\cite{tHooft:1979rat}).\footnote{We are grateful to Zhengdi Sun for helpful discussions on this point.} 

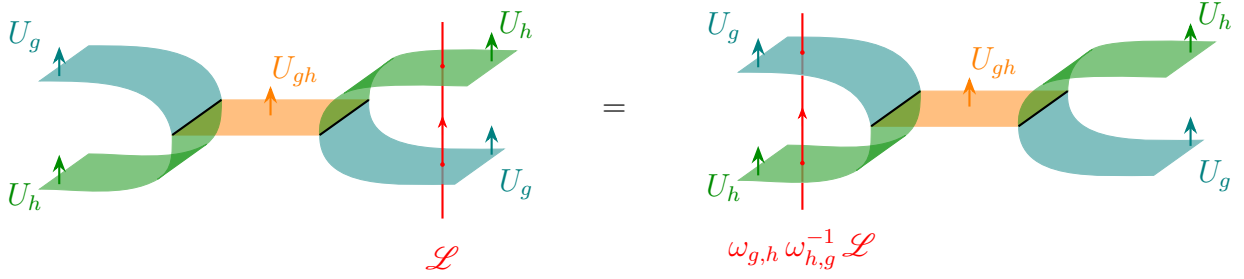
\begin{figure}[t!]
   \centering
    \begin{tikzpicture}[scale=0.65]
    
            \filldraw[orange,opacity=0.5] (-1,0) -- (0,0.715) -- (-3,0.715) -- (-4,0) -- (-1,0);

             \draw[red,thick, opacity=0.5] (1.5,1) -- (1.5,1.4);    
                
        \filldraw[dgreen,opacity=0.5] (-0.5,0.85) to[out=30, in=180, looseness=0.5] (2,1) -- (3,1.715) to[out=180, in=30, looseness=0.5] (.5,1.565) -- (-.5,0.85);
        
        \filldraw[dgreen,opacity=0.5] (-1,0) to[out=95, in=210, looseness=1] (-.5,0.85) -- (0.5,1.565) to[out=210,in=90, looseness=1] (0,0.715) -- (-1,0);
        
        \draw[thick, color = dgreen, -Stealth](2.5,1.5) --(2.5,2.1);
        \draw[thick, color = teal, -Stealth](2.5,-.4) --(2.5, 0.2);
        \draw[thick, color = orange, -Stealth](-2,0.4) --(-2, 1);

        \draw[red,thick, opacity=0.5] (1.5,-1) --(1.5,-0.6);
        
        \filldraw[teal,opacity=0.5] (-1,0) to[out=-90, in=180, looseness=1] (1.75,-1) -- (2.75,-.285) to[out=180, in=-95, looseness=1] (0,0.715) -- (-1,0);

        \draw[thick] (-1,0) -- (0,0.715);
       % \draw[thick, -Stealth] (-0.4,0.42) -- (-0.5,0.34);
        \node at (-1.5,1.3) {$\color{orange}{U_{gh}}$};
        \node at (3,-1) {$\color{teal}{U_g}$};
        \node at (3,2.25) {$\color{dgreen}{U_h}$};

        \filldraw[teal,opacity=0.5] (-4,0) 
        to [out=100, in=0, looseness=1] 
        (-6.7,1.1) --  (-5.7,1.815)
        to [out=0, in=100, looseness=1] 
        (-3,0.715) -- (-4,0);
        
         \filldraw[dgreen,opacity=0.5] (-3.3, -0.185) -- (-4.3,-.9)
        to [out=-150, in=0, looseness=.8] 
        (-6.7,-1.1) --  (-5.7, -.385)
        to [out=0, in=-150, looseness=.8] 
        (-3.3, -0.185) ;

        \filldraw[dgreen,opacity=0.5] (-4,0) 
        to [out=-90, in=30, looseness=1] 
        (-4.3,-.9) -- (-3.3, -0.185)
        to [out=40, in=-90, looseness=1] 
        (-3, 0.715) -- (-4,0);
        
        \draw[thick] (-4,0) -- (-3,0.715);
       % \draw[thick, -Stealth] (-3.4,0.43) -- (-3.5,0.36);

         \draw[thick, color = teal, -Stealth](-6.3,1.2) -- (-6.3,1.8);
        \draw[thick, color = dgreen, -Stealth](-6.3,-1) --(-6.3, -.4);
        \node at (-7,-1.251) {$\color{dgreen}{U_h}$};
        \node at (-7,2) {$\color{teal}{U_g}$};

        \draw[red,thick] (1.5,-1.7) -- (1.5,-1);
    
        \filldraw[red] (1.5,-0.6) circle (1pt);
         \draw[red, thick] (1.5,-0.6) -- (1.5,1);
         \draw[red, -Stealth] (1.5,0) -- (1.5, 0.4);
       
        \filldraw[red] (1.5,1.4) circle (1pt);
        \draw[red, thick] (1.5,1.4) -- (1.5, 2.3);
        \node at (1.5,-2.5) {$\color{red}{\scrL}$};
         %   \draw[-Stealth] (4,0.5) -- (6,0.5);
        \node at (5,0.5) {$=$};
    \end{tikzpicture}
    \hspace{.5cm}
    \begin{tikzpicture}[scale=0.65]
               \filldraw[orange,opacity=0.5] (-1,0) -- (0,0.715) -- (-3,0.715) -- (-4,0) -- (-1,0);
        \filldraw[dgreen,opacity=0.5] (-0.5,0.85) to[out=30, in=180, looseness=0.5] (2,1) -- (3,1.715) to[out=180, in=30, looseness=0.5] (.5,1.565) -- (-.5,0.85);
        
        \filldraw[dgreen,opacity=0.5] (-1,0) to[out=95, in=210, looseness=1] (-.5,0.85) -- (0.5,1.565) to[out=210,in=90, looseness=1] (0,0.715) -- (-1,0);
        
        \draw[thick, color = dgreen, -Stealth](2.5,1.5) --(2.5,2.1);
        \draw[thick, color = teal, -Stealth](2.5,-.4) --(2.5, 0.2);
        \draw[thick, color = orange, -Stealth](-2,0.4) --(-2, 1);
        
        \filldraw[teal,opacity=0.5] (-1,0) to[out=-90, in=180, looseness=1] (1.75,-1) -- (2.75,-.285) to[out=180, in=-95, looseness=1] (0,0.715) -- (-1,0);

        \draw[thick] (-1,0) -- (0,0.715);
      %  \draw[thick, -Stealth] (-0.4,0.42) -- (-0.5,0.34);
        \node at (-1.5,1.3) {$\color{orange}{U_{gh}}$};
        \node at (3,-1) {$\color{teal}{U_g}$};
        \node at (3,2.25) {$\color{dgreen}{U_h}$};

        \draw[red,thick, opacity=0.5] (-5.4,1.06) -- (-5.4,1.5);

        \filldraw[teal,opacity=0.5] (-4,0) 
        to [out=100, in=0, looseness=1] 
        (-6.7,1.1) --  (-5.7,1.815)
        to [out=0, in=100, looseness=1] 
        (-3,0.715) -- (-4,0);
        
         \draw[red, thick, opacity=0.5] (-5.4,-1.13) --(-5.4,-0.75);

         \filldraw[dgreen,opacity=0.5] (-3.3, -0.185) -- (-4.3,-.9)
        to [out=-150, in=0, looseness=.8] 
        (-6.7,-1.1) --  (-5.7, -.385)
        to [out=0, in=-150, looseness=.8] 
        (-3.3, -0.185) ;

        \filldraw[dgreen,opacity=0.5] (-4,0) 
        to [out=-90, in=30, looseness=1] 
        (-4.3,-.9) -- (-3.3, -0.185)
        to [out=40, in=-90, looseness=1] 
        (-3, 0.715) -- (-4,0);
        
                \draw[thick] (-4,0) -- (-3,0.715);
       % \draw[thick, -Stealth] (-3.4,0.43) -- (-3.5,0.36);

        \draw[thick, color = teal, -Stealth](-6.3,1.2) -- (-6.3,1.8);
        \draw[thick, color = dgreen, -Stealth](-6.3,-1) --(-6.3, -.4);
        \node at (-7,-1.251) {$\color{dgreen}{U_h}$};
        \node at (-7,2) {$\color{teal}{U_g}$};

        \draw[red,thick] (-5.4,-1.7) -- (-5.4,-1.15);
       
        \filldraw[red] (-5.4,-0.75) circle (1pt);
         \draw[red, thick] (-5.4,-0.75) -- (-5.4,1.03);
         \draw[red, -Stealth] (-5.4,0) -- (-5.4, 0.4);
       
        \filldraw[red] (-5.4,1.5) circle (1pt);
        \draw[red, thick] (-5.4,1.5) -- (-5.4, 2.3);

                \node at (-5.4,-2.5) {$\color{red}{\omega_{g, h}\,\omega_{h,g}^{-1}\, \scrL}$};

        \node[] at (0, -1.8) {};
    \end{tikzpicture}
       \caption{Two 0-form symmetry defects labeled by elements that commute in $G^\0$ may only commute up to a phase when pierced by a line.}
     \label{fig:commutator} 

\end{figure}

Finally, there are certain combinations of phases which are scheme-independent. For instance, consider a pair of group elements $g,h \in G^\0$ which commute, $g h = hg$. The associated symmetry operators $U_g, \, U_h$ commute when acting on local operators, but not necessarily when acting on lines. Specifically, by fusing $U_g \times U_h \to U_{gh} = U_{hg}$ and then splitting $U_{hg} \to U_h\times U_g$ as shown in Figure~\ref{fig:commutator}, one finds that in the presence of a line,
\begin{equation} \label{eq:commutator} 
U_g \times U_h =   \chi_{g,h}\,U_h\times U_g \,, \  \text{ where }\ \chi_{g,h} \equiv \frac{\omega_{g,h}}{\omega_{h,g}}\,.    
\end{equation}
One can easily verify that this phase is scheme-independent. It will play an important role in Section~\ref{sec:selectionrulesanddegeneracies}.

We can summarize the above discussion as follows: take a theory with 0-form global symmetry $G^\0$ and consider a line operator $\scrL$ which is neutral under any 1-form symmetries. Then, 
\begin{itemize}
    \item There is a possible `symmetry fractionalization' class characterized by an element of $H^2(BG^\0,U(1))$.
    \item The cohomology class of the symmetry fractionalization is scheme-independent, but fixing the representative phases $\omega_{g,h}$ requires a choice of counterterms on the line $\scrL$. 
\end{itemize}

%%%%%%%%%%%%%%%%%%%%%%%%%%%%%%%%%%%%%%%%%%%%%%%%%%%%
\subsection{Lines Charged Under 1-Form Symmetries}
%%%%%%%%%%%%%%%%%%%%%%%%%%%%%%%%%%%%%%%%%%%%%%%%%%%%
\label{sec:1formsymfrac}

We now comment on the notion of symmetry fractionalization for lines which \emph{are} charged under 1-form symmetries. For simplicity of the following discussion we consider the case when the 1-form symmetry is $\ZZ_N$ --- the extension to a general invertible 1-form symmetry $G^\1$ is straightforward. Let $\scrL_q$ denote a line with charge $q$ (defined modulo $N$) under the 1-form symmetry. If $q$ is non-trivial, the 1-form symmetry prevents the line $\scrL_q$ from ending on a local operator~\cite{Gaiotto:2014kfa,Rudelius:2020orz}. As a result, we cannot unambiguously fix the action of $G^\0$ on $\scrL_q$ (specifically, the action of codimension-2 junctions of $G^\0$ symmetry operators on the lines).

Indeed, on top of the possibility of adding local counterterms $\alpha_g \in U(1)$ as discussed above, there is an additional choice we must make: how to decorate the codimension-2 junctions of 0-form symmetry operators with the topological 1-form symmetry operators. Let $V_\lambda$ denote the codimension-2 1-form symmetry operator labeled by the group element $\lambda \in \ZZ_N$, and consider the trivalent junction of 0-form symmetry operators $U_g$ and $U_h$ fusing to $U_{gh}$. To each such fusion junction labeled by pairs of group elements $(g,h)$ we can attach a 1-form symmetry operator $V_{\eta_{g,h}}$, where $\eta: G^\0 \times G^\0 \to \ZZ_N$ is a function obeying the consistency relation $\eta_{h,k}\, \eta_{g,hk} = \eta_{g,h}\,\eta_{gh,k}$ (i.e. it forms a group 2-cocycle --- here we are using the multiplicative notation for $\ZZ_N$).\footnote{This condition is equivalent to having a trivial Postnikov class characterizing the possible combining of the 0- and 1-form symmetries into a 2-group~\cite{Barkeshli:2014cna,Benini:2018reh,Cordova:2018cvg}. } Different choices of $\eta$ will lead to different projective actions of $G^\0$ on the set of lines charged under the $\ZZ_N$ 1-form symmetry. Dressing the 0-form junction in the above way modifies the projective action on the set of lines in a way which is correlated with their 1-form charge,
\begin{equation}
\omega_{\scrL_q;g,h} \to \eta_{g,h}^q \, \omega_{\scrL_q;g,h}  \,, \quad \omega \in U(1)\,, \ \eta \in \ZZ_N\,. 
\end{equation}
Crucially, two lines $\scrL_q,\,\scrL'_q$ with the same 1-form charge may transform by different projective phases $\omega_{\scrL_q} \not= \omega_{\scrL_q'}$, but the \emph{change} in the projective phase induced by modifying the $G^\0$ junction is the same across all line operators with the same charge. 

The above process of activating a 0-form symmetry through a 1-form symmetry has a complementary description in terms of the background gauge fields for the two symmetries~\cite{Hsin:2019fhf,Delmastro:2022pfo,Brennan:2022tyl}. Inserting topological symmetry operators for the 0-form $G^\0$ and 1-form $G^\1$ symmetries is equivalent to turning on flat background gauge fields $A^\1 $ and $B^{(2)} $. Concretely, we triangulate the manifold and assign a $G^\0$ element $A^\1_{ij}$ to each 1-simplex (link) and a $G^\1$ element $B^{(2)}_{ijk}$ to each 2-simplex (plaquette).  Inserting the 1-form symmetry operator at the 0-form junction is equivalent to sourcing
\begin{equation}
   B^{(2)}_{ijk} = \eta(A^\1_{ij}, A^\1_{jk}) \, \quad \text{or} \quad  B^{(2)} = (A^\1)^*\eta\,,
\end{equation}
where $(A^\1)^*\eta$ is the pullback of $\eta$ by the gauge field $A^\1$. This relation means that a generic background for the 0-form symmetry will also activate the 1-form symmetry background, in a way controlled by the 2-cocycle $\eta$. For this relation to be gauge-invariant, the background field for the 1-form symmetry must shift under 0-form background transformations. This reflects the fact that when we topologically deform a junction of 0-form symmetry operators, the 1-form symmetry operator at the junction moves accordingly.

The fact that we have the freedom to modify the action of the 0-form symmetry through a 1-form symmetry via the above `junction decoration' procedure means that the pattern of symmetry fractionalization is not fixed, but instead is a choice~\cite{Barkeshli:2014cna,Delmastro:2022pfo,Brennan:2022tyl,Bartsch:2023pzl}. This does not mean that the different choices are equivalent --- the main point of Ref.~\cite{Delmastro:2022pfo} is that choosing different symmetry fractionalization in general may change the 't Hooft anomalies of the $G^\0$ symmetry (however, this does not mean that different choices of symmetry fractionalization can lead to different IR \emph{dynamics}, since it is the anomaly of the 1-form symmetry that guarantees that any $G^\0$ anomalies activated by symmetry fractionalization can automatically be matched).\footnote{Similarly, changing the choice of fractionalization class does not change the worldline anomalies of charged lines. Any would-be change in the worldline anomaly due to a change in symmetry fractionalization can be canceled by a modified transformation of the background gauge field for the 1-form symmetry as described above.} A particular symmetry fractionalization (and anomaly) pattern can only be fixed by breaking the 1-form symmetry completely. In gauge theories, this entails introducing new gauge-charged degrees of freedom whose quantum numbers pick out a particular fractionalization class. 

The inequivalent choices of the induced symmetry fractionalization are labeled by cohomology classes $[\eta] \in H^2(B G^\0, G^\1)$, i.e. central extensions of $G^\0$ by $G^\1$. This is in general distinct from the group $H^2( BG^\0, U(1))$ of projective representations specifying the possible worldline anomalies. More specifically, it can happen that a symmetry $G^\0$ has no projective representations, but does admit central extensions by a finite group $G^\1$. For instance, neither $U(1)$ nor any of its subgroups admit projective representations, so it is not possible for a line to carry a worldline anomaly for such symmetries. However, since e.g. $H^2(BU(1),\ZZ_N) = \ZZ_N$, it still makes sense to talk about the change in the fractionalization of a $U(1)$ 0-form symmetry via a 1-form $\ZZ_N$ symmetry.

To summarize, in a theory with 0-form symmetry $G^\0$ and 1-form symmetry $G^\1$, 
\begin{itemize}
    \item The freedom to change symmetry fractionalization of lines charged under $G^\1$ is characterized by a class in $H^2(BG^\0,G^\1)$.\footnote{In other words, the symmetry fractionalization classes are a torsor over the group $H^2(BG^\0,G^\1)$.} 
    \item The symmetry fractionalization is not fixed by any consistency condition as long as the 1-form symmetry remains unbroken. However, the fractionalization class can be chosen to match any class that is fixed by a specific deformation which breaks the 1-form symmetry explicitly. 
\end{itemize}

%%%%%%%%%%%%%%%%%%%%%%%%%%%%%%%%%%%%%%%%%%%%%%%%%%%%
\subsection{Topological Lines and Mixed Anomalies}
\label{sec:topological_lines}
%%%%%%%%%%%%%%%%%%%%%%%%%%%%%%%%%%%%%%%%%%%%%%%%%%%%

In the previous subsection, we discussed the freedom of changing the symmetry fractionalization of lines charged under a 1-form symmetry $G^\1$. In this subsection, we consider the case where the lines are uncharged under any 1-form symmetry, but are themselves topological (and hence generate a $(d-2)$-form symmetry $G^{(d-2)}$ which we assume to be group-like). In this case, it is well-known that the projective action of $G^\0$ on the generators of $G^{(d-2)}$ encodes the mixed anomaly between these symmetries~\cite{Tachikawa:2017gyf,Bhardwaj:2022dyt,Bartsch:2023pzl,Brennan:2022tyl}. 

For simplicity, we consider the case when the line $\scrL$ generates a discrete $\ZZ_N^{(d-2)}$ symmetry. In principle, one could imagine adding a local counterterm to $\scrL$ which assigns an arbitrary $U(1)$ phase $\alpha_g$ to the point-like junction of $\scrL$ and the symmetry operator $U_g$. However, there is an additional constraint coming from fusion --- roughly speaking, the fact that 
\begin{equation}\label{linefusionconstraint}
\underbrace{\scrL \times \cdots \times \scrL}_{N \text{ times}} = \mathbbm{1} ~,
\end{equation}
suggests that the phases $\omega_{g,h}$ and allowed counterterms are $\alpha_g$ must be $\ZZ_N$-valued.\footnote{Similar constraints can also be derived for non-topological supersymmetric line operators in SUSY theories without 1-form symmetry. For example, in $4d$ $\CN=2$ gauge theories, 2-point functions of parallel $\half$-BPS line operators are independent of the separating distances due to the SUSY algebra \cite{Gaiotto:2010be} which endows them with a fusion constraint. Generalizing such fusion constraints to generic non-topological lines requires more thought, see~\cite{Gagliano:2025gwr} for related recent work.}  A more careful argument is presented in Appendix~\ref{app:fusion}, where we show that the scheme-independent data one can track in this setting is captured by a central extension $[\widetilde\omega] \in H^2(BG^\0, \ZZ_N)$ rather than a bona fide projective representation $H^2(BG^\0, U(1))$. As discussed in the previous Section, this distinction is especially important when $G^\0$ has no projective representations but admits central extensions by $\ZZ_N$.\footnote{A classic example of this type arises in the fractional quantum hall effect, where the electric charge of anyons is a fraction $1/N$. See for example~\cite{Chen:2016fxq} for a review which emphasizes constraints on fractionalization coming from the fusion of anyons.  } 

To see how this gives rise to a mixed anomaly, note that we can cancel the phase $\widetilde\omega$ on the line $\scrL(\gamma)$ by attaching it to an `anomaly-inflow' surface\footnote{We use quotes here because it may be that $H^2(BG^\0,U(1))$ is trivial, even when the bulk anomaly in Eq.~\eqref{bulkanomalyZNtopological} is non-trivial. We reserve the phrase `worldline anomaly' for the situation where the line defect Hilbert space carries a bona fide projective representation, which only happens when $H^2(BG^\0,U(1))$ is non-trivial. } 
\begin{equation}
    \scrL(\gamma) \, \exp\left( \frac{2\pi i}{N} \int_\sigma (A^\1)^*\widetilde\omega \right) \,, 
\end{equation}
where $A^\1$ is the background gauge field for $G^\0$ and $\partial\sigma = \gamma$. Then we use the fact that inserting $\scrL(\gamma)$ is equivalent (via Poincar\'e duality) to turning on a particular background field for $G^{(d-2)}$, namely $B^{(d-1)} = \delta^{(d-1)}(\gamma)$. The prescription to attach a disk to the line $\gamma$ is then equivalent to writing the following anomaly inflow action in $(d+1)$-dimensions (here we normalize background fields so that the integral computes an integer)
\begin{equation}\label{bulkanomalyZNtopological}
    \mathcal A = \frac{2\pi i}{N}\int_{M_{d+1}}  (A^\1)^*\widetilde\omega \cup  B^{(d-1)}\,.
\end{equation}

As a simple example of a situation where a symmetry is fractionalized on a topological line, consider a topological $\ZZ_N$ gauge theory in 3d coupled to massive `electric' matter. We take the matter to be a complex scalar field $\phi$, so the total action is 
\begin{equation}
    S = \frac{iN}{2\pi} \int a \wedge db + \int d^3x\, |(\partial - i a)\phi|^2 + m^2 |\phi|^2 \,,
\end{equation}
where $a$ and $b$ are $U(1)$ gauge fields. There is a gauge redundancy $a \to a + d\lambda$ with $\phi \to e^{i\lambda}\phi$. The equation of motion of $b$ sets $da = 0$, so that electric Wilson lines 
\begin{equation}
W_q(\gamma) = e^{i q\oint_\gamma a}~,
\end{equation}
are topological, while summing over quantized fluxes $\frac{1}{2\pi}\oint db \in \ZZ$ restricts the holonomies to be $N$th roots of unity, i.e. $W_q^N = 1$. The electric Wilson line generates an exact $\ZZ_N^\1$ `magnetic' 1-form symmetry that acts on the magnetic Wilson line operators $e^{iq\oint b}$. 

The theory also enjoys a $G^\0 = U(1)$ global symmetry generated by $e^{i \theta \oint \frac{db}{2\pi}}$. We can define monopole operators of the $b$ gauge field $\mathcal M_b(x)$ which enforce $\oint_{S^2} db = 2\pi$ for any sphere surrounding the point $x$. Because of the BF-term, such an operator has charge $N$ under the gauge symmetry of the $a$ gauge field. In addition, $\phi$ has $-\frac{1}{N}$ magnetic charge. To see this, one can turn on a background gauge field $A^{(1)}$ for the global symmetry 
\begin{equation}
     \frac{i}{2\pi} \int A^\1 \wedge db\,.
\end{equation}
Then, the action is invariant under the background gauge transformations
\begin{equation}
    A^\1 \rightarrow A^\1  + d\Lambda^\0, \quad 
    a \rightarrow a  - \frac{1}{N}d\Lambda^\0, \quad
    \phi \rightarrow e^{-\frac{i}{N} \Lambda^\0} \phi~.
\end{equation}
Hence, the operator $\phi^N \mathcal M_b^\dagger$ is gauge-invariant and has charge 1 under the $U(1)$ symmetry, while  $\phi(x) e^{i \int_x a}$~ is correlated with a `fractionalized' action of the global symmetry on the Wilson line. The $U(1)$ fractionalization can also be seen from the equation of motion for $a$: $\frac{N db}{2\pi}=\star j_\phi$ where $j_\phi$ is the number current for $\phi$. 
We conclude that local operators have integer charge under the global symmetry, while operators attached to  Wilson lines have fractional charge. This fractionalization is described by the following non-trivial group 2-cocycle in $H^2(BU(1),\ZZ_N)$,
\begin{equation}
    \widetilde\omega_q(\alpha,\beta) = \exp\left( \frac{i q}{N} \left([\alpha] + [\beta] - [\alpha + \beta] \right) \right)\,,
\end{equation}
where $[\alpha] \equiv \alpha \text{ mod } 2\pi$. Note that $\widetilde\omega_q(\alpha,\beta)$ is actually exact if viewed as a $U(1)$ cocycle, since it can be written as $\frac{\nu_q(\alpha)\nu_q(\beta)}{\nu_q(\alpha+\beta)}$ where $\nu_q(\alpha) = e^{\frac{iq}{N}[\alpha]}$ are $U(1)$ phases. However, $\widetilde\omega_q(\alpha,\beta)$ is not the differential of any $\ZZ_N$-valued 1-cochain. The fusion rule obeyed by the topological $\ZZ_N$ Wilson line restricts the counterterms affecting $\widetilde\omega$ to be valued in $\ZZ_N$, so that we cannot cancel the above fractionalized action. 

In fact, the fractional $U(1)$ charge of the symmetry operator for the magnetic $\ZZ_N^\1$ symmetry is an indication that there is a mixed anomaly between these symmetries. The anomaly is characterized by the inflow action 
\begin{equation} \label{eq:U(1)ZN_anomaly}
    \mathcal A = \frac{i}{2\pi} \int dA^\1 \wedge B^{(2)}_m\,.
\end{equation}
where $A^\1$ and $B^{(2)}_m$ are the background gauge fields for the 0-form $U(1)$ and 1-form $\ZZ_N$ magnetic symmetries, respectively. 

Now suppose $m^2 > 0$. At long distances, i.e. at energies below the mass scale of the heavy charged matter, we have a pure $\ZZ_N$ gauge theory with the aforementioned $\ZZ_N^\1$ magnetic 1-form symmetry generated by the electric Wilson line, and an emergent $\ZZ_N^\1$ electric 1-form symmetry generated by the magnetic Wilson line. These symmetries have a mixed anomaly with a simple inflow action
\begin{equation}
    \mathcal A = \frac{iN}{2\pi} \int B^{(2)}_e \wedge B^{(2)}_m\,.
\end{equation}
One can ask how the long distance infrared effective theory matches the anomaly \eqref{eq:U(1)ZN_anomaly} computed in the UV. The answer is that we \emph{choose} a symmetry fractionalization pattern \cite{Delmastro:2022pfo} of the UV $U(1)$ 0-form symmetry by activating the 1-form electric symmetry of the IR via 
\begin{equation}
  \int_\Sigma  B^{(2)}_e = \frac{1}{N} \oint_\Sigma dA^\1~,
\end{equation}
for all 2-cycles $\Sigma$. 

The theory considered above can be viewed as a rough particle-vortex dual of the standard abelian Higgs model with a charge-$N$ Higgs field~\cite{Nguyen:2024ikq}. In the dual frame, the $G^\0 = U(1)$ symmetry is the 0-form `topological' symmetry due to the absence of dynamical monopole-instantons, and the exact $\ZZ_N^\1$ symmetry is due to the fact that the matter fields have charge $N$. 
The $U(1)$ topological symmetry is fractionalized on the Gukov-Witten operator, which ends on a fractional monopole operator. In the Higgs phase such a fractional monopole operator creates an Abrikosov-Nielsen-Olesen vortex with flux $2\pi/N$ (and hence carries fractional magnetic charge).

%%%%%%%%%%%%%%%%%%%%%%%%%%%%%%%%%%%%%%%%%%%%%%%%%%%%
\subsection{The Disk Operator}
%%%%%%%%%%%%%%%%%%%%%%%%%%%%%%%%%%%%%%%%%%%%%%%%%%%%
\label{sec:diskoperator}

\begin{figure}[t!]
    \centering
    \begin{tikzpicture}[scale=1]      
 \filldraw[teal,opacity=0.3] (0,0) [partial ellipse=180:360:1.2] arc (360:180:1.2 and 0.4);
     \filldraw[orange,opacity=.4] (0,0) [partial ellipse=0:360:1.2 and 0.4];
         \filldraw[dgreen,opacity=0.3] (0,0) [partial ellipse=0:180:1.2] arc (180:360:1.2 and 0.4);
        
        \node[orange] at (-1.75,0) {$U_{gh}$};
        \node[dgreen] at (-1.5,0.75) {$U_h$};
        \node[teal] at (-1.5,-0.75) {$U_g$};

        \draw[thick,black] (0,0) [partial ellipse=0:-180:1.2 and 0.4];
        \draw[thick,dashed,black] (0,0) [partial ellipse=0:180:1.2 and 0.4];
        \draw[-Stealth,thick] (2,0) -- (3.2,0);
        
        \filldraw[olive,opacity=.5] (5,0) [partial ellipse=0:360:1.2 and 0.4];
        \draw[black,thick] (5,0) [partial ellipse=0:360:1.2 and 0.4];
        \node[olive] at (4.9,0.75) {$\scrD_{g,h}$};
    \end{tikzpicture}
    \caption{Collapsing $G^{(0)}$ defects to create a disk operator.}
    \label{fig:squash}
 \end{figure}
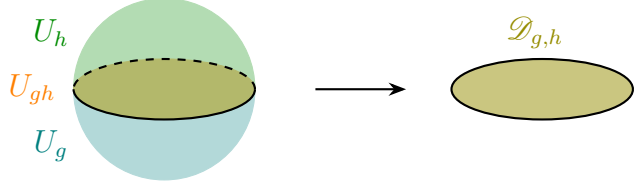

In the previous sections, we considered the action of $G^{(0)}$-junctions on line operators. One can imagine folding such a junction by bringing the three $G^{(0)}$ defects on top of each other. After folding, the bulk of the defect acts trivially on genuine local operators but can still act non-trivially on open and closed lines.  This folding process produces an operator supported on an \emph{open} codimension-1 surface whose bulk \emph{and} boundary are topological, and which can be used to detect the symmetry fractionalization properties of line operators. Alternatively, we can construct such an operator on a disk by squashing $G^{(0)}$ defects as in Figure~\ref{fig:squash}. We refer to this topological operator as \emph{the disk operator}. In this section we determine what labels such disk operators, and present a concrete realization of them in a gauge theory with fundamental matter fields. 

Up until this point, we have phrased the entire discussion in terms of the 0-form symmetry group $G^\0$ that acts faithfully on genuine local operators and projectively on (fractionalized) endable line operators. Projective representations of $G^\0$ can be thought of as ordinary representations of a centrally-extended group $\hatG^\0$, where $G^\0 = \hatG^\0/\Gamma$ and $\Gamma$ is a finite subgroup of the center of $\hatG^\0$ which we also assume is finite. As discussed previously, in the gauge theory context it is more natural to discuss symmetry fractionalization in terms of $\hatG^\0$ rather than $G^\0$ itself, since it is $\hatG^\0$ which acts on the gauge-non-invariant matter fields (while $\Gamma$ coincides with gauge transformations). 

\begin{figure}[t!] %  figure placement: here, top, bottom, or page
   \centering
    \begin{tikzpicture}[scale=0.75]

       \filldraw[orange,opacity=0.5] (-1,0) -- (0,0.715) -- (-3,0.715) -- (-4,0) -- (-1,0);
        \filldraw[dgreen,opacity=0.5] (-0.5,0.85) to[out=30, in=180, looseness=0.5] (1.75,1) -- (2.75,1.715) to[out=180, in=30, looseness=0.5] (.5,1.565) -- (-.5,0.85);
        \filldraw[dgreen,opacity=0.5] (-1,0) to[out=95, in=210, looseness=1] (-.5,0.85) -- (0.5,1.565) to[out=210,in=90, looseness=1] (0,0.715) -- (-1,0);
        
        \draw[thick, color = dgreen, -Stealth](2.5,1.6) --(2.5,2.1);
        \draw[thick, color = teal, -Stealth](2.5,-0.4) --(2.5, 0.1);
        \draw[thick, color = orange, -Stealth](-3.6,0.1) --(-3.6, 0.6);
        
        \filldraw[teal,opacity=0.5] (-1,0) to[out=-90, in=180, looseness=1] (1.75,-1) -- (2.75,-.285) to[out=180, in=-95, looseness=1] (0,0.715) -- (-1,0);
     
        \draw[thick] (-1,0) -- (0,0.715);
        \node at (-4.2,0.5) {$\color{orange}{U_{gh}}$};
        \node at (3,-1) {$\color{teal}{U_g}$};
        \node at (3,2.25) {$\color{dgreen}{U_h}$};
       % \draw[thick, -Stealth] (-0.4,0.42) -- (-0.5,0.34);

        \draw[-Stealth] (4,0.5) -- (5.5,0.5);
        \node at (4.75,0.8) {\footnotesize{lift}};

          \filldraw[orange,opacity=0.5] (9.5,0) -- (10.5,0.715) -- (7.5,0.715) -- (6.5,0) -- (9.5,0);
            \filldraw[olive,opacity=0.5] (9.5,0) -- (10.5,0.715) -- (7.5,1.95) -- (6.5,1.2) -- (9.5,0);
            
        \filldraw[dgreen,opacity=0.5] (10,0.85) to[out=30, in=180, looseness=0.5] (12.25,1) -- (13.25,1.715) to[out=180, in=30, looseness=0.5] (11,1.565) -- (10,0.85);
        \filldraw[dgreen,opacity=0.5] (9.5,0) to[out=95, in=210, looseness=1] (10,0.85) -- (11,1.565) to[out=210,in=90, looseness=1] (10.5,0.715) -- (9.5,0);
        \filldraw[teal,opacity=0.5] (9.5,0) to[out=-90, in=180, looseness=1] (12.25,-1) -- (13.25,-0.285) to[out=180, in=-95, looseness=1] (10.5,0.715) -- (9.5,0);

      %  \draw[thick, -Stealth] (10,0.35) -- (9.9,0.27);
        \draw[thick] (9.5,0) -- (10.5,0.715);
        \node at (7,-.75) {$\color{orange}{\widehat{U}_{\widehat{g}\, \widehat{h}   }}$};
        \node at (13.5,-1) {$\color{teal}{\widehat{U}_{\widehat{g}}}$};
        \node at (13.5,2.25) {$\color{dgreen}{\widehat{U}_{\widehat{h}}}$};
        \node at (7.2,2.3) {$\color{olive}{\mathscr{D}_{c_{g,h}}}$};

        \draw[thick, color = teal, -Stealth](13,-0.4) --(13,0.1);
        \draw[thick, color = dgreen, -Stealth](13,1.6) --(13,2.1);
        \draw[thick, color = orange, -Stealth](6.9,0.1) --(6.9, 0.6);
        \draw[thick, color = olive, -Stealth](6.8,1.2) --(6.6, 0.7);
        
        % \draw[black,opacity=0.5] (0,0) to[out=90, in=210, looseness=1] (0.5,-0.85) -- (1.5,-1.565) to[out=210,in=90, looseness=1] (1,0.715) -- (0,0);
         
    \end{tikzpicture}
       \caption{The disk operator appears when we lift the $G^\0$ junction to $\hatG^\0$.
     \label{fig:liftedjunction}}
\end{figure}
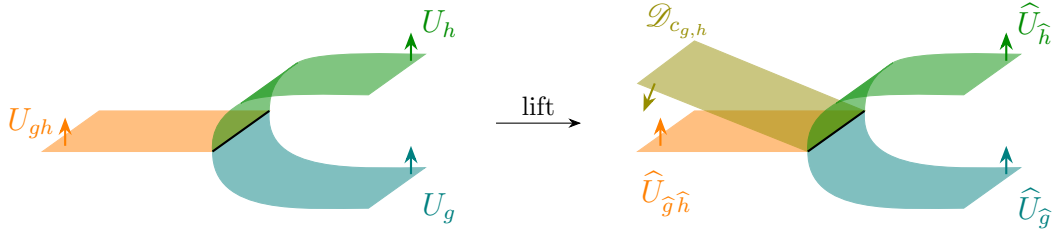

It is most straightforward to describe the disk operator in terms of the extended group $\hatG^\0$. To begin, we choose a lift $\hatg \in \hatG^\0$ for each $g \in G^\0$. In general, the lifted group elements only satisfy the composition law in $\hatG^\0$ up to elements of $\Gamma$:
\begin{equation}
\hatg\, \hath = c_{g,h} \, \hatgh\,, \quad c \in \Gamma\,.
\end{equation}
Here $c_{g,h}$ is a representative of $H^2(BG^\0,\Gamma)$. Now we lift all symmetry operators $U_g$ to operators $\hatU_{\hatg}$ labeled by elements of the extended group. Such operators act in the same way as the $G^\0$ operators on genuine local operators. But because the lifted group elements fail to satisfy the group multiplication in $\hatG^\0$, lifting a trivalent fusion junction in $G^\0$ to $\hatG^\0$ results in  
\begin{equation}
U_g \times U_h  =  U_{gh} \quad  \stackrel{\text{lift}}{\longrightarrow}
 \quad   \hatU_{\hatg} \times \hatU_{\hath} =   \hatU_{\hatgh} \, = \,  \hatU_{c_{g,h}^{-1}\, \hatg \, \hath} \,.
\end{equation}
Now we define \emph{the disk operator}
\begin{equation}
\scrD_{c} \equiv \hatU_{c}~,
\end{equation}
which is nothing but the symmetry operator for the group element $c \in \Gamma \subset \hatG^\0$. Since the faithfully-acting 0-form symmetry is $G^\0$, the disk operator acts trivially on any genuine local operator. This also implies that (as the name suggests) the disk operator can be defined on codimension-1 surfaces with boundary\footnote{Note that we will refer to this operator as a `disk' operator even when it is supported on a closed codimension-1 surface.} --- in which case it can be thought of as the twist defect for the trivially-acting subgroup $\Gamma$.\footnote{Such `non-effectively-acting' symmetry operators were considered in a 2d context in~\cite{Robbins:2022wlr}.} While it is transparent to local operators, the disk acts on lines, and  is crucial for reproducing symmetry fractionalization. Using the disk operator, we can rewrite the fusion of the lifted symmetry operators as (see Figure~\ref{fig:liftedjunction})
\begin{equation}
 \hatU_{\hatg} \times \hatU_{\hath} =    \scrD_{c_{g,h}}^{-1} \, \hatU_{\hatg\, \hath} \,.
\end{equation}
In other words, whenever $c_{g,h}$ is non-trivial, the $G^\0$ junction lifts to a proper $\hatG^\0$ junction plus a disk operator.   

%%%%%%%
%%%%%%%

We can match the action of the $G^\0$ junction on lines in the lifted setup using the disk operator. The disk operator $\scrD_c$ can act on a line $\scrL$ by a phase when the two operators intersect at a point (analogous to Figure~\ref{fig:projective_action}). In other words, we have the action 
\begin{equation} \label{eq:diskaction}
    \scrD_c(D) \, \scrL(\gamma) = \rho_c^{\text{Int}(D,\gamma)}\, \scrL(\gamma)\,,
\end{equation}
where $\rho : \Gamma \to U(1)$ is a one-dimensional representation (depending on the line $\scrL$) of the abelian group $\Gamma$. Loosely speaking, one can say that the line $\scrL$ is `charged' under the disk operator $\scrD_c$, in the sense that pulling the boundary of the disk across the line multiplies it by a phase. 

\begin{figure}[t!]
    \centering
    \begin{tikzpicture}[scale=1.2]
          \draw[thick, red] (-0.95,-1)  -- (-0.95,-0.35);	
            \draw[thick, red,opacity=0.5] (-0.95,-0.35)  -- (-0.95,-0.15);
            \draw[fill = olive,fill opacity=0.5] (0,0) ++ (100:2 and -.15) arc (0:360:.6 and .2);
            \draw[thick, red] (-0.95,0.7)  -- (-0.95,-0.15);
            \draw[fill, red] (-0.95,-0.15) circle (0.5pt);
            \draw[thick, red, -Stealth] (-0.95,0.4)  -- (-0.95,0.41);
            \draw[-Stealth]  (-0.95,-0.35) -- (-0.85,-0.35);
            \node[red] at (-0.5, 0.7) {$\scrL$};
            \node[red] at (-0.6,-1.2) {$\mathcal{O}$};
            \draw[fill, red] (-0.95, -1) circle (1pt); 
            \node[olive] at (0,-0.15) {$\scrD_{c}$};
        
            \draw[thick, red] (2.75,2.6)  -- (2.75,0.9);
            \draw[thick, red, -Stealth] (2.75,2)  -- (2.75,2.1);
            \draw[fill, red] (2.75, 0.9) circle (1pt); 
            \node[red] at (3.15,0.7) {$\mathcal{O}$};
            \node[red] at (3.4, 2.6) {$\rho_c\, \scrL$};
            \draw[fill = olive,fill opacity=0.5] (4.75,1.9) ++ (100:2 and -.15) arc (0:360:.6 and .2);            
            \node[olive] at (4.5, 1.4) {$\scrD_{c}$};
            \draw[-Stealth] (3.8,1.55) -- (3.9,1.55);

            \draw[thick, red, thick] (2.75,-0.8)  -- (2.75,-2.5);
            \draw[thick, red, -Stealth] (2.75,-1.4)  -- (2.75,-1.3);
            \draw[fill, red] (2.75, -2.5) circle (1pt); 
            \node[red] at (3.4,-2.5) {$\widehat{R}_c\,\mathcal{O}$};
            \node[red] at (3.25, -1) {$\scrL$};
            \draw[fill = olive,fill opacity=0.5] (3.7,-3) ++ (100:2 and -.15) arc (0:360:.6 and .2);            
            \node[olive] at (3.45,-3.5) {$\scrD_{c}$};
            \draw[-Stealth] (2.75,-3.35) -- (2.85,-3.35);

            \draw[thick, red, thick] (7.5,-1)  -- (7.5,0.7);
            \draw[thick, red, -Stealth] (7.5,0.2)  -- (7.5, 0.21);
            \draw[fill, red] (7.5, -1) circle (1pt); 
            \node[red] at (7.9,-1) {$\mathcal{O}$};
            \node[red] at (7.9, 0.7) {$\scrL$};
            \node[red] at (6.5,-.2) {$\rho_c= \widehat{R}_c$};
            
      %      \draw[-Stealth] (-.5,1.75) to[in=170, out=40,looseness=1] (1.75,2.25);
       %     \draw[-Stealth] (-.5,-2) to[in=-170, out=-40,looseness=1] (1.75,-2.5);
        %    \draw[-Stealth] (5,-2.5) to[in=-140, out=-10,looseness=1] (7.25,-2);
         %   \draw[-Stealth] (5,2.25) to[in=140, out=10,looseness=1] (7.25,1.75);
        
        	\node[rotate = 30] at (1,1.5) {$=$};
			\node[rotate = -30] at (1,-2) {$=$};
			\node[rotate = -30] at (5.5,1.5) {$=$};
			\node[rotate = 30] at (5.5,-2) {$=$};
         
        \end{tikzpicture}
        \caption{The two ways of un-piercing the disk operator --- through the line (above) or through the end point (below) --- must lead to the same action on the open line. This relates the action $\rho_c$ of the disk on closed lines to the action $\widehat{R}_c$ on its endpoints. \label{fig:DiskOpenLineAction}}
\end{figure}
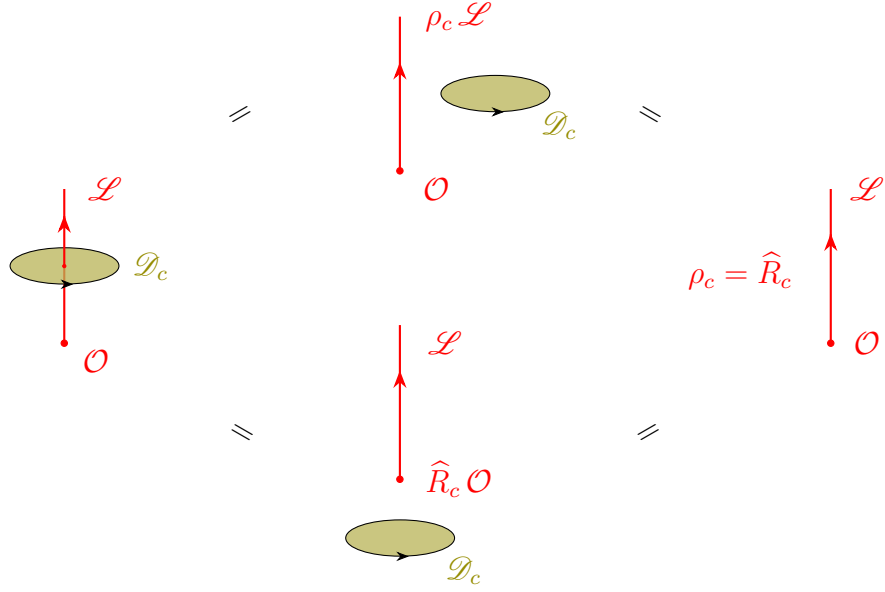

Next, we note that the lifted symmetry operators have an action on open lines which is analogous to Figure~\ref{fig:endpoint} --- except now the endpoints can be regarded as transforming in linear representations of $\hatG^\0$, so that $\CO \to \widehat{R}_{\hatg}\cdot \CO$, with\footnote{In general, this equation will only hold up to an exact 2-cocycle. Here we have chosen counterterms on the line such that $\widehat{R}$ is a true group homomorphism.}
\begin{equation} \label{eq:linearGhataction}
 \widehat{R}_{\hatg}\cdot \widehat{R}_{\hath}\cdot \CO = \widehat{R}_{\hatg\, \hath}\cdot \CO   \,.
\end{equation}
Suppose we start in a configuration where an open line $\scrL$ pierces the disk $\scrD_c$. There are two ways of removing the disk --- by moving its boundary across the line, or by deforming its interior across the endpoint $\CO$ (see Figure~\ref{fig:DiskOpenLineAction}). The former move multiplies the line $\scrL$ by $\rho_c$, while the latter move multiplies the endpoint operator by $\widehat{R}_c$. Consistency requires these two phases to be equal, 
\begin{equation}
    \rho_c = \widehat{R}_c\,.
\end{equation}

Now we return to the lifted junction. If we assume we have chosen counterterms such that Eq.~\eqref{eq:linearGhataction} holds, the valid $\hatG^\0$ junctions themselves have a trivial action on the line $\scrL$. With this choice of scheme, the projective action is then completely captured by the action \eqref{eq:diskaction} of the disk operator on $\scrL$. This allows us to read off the projective phase from Section~\ref{sec:no1formsymfrac}, 
\begin{equation}
    \omega_{g,h} = \rho_{c_{g,h}}\,. 
\end{equation}
The disk operator is therefore a convenient way of encoding the projective transformation properties of a line.  Let us next comment on the ambiguity of this action. 

First, as we have already mentioned repeatedly, the action of the disk on lines is subject to counterterm ambiguities. Above we partially fixed a scheme such that the open line operators transform in true, linear representations of $\hatG^\0$. This leaves the freedom to redefine $\widehat{R} \to \lambda\, \widehat{R}$ where $\lambda$ is any one-dimensional representation of $\hatG^\0$. Such a one-dimensional representation need not represent $\Gamma \subset \hatG^\0$ trivially, so this change in scheme affects the phase in Eq.~\eqref{sec:no1formsymfrac}.\footnote{Note that, the one dimensional representations of $\hatG^\0$ that are not representations of $G^\0$ (classified by $\Gamma$-valued elements of $H^1(B\hatG^\0,U(1))$) are  not necessarily projective representations of $G^\0$ (classified by elements of $H^2(BG^\0,U(1))$).} But any such $\lambda$ represents the 2-cocycle $c_{g,h}$ as something which is exact, namely
\begin{equation}
  \lambda_{\hatg}\, \lambda_{\hath} = \lambda_{\hatg \, \hath} = \lambda_{c_{g,h}\, \hatgh} =   \lambda_{c_{g,h}}\, \lambda_{\hatgh} \,. 
\end{equation}
Therefore, the effect of these counterterms is to multiply the projective phase $\omega$ by something exact
\begin{equation}
    \omega_{g,h} \to \omega_{g,h} \, \frac{\lambda_{\hatg}\,\lambda_{\hath}}{\lambda_{\hatgh}}\,.
\end{equation}
Correspondingly, in order for $\omega_{g,h}$ to be cohomologically non-trivial, the dimension of $\widehat{R}$ must be greater than one. This in turn, requires $\hatG^\0$ to be non-abelian. 

A second source of ambiguity enters in the choice of lift from $G^\0$ to $\hatG^\0$, which is not unique. Changing this lift affects all lines uniformly, through the change in $c_{g,h} \to c_{g,h} \, (\delta b)_{g,h}$. As above, this multiplies $\omega$ by something exact
\begin{equation}
    \omega_{g,h} \to \omega_{g,h} \, \frac{\rho_{b_g}\, \rho_{b_h}}{\rho_{b_{gh}}}\,,
\end{equation}
without affecting the cohomology class $[\omega]$. 

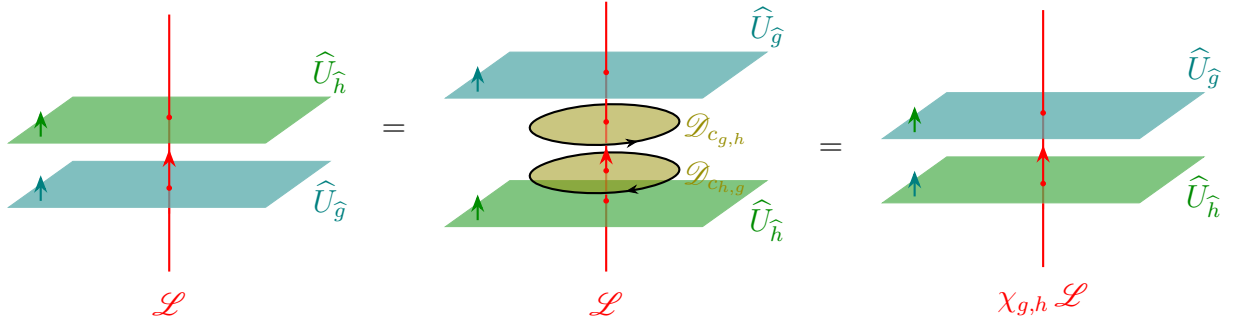
\begin{figure}
\centering
  \begin{tikzpicture}[scale=.85]

   \draw [thick, red, opacity=0.5] (5.5,0.0) -- (5.5,0.4);
        \filldraw[dgreen ,opacity=0.5] (7,0) -- (8,0.715) -- (4,0.715) -- (3,0) -- (7,0);
        \draw[thick, color = dgreen, -Stealth](3.5,0.1) --(3.5, 0.5);

         \draw [thick, red, opacity=0.5] (5.5,-0.7) -- (5.5,-1.1);
        \filldraw[teal ,opacity=0.5] (7,-1) -- (8,-0.285) -- (4,-0.285) -- (3,-1) -- (7,-1);
         \draw[thick, color = teal, -Stealth](3.5, -0.9) --(3.5, -0.5);
        \node[color = dgreen] at (8,1.1) {$\hatU_{\hath}$};
        \node[color = teal] at (8,-0.9) {$\hatU_{\hatg}$};

        \draw [thick, red] (5.5,0.4) -- (5.5,2);
        \draw[fill, red] (5.5,0.4) circle (1pt); 
       
        \draw[fill, red] (5.5, -0.7) circle (1pt);
        \draw [thick, red] (5.5,-0.7) -- (5.5, 0);
       
        \draw[thick, red] (5.5, -1.01) -- (5.5, -2);       
        \draw [thick, red, -Stealth] (5.5,-0.2) -- (5.5,-0.1);
        \node[red] at (5.5, -2.5) {$\scrL$};
         \node at (9, 0.2) {$=$};
    \end{tikzpicture}
\hspace{.1cm}
  \begin{tikzpicture}[scale=.85]

          \draw [thick, red, opacity=0.5] (5.5,-0.64) -- (5.5,-1);
  
      \filldraw[dgreen ,opacity=0.5] (7,-1) -- (8,-0.285) -- (4,-0.285) -- (3,-1) -- (7,-1);
         \draw[thick, color = dgreen, -Stealth](3.5, -0.9) --(3.5, -0.5);

       \draw[thick,red, opacity=0.5] (5.5,-.45) -- (5.5,-.13);
           
           \draw[black,thick, fill= olive, fill opacity=0.5] (4.6,0.75) to[out=20, in=20, looseness=1.5](6.35,.425) to[out=200, in=200, looseness=1.5] (4.6,0.75); 
           
             \draw [thick, red, opacity=0.5] (5.5,0.3) -- (5.5,0.65);

            \draw[black,thick, fill= olive, fill opacity=0.5]  (4.6,0) to[out=20, in=20, looseness=1.5](6.35,-.325) to[out=200, in=200, looseness=1.5] (4.6,0); 
    
                \draw [thick, red, opacity=0.5] (5.5,1) -- (5.5,1.4);
             
        \filldraw[teal ,opacity=0.5] (7,1) -- (8,1.715) -- (4,1.715) -- (3,1) -- (7,1);
        \draw[thick, color =teal, -Stealth](3.5,1.1) --(3.5, 1.5);

    	\node[olive] at (7.2,0.5) {$\scrD_{c_{g,h}}$};
		\node[olive] at (7.2,-.25) {$\scrD_{c_{h,g}}$};
        \node[color = teal] at (8,2.1) {$\hatU_{\hatg}$};
        \node[color = dgreen] at (8,-0.9) {$\hatU_{\hath}$};

		\draw [thick, red] (5.5,1.4) -- (5.5,2.5);
		\draw[fill, red] (5.5,1.4) circle (1pt);
    
        \draw [thick, red] (5.5,0.65) -- (5.5,.99);
        \draw[fill, red] (5.5,0.63) circle (1pt);

        \draw [thick, red] (5.5,-0.1) -- (5.5, 0.26);
        \draw [thick, red, -Stealth] (5.5,0.15) -- (5.5,0.24);
         \draw[fill, red] (5.5, -0.13) circle (1pt);
         \draw[thick,red] (5.5,-0.6) -- (5.5,-0.5);
        \draw[fill, red] (5.5, -0.6) circle (1pt);
        \draw[thick, red] (5.5, -1) -- (5.5, -1.7);       
        
        \draw[-Stealth] (5.85,-.443) -- (5.8,-.452);
        \draw[-Stealth] (5.95,.32) -- (6,.329);

        \node[red] at (5.5, -2.2) {$\scrL$};
         \node at (9, 0.2) {$=$};

    \end{tikzpicture}
    \hspace{.1cm}
  \begin{tikzpicture}[scale=.85]

    \draw [thick, red, opacity=0.5] (5.5,0) -- (5.5,0.4);
        \filldraw[teal ,opacity=0.5] (7,0) -- (8,0.715) -- (4,0.715) -- (3,0) -- (7,0);
        \draw[thick, color = dgreen, -Stealth](3.5,0.1) --(3.5, 0.5);
        
         \draw [thick, red, opacity=0.5] (5.5,-0.7) -- (5.5,-1.1);
        \filldraw[dgreen ,opacity=0.5] (7,-1) -- (8,-0.285) -- (4,-0.285) -- (3,-1) -- (7,-1);
         \draw[thick, color = teal, -Stealth](3.5, -0.9) --(3.5, -0.5);
        \node[color = teal] at (8,1.1) {$\hatU_{\hatg}$};
        \node[color = dgreen] at (8,-0.9) {$\hatU_{\hath}$};

        \draw [thick, red] (5.5,0.4) -- (5.5,2);
        \draw[fill, red] (5.5,0.4) circle (1pt); 
      
        \draw[fill, red] (5.5, -0.7) circle (1pt);
        \draw [thick, red] (5.5,-0.7) -- (5.5, 0);
       
        \draw[thick, red] (5.5, -1.01) -- (5.5, -2);       
        \draw [thick, red, -Stealth] (5.5,-0.2) -- (5.5,-0.1);
        \node[red] at (5.5, -2.5) {$\chi_{g,h}\,\scrL$};
    \end{tikzpicture}
    \caption{Symmetry operators which commute in the absence of lines may not commute when pierced by line operators. Equivalently, $G^\0$ elements which commute may only commute up to $\Gamma$ when lifted to $\hatG^\0$. In terms of symmetry defects this non-commutativity is captured by a pair of disk operators, which can act non-trivially on lines by a phase. \label{fig:commutator_disk}}
\end{figure}

Various manipulations involving the symmetry operators $U_g$ can be performed using the lifted symmetry operators $\hatU_{\hatg}$ together with the disk operators $\scrD_c$. Take, for instance, the fact that $G^\0$ symmetry operators which commute in their action on local operators may not commute in their action on lines (recall Eq.~\eqref{eq:commutator} and Figure~\ref{fig:commutator}). This is reflected in the fact that the lifts $\hatg,\hath$ of commuting elements in $G^\0$ will in general only commute up to $\Gamma$. This means that the lifted symmetry operators commute up to disk operators, which can act on lines. See Figure~\ref{fig:commutator_disk}. 

%%%%%%%%%%%%%%%%%%%%%%%%%%%%%%%%%%%%%%%%%%%%%%%%%%%%
\subsection{Explicit Realization: Scalar QED with $N$ Flavors}
%%%%%%%%%%%%%%%%%%%%%%%%%%%%%%%%%%%%%%%%%%%%%%%%%%%%
\label{sec:qed_junction} 

To ground the above discussion, we now turn to a specific setting where we can explicitly realize the topological junctions of 0-form symmetry operators, as well as the `disk operator' described in the previous section. Consider a $U(1)$ gauge theory with $N$ complex scalar fields $\Phi^I$, $I = 1,\ldots, N$ with charge $1$ under the gauge group. We assume the scalar fields have equal masses and we choose a potential such that the Lagrangian is invariant under the $SU(N)$ transformation that rotates them. However, the 0-form symmetry that acts faithfully on gauge-invariant operators is $PSU(N) = SU(N)/\ZZ_{N}$. A suitable potential can drive the scalars to condense, Higgsing the gauge field and (for spacetime dimension $d>2$) spontaneously breaking $PSU(N) \to S[U(1) \times U(N-1)]$, leading to a sigma model with $\mathbb{CP}^{N-1}$ target space. The total symmetry structure of the theory is
\begin{equation}
\label{eq:SymSctr}
U(N) = \frac{U(1)_g \times SU(N)}{\IZ_{N}}\,. 
\end{equation} 
As a result, the $U(1)$ gauge field violates the standard Dirac quantization condition in the presence of particular backgrounds for $PSU(N)$: 
\begin{equation}\label{eq:fractional_flux}
    \int_\Sigma \frac{da}{2\pi} = \frac{1}{N}\int_\Sigma w_2(A)  \text{ mod } \IZ\,,
\end{equation}
where $A$ is the $PSU(N)$ background gauge field and $w_2(A)$ is the class obstructing its lift to a $SU(N)$ bundle. The main purpose of this section is to explain the origin of this constraint in a concrete setting, and explain its connection to the junctions of the topological symmetry operators generating the $PSU(N)$ symmetry. 

To proceed, we cover the spacetime manifold $M$ with open sets $\{\mathcal U_i \}$ labeled by an ordered list $i_1 < i_2 < \cdots$ and choose an associated partition of spacetime into closed regions (i.e. simplices) $\{\sigma_i \}$ such that $\sigma_i \subset \mathcal U_i$, and the overlaps $\sigma_{i_1\cdots i_k} \equiv \sigma_{i_1} \cap \cdots \cap \sigma_{i_k} \subset \mathcal U_{i_1} \cap \cdots \cap \mathcal U_{i_k}$ (note that $\sigma_{i_1\cdots i_k}$ are codimension $k-1$). We describe the gauge bundle in terms of differential cohomology~\cite{Alvarez:1984es,Freed:2006yc,Freed:2006ya,Kapustin:2014gua, MooreDiffCoh}. The data are a real-valued 1-form gauge field $a_i$ on every patch $\mathcal U_i$, real-valued transition functions $\phi_{ij}$ on double overlaps, and constants $m_{ijk}$ on triple overlaps such that 
\begin{equation}
\begin{split}
    (\delta a)_{ij} &= a_j - a_i =  d\phi_{ij}\,, \quad (\delta\phi)_{ijk} = \phi_{jk} - \phi_{ik} +\phi_{ij} = 2\pi m_{ijk}\,, \\
    (\delta m)_{ijk\ell} &= m_{jk\ell} - m_{ik\ell} + m_{ij\ell} - m_{ijk} = 0 \,. 
\end{split}
\end{equation} 
The last equality holds because the coboundary operator $\delta$ is nilpotent, $\delta^2 = 0$. The first Chern number is given as a sum of $m_{ijk}$ over a closed surface, 
\begin{equation}
\int_\Sigma \frac{da}{2\pi}  = \sum_{i< j < k} \left. (-1)^{s_{ijk}(\Sigma)}\, m_{ijk} \right|_{\sigma_{ijk} \cap \Sigma}\,,
\end{equation}
where $s_{ijk}(\Sigma) = 0,1$ depending on whether traversing $\mathcal U_i \to \mathcal U_j \to \mathcal U_k$ agrees with the orientation of $\Sigma$. In a pure $U(1)$ gauge theory it is conventional to take $m_{ijk} \in \ZZ$.

In the absence of a background gauge field for the global symmetry, the numbers $m_{ijk}$ are integer valued. Small gauge transformations act as 
\begin{equation}
    a_i \rightarrow a_i + d\lambda_i, \quad \phi_{ij} \rightarrow \phi_{ij} +(\delta \lambda)_{ij} ~,
\end{equation}
while large gauge transformations act as
\begin{equation}
\label{LargeGauge}
    \phi_{ij} \rightarrow \phi_{ij} + 2\pi \ell_{ij},\quad m_{ijk} \to m_{ijk} + (\delta\ell)_{ijk} \quad \ell_{ij} \in \mathbb{Z}~.
\end{equation}
The gauge-invariant Wilson line on a curve $\gamma$ is defined by summing integrated segments of the local 1-form gauge field $a$ in each path which end on the transition functions on the double overlaps, 
\begin{equation} \label{eq:gaugeinvariantW} 
    W_q(\gamma) = \exp\left(i q \sum_i \int_{\gamma \cap \sigma_i } a_i\right)\exp\left(i q \sum_{i<j}\left.(-1)^{s_{ij}(\gamma)}\, \phi_{ij}\right|_{\gamma \cap \sigma_{ij}} \right) \,,
\end{equation}
where $s_{ij}(\gamma) =0,1$ depending on whether traversing $\mathcal U_i \to \mathcal U_j$ agrees with the direction of $\gamma$. Of course, gauge-invariance requires $\gamma$ to be closed and $q \in \ZZ$. 

Let us now turn on a background gauge field for the global symmetry. For the purposes of our discussion, we only turn on flat background fields where the local $PSU(N)$ field strength vanishes. Then all of the data of the background field can be captured by transition functions which glue the matter fields from one patch to another (and a chain of consistency conditions on higher overlaps). 

Though the faithfully acting symmetry is $PSU(N)$, the gauge non-invariant fields in the Lagrangian transform under its simply-connected cover $SU(N)$. As a result, the gluing conditions for the Higgs fields $\Phi_i$ in each patch involve \emph{lifts}  of $G_{ij}\in PSU(N)$ transition functions to $G_{ij}\in SU(N)$, 
\begin{equation}
    \Phi_i = e^{-i \phi_{ij}}\, \hatG_{ij}\, \Phi_j \, ,
\end{equation}
where we are suppressing the $SU(N)$ indices $I,J$. 
Note that for consistency $\phi_{ji} = -\phi_{ij}$ and $\hatG_{ji} = \hatG_{ij}^\dagger$, and the above equation is invariant under the gauge redundancies of each patch provided $\Phi_i \to e^{i \lambda_i }\,\Phi_i$. We can use the transition functions to effectively insert topological symmetry operators for the $G^\0$ global symmetry. To see this, consider activating $\hatG_{ij}$ on a sequence of double-overlap regions such that the associated $\sigma_{ij}$'s form a closed codimension-1 surface. Then, gauge-invariant charged operators such as $\Phi^I (\Phi^\dagger)^J$ will be related on the two sides of the interface by the $G^\0$ action. More precisely, we are in the situation described in Section~\ref{sec:diskoperator} and are describing the $G^\0$ symmetry operators in terms of a lift to $\hatG^\0$. In addition to gauge redundancies, we can make a field redefinition  
\begin{equation} \label{eq:Phi_redef} 
    \Phi_i = \hatU_i\, \Phi_i'\,,
\end{equation}
with $\hatU_i \in SU(N)$. Then the redefined fields obey
\begin{equation}
\Phi_i' = e^{-i \phi_{ij}}\, \hatG_{ij}'\, \Phi_j'\,, \text{ where } \hatG_{ij}' = \hatU_i^\dagger \, \hatG_{ij}\, \hatU_j\,. 
\end{equation}
Hence, by performing a field redefinition of the Higgs fields $\Phi$ in a spacetime region $X$, we can effectively deform the network of codimension-1 defects by a boundary $\partial X$. This is the sense in which the symmetry defects are topological. Below we will analyze the consistency conditions that must be obeyed by a network of topological defects described by a collection of $\hatG_{ij}$. 

By relating $\Phi_k$ to $\Phi_i$ in two different ways on the triple overlap $\CU_i \cap \CU_j \cap \CU_k$, we obtain the cocycle condition for the gauge and flavor transition functions:
\begin{equation}
\hatG_{ik}^\dagger\, \hatG_{ij}\, \hatG_{jk} = \mathbbm{1}\, e^{i (\delta\phi)_{ijk}}\,. 
\end{equation}
Taking the determinant, we find that $e^{i (\delta\phi)_{ijk}} \in \ZZ_{N}$. Hence, $\delta \phi$ can be decomposed
\begin{equation} \label{eq:modified_coboundary}
    (\delta\phi)_{ijk} = \frac{2\pi}{N} \Lambda_{ijk} + 2\pi m_{ijk}\,,
\end{equation}
with $\Lambda_{ijk}, m_{ijk} \in \ZZ$. Note that we sum over $m_{ijk}$ in the path integral, while $\Lambda_{ijk}$ mod $N$ is fixed by the global symmetry background, $\hatG_{ik}^\dagger\, \hatG_{ij}\, \hatG_{jk} = \mathbbm{1}\,e^{\frac{2\pi i}{N} \Lambda_{ijk}}$ (shifting $\Lambda \to \Lambda + N$ can be absorbed by $m \to m - 1$). The modified cocycle condition implies that 
\begin{equation}
    \int_\Sigma \frac{da}{2\pi} =\frac{1}{N} \sum_{i<j<k}  \left. (-1)^{s_{ijk}(\Sigma)} \Lambda_{ijk}\right|_{\sigma_{ijk} \cap \Sigma } \text{ mod } \IZ\,.
\end{equation}
Comparing with Eq.~\eqref{eq:fractional_flux}, we can identify $\Lambda_{ijk}$ mod $N$ with $w_2(A)$, the obstruction class for lifting the $PSU(N)$ bundle to an $SU(N)$ bundle. That this obstruction is cohomological in nature is reflected in the fact that we can perform a redefinition of the transition functions
\begin{equation} \label{eq:phi_redef} 
\phi_{ij} = \phi_{ij}' + \frac{2\pi}{N}V_{ij}\,,
\end{equation}
so that the primed quantities see the \emph{background} gauge-transformed 
\begin{equation}\label{eq:large_bgd_gauge}
    \hatG_{ij}'  = e^{-\frac{2\pi i}{N} V_{ij}}\, \hatG_{ij}\,,\quad  \Lambda_{ijk}' = \Lambda_{ijk} + (\delta V)_{ijk}\,,
\end{equation}
where $V_{ij} \in \ZZ$ parameterizes (mod $N$) the trivially-acting center of $SU(N)$. Finally, By taking the coboundary of the modified cocycle condition \eqref{eq:modified_coboundary} obeyed by $\phi_{ij}$, we find 
\begin{equation}
(\delta \Lambda)_{ijk\ell} = 0 \text{ mod } N\,,
\end{equation}
so $\Lambda$ is closed mod $N$.\footnote{Using $\Lambda$ we can define the following integer on quadruple overlaps
\begin{equation}
\beta_{ijk\ell} \equiv \frac{1}{N} (\delta \Lambda)_{ijk\ell} \,.
\end{equation}
Clearly $\delta\beta = 0$, but $\beta$ is not necessarily equal to the coboundary of some integer 2-cochain (while $N\beta$ is). Hence, $\beta_{ijk\ell}$ (the Bockstein of $\Lambda_{ijk}$) represents a class in $H^3(M, \ZZ_{N})$.}

We now explain how to fit this concrete realization of $PSU(N)$ symmetry backgrounds into the general discussion in Section~\ref{sec:diskoperator}. To do this, we fix a choice of lift $\widehat{g} \in SU(N)$ for every element $g =\in PSU(N)$, which we use in the transition functions for the matter fields. In general, the lifts satisfy $\widehat{g}\,\widehat{h} = c_{g,h}\, \widehat{gh}$, with $c_{g,h} \in \ZZ_{N}$. Now consider a triple overlap $\CU_i \cap \CU_j \cap \CU_k$ of patches where we set $\hatG_{ij} =\hatg, \hatG_{jk} = \hath, \hatG_{ik} = \hatgh$. In such a configuration we have 
\begin{equation}
e^{\frac{2\pi i}{N} \Lambda_{ijk}} = c_{g,h}\,,
\end{equation}
which for non-trivial $c_{g,h} \in \ZZ_N$ indicates that the dynamical gauge field obeys a twisted cocycle condition at the junction of the three 0-form symmetry operators. 

\begin{figure}[t!]
    \centering
    \begin{tikzpicture}
        \draw[draw=white,fill=white,
            path picture={
                \node[] at (0,0.5)
                {
                    \includegraphics[scale=0.135]{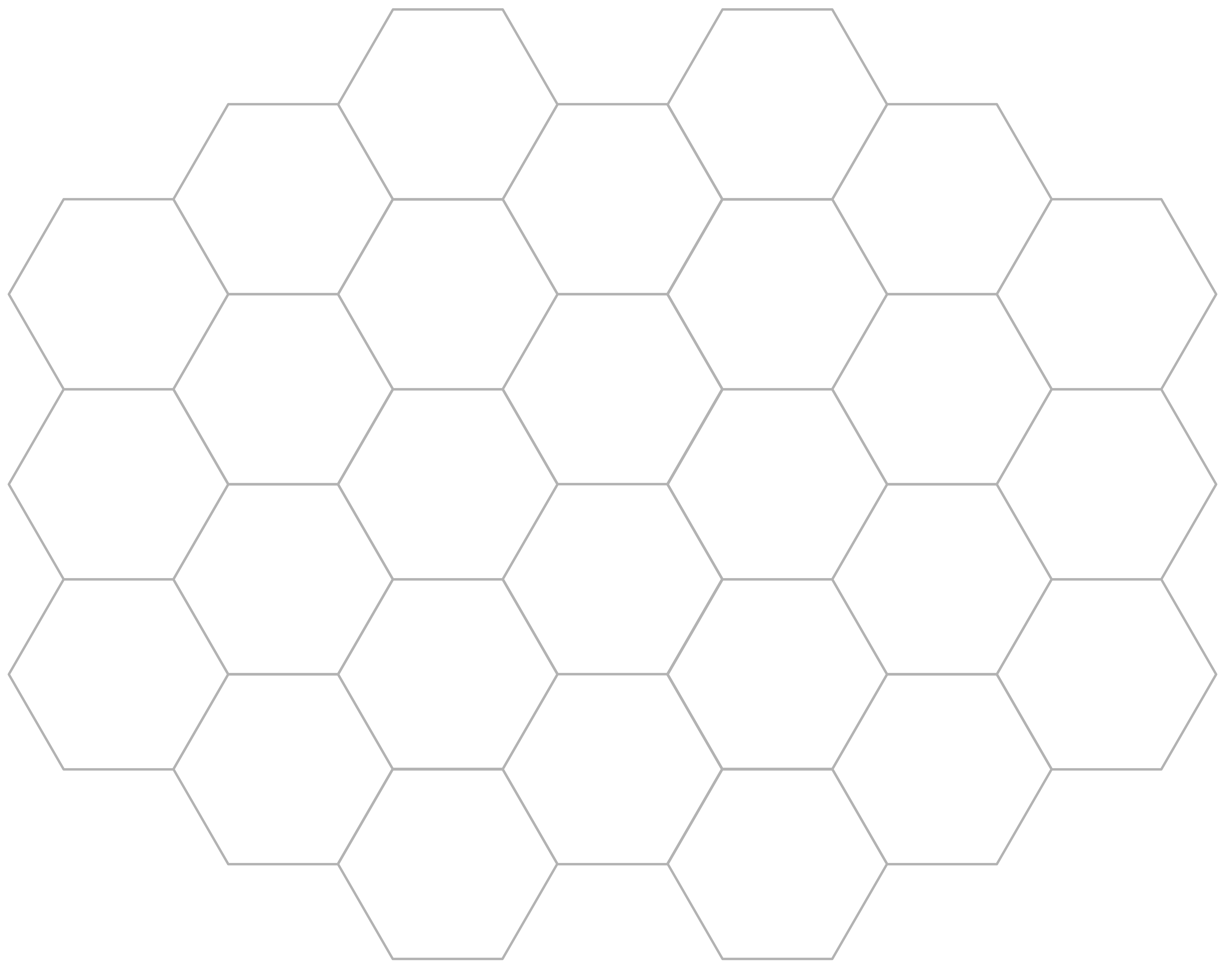}
                };
            }
            ] (0,0) circle (2.1);  
        \node[anchor = south west, opacity=0.5] at (-0.4,0.8) {$\scriptstyle \sigma_1$};
        \node[anchor = south west, opacity=0.5] at (-1.3,0.25) {$\scriptstyle \sigma_2$};
        \node[anchor = south west, opacity=0.5] at (0.7,0.25) {$\scriptstyle \sigma_3$};
        \node[anchor = south west, opacity=0.5] at (-0.4,-0.3) {$\scriptstyle \sigma_4$};
        \node[anchor = south west, opacity=0.5] at (-1.3,-0.85) {$\scriptstyle \sigma_5$};
        \node[anchor = south west, opacity=0.5] at (0.7,-0.85) {$\scriptstyle \sigma_6$};
        \node[anchor = south west, opacity=0.5] at (-0.3,-1.4) {$\scriptstyle \sigma_7$};

        \draw[color = olive,line width=1.5]  
        (-0.34, -0.62)--
        (-0.67, -0.06) --
        (-0.34, 0.49)-- 
        (0.3, 0.49) --
        (0.62,-0.07);

        \draw[color = olive, -Stealth] (-0.1, 0.49)-- (-0.1, 0.8);

        \fill[black] (0.62,-0.07) circle (1.5pt);
        \fill[black] (-0.34, -0.62) circle (1.5pt);
        
         \draw[red, thick] plot [smooth] coordinates {(0.2,0) (0.1,0.5)  (0.2,1.2)  (0.8,2)};
         \draw[red, thick, -Stealth] (.23,1.25) -- (0.29,1.35);
         \node[red] at (0.9,2.3) {$W$};
        \fill[red] (0.2, 0) circle (1.5pt);
        \node[red] at (0.2,-0.3) {$\Phi$};
        
        %\node[red, anchor = west] at (.8,1.8) {$W_q$};
		%\node[red, anchor = west] at (-.6,.1) {$\Phi^q$};
		
		\node[rotate = 30] at (2.7,2) {$=$};
		\node at (2.6,2.75) {$\scriptstyle \phi_{14} \,\to \,\phi_{14} -\frac{2\pi}{N}$};
		\node at (2.6,3.5) {$\scriptstyle \phi_{34} \,\to\, \phi_{34} -\frac{2\pi}{N}$};
		
		\node[rotate=-30] at (2.7,-2) {$=$};
		\node at (2.4,-2.75) {$\scriptstyle \Phi_4 \,\to\, e^{\frac{2\pi i}{N}}\, \Phi_4$};
		
		          \draw[draw=white,fill=white,
            path picture={
                \node[] at (6,3.5)
                {
                    \includegraphics[scale=0.135]{honeycomb.png}
                };
            }
            ] (6,3) circle (2.1);  
        \node[anchor = south west, opacity=0.5] at (5.6,3.8) {$\scriptstyle \sigma_1$};
        \node[anchor = south west, opacity=0.5] at (4.7,3.25) {$\scriptstyle \sigma_2$};
        \node[anchor = south west, opacity=0.5] at (6.7,3.25) {$\scriptstyle \sigma_3$};
        \node[anchor = south west, opacity=0.5] at (5.6,2.7) {$\scriptstyle \sigma_4$};
        \node[anchor = south west, opacity=0.5] at (4.7,2.15) {$\scriptstyle \sigma_5$};
        \node[anchor = south west, opacity=0.5] at (6.7,2.15) {$\scriptstyle \sigma_6$};
        \node[anchor = south west, opacity=0.5] at (5.7,1.6) {$\scriptstyle \sigma_7$};

        \draw[color = olive, line width=1.5] (5.66,2.38)-- (5.33,2.93) -- (5.66,3.49);
        \draw[color = olive, -Stealth] (5.5, 3.2)-- (5.25, 3.35);
        \fill[black] (5.66,2.38) circle (1.5pt);
        \fill[black] (5.66,3.49) circle (1.5pt);

       \draw[red, thick] plot [smooth] coordinates {(6.2,3) (6.1,3.5)  (6.2,4.2)  (6.8,5)};
       \draw[red, thick, -Stealth] (6.23,4.25) -- (6.29,4.35);
       \node[red] at (6.8,5.3) {$e^{\frac{2\pi i}{N}}\, W$};
        \fill[red] (6.2, 3) circle (1.5pt);
        \node[red] at (6.2,2.7) {$\Phi$};
   
	   	\node at (9.7,2.75) {$\scriptstyle \phi_{24} \,\to\, \phi_{24} -\frac{2\pi}{N}$};
	    \node at (9.7,3.5) {$\scriptstyle \phi_{45} \,\to\, \phi_{45} +\frac{2\pi}{N}$};

		\node[rotate=-30] at (9.1,2) {$=$};

          \draw[draw=white,fill=white,
            path picture={
                \node[] at (6,-2.5)
                {
                    \includegraphics[scale=0.135]{honeycomb.png}
                };
            }
            ] (6,-3) circle (2.1);  
        \node[anchor = south west, opacity=0.5] at (5.6,-2.2) {$\scriptstyle \sigma_1$};
        \node[anchor = south west, opacity=0.5] at (4.7,-2.75) {$\scriptstyle \sigma_2$};
        \node[anchor = south west, opacity=0.5] at (6.7,-2.75) {$\scriptstyle \sigma_3$};
        \node[anchor = south west, opacity=0.5] at (5.6,-3.3) {$\scriptstyle \sigma_4$};
        \node[anchor = south west, opacity=0.5] at (4.7,-3.85) {$\scriptstyle \sigma_5$};
        \node[anchor = south west, opacity=0.5] at (6.7,-3.85) {$\scriptstyle \sigma_6$};
        \node[anchor = south west, opacity=0.5] at (5.7,-4.4) {$\scriptstyle \sigma_7$};

        \draw[color = olive, line width=1.5] (5.66,-3.62)-- (6.30,-3.62) -- (6.63,-3.06);
        \draw[color = olive, -Stealth] (5.9,-3.62) -- (5.9,-3.35);
        \fill[black] (6.63,-3.06) circle (1.5pt);
        \fill[black] (5.66,-3.62) circle (1.5pt);

       \draw[red, thick] plot [smooth] coordinates {(6.2,-3) (6.1,-2.5)  (6.2,-1.8)  (6.8,-1)};
             \draw[red, thick, -Stealth] (6.23,-1.75) -- (6.29,-1.65);
       \node[red] at (6.9,-0.7) {$W$};
        \fill[red] (6.2, -3) circle (1.5pt);
        \node[red] at (5.8,-3.3) {$e^{\frac{2\pi i}{N}}\, \Phi$};
 	
	\node[rotate=30] at (9.1,-2) {$=$};

	      \node at (9.7,-2.75) {$\scriptstyle \phi_{46} \,\to\, \phi_{46} -\frac{2\pi}{N}$};
	    \node at (9.7,-3.5) {$\scriptstyle \phi_{47} \,\to\, \phi_{47} -\frac{2\pi}{N}$};
      
        \draw[draw=white,fill=white,
            path picture={
                \node[] at (12,0.5)
                {
                    \includegraphics[scale=0.135]{honeycomb.png}
                };
            }
            ] (12,0) circle (2.1);  
        \node[anchor = south west, opacity=0.5] at (11.6,0.8) {$\scriptstyle \sigma_1$};
        \node[anchor = south west, opacity=0.5] at (10.7,0.25) {$\scriptstyle \sigma_2$};
        \node[anchor = south west, opacity=0.5] at (12.7,0.25) {$\scriptstyle \sigma_3$};
        \node[anchor = south west, opacity=0.5] at (11.6,-0.3) {$\scriptstyle \sigma_4$};
        \node[anchor = south west, opacity=0.5] at (10.7,-0.85) {$\scriptstyle \sigma_5$};
        \node[anchor = south west, opacity=0.5] at (12.7,-0.85) {$\scriptstyle \sigma_6$};
        \node[anchor = south west, opacity=0.5] at (11.6,-1.4) {$\scriptstyle \sigma_7$};

         \draw[red, thick] plot [smooth] coordinates {(12.2,0) (12.1,0.5)  (12.2,1.2)  (12.8,2)};
          \draw[red, thick, -Stealth] (12.23,1.25) -- (12.29,1.35);
         \node[red] at (12.8,2.3) {$W$};
        \fill[red] (12.2, 0) circle (1.5pt);
         \node[red] at (12.2,-0.3) {$\Phi$};
         \node[red] at (11.8,1.5) {$e^{\frac{2\pi i}{N}}$};
     
         \end{tikzpicture}
    \caption{We can deform the bulk of the disk operator by performing field redefinitions of the matter fields $\Phi_i$, and deform the bulk and boundary of the disk operator by redefining the gauge field transition functions $\phi_{ij}$. Shrinking the disk in the presence of an open Wilson line in two different ways leads to the same phase, as in Figure~\ref{fig:DiskOpenLineAction}.  }
    \label{fig:BGTransf}
\end{figure}

Since we are explicitly working with the lift to $SU(N)$, the disk operator is particularly simple to describe: we turn on $\hatG_{ij} = e^{\frac{2\pi i}{N}}$ on a set of double overlaps $\CU_{ij}$ whose associated $\sigma_{ij}$ form a $(d-1)$-dimensional disk $D^{d-1}$. It is not hard to see that it is topological. First, we can deform the interior of the disk by performing field redefinitions $\Phi_i = \hatU_i \, \Phi_i'$ with $\hatU_i = e^{\frac{2\pi i}{N}}$ for all $j$ lying in a $d$-dimensional ball whose boundary is constructed by gluing two disks $D^{d-1},D'^{d-1}$ by their common boundary. This changes the location of the interior of the disk to $D'^{d-1}$, with $\partial D' = \partial D$ fixed. On the other hand, we can deform the bulk \emph{and} boundary of the disk by performing field redefinitions of the gauge field transition functions as in Eq.~\eqref{eq:phi_redef}. The disk operator can be completely removed by shifting $\phi_{ij} = \phi_{ij}' + \frac{2\pi }{N}$ on all double-overlaps where the disk operator is located. The action of the disk operator on the the gauge-invariant Wilson line \eqref{eq:gaugeinvariantW} comes precisely from this transformation of the gauge field transition functions.

As an example, consider the configuration on the left of Figure~\ref{fig:BGTransf} where all $\hatG_{ij}$ are unity except $\hatG_{34} = \hatG_{14} = \hatG_{24} = e^{-\frac{2\pi i}{N}}$ and $\hatG_{45} = e^{\frac{2\pi i}{N}}$. This defines an open, codimension-1 defect --- a disk operator. At the boundaries of the defect, we have
\begin{equation}
   e^{\frac{2\pi i}{N} \Lambda_{346}} = e^{-\frac{2\pi i}{N}}\,, \quad  e^{\frac{2\pi i}{N} \Lambda_{457}} = e^{\frac{2\pi i}{N}}\,. 
\end{equation}
The bulk of the disk can be deformed, keeping the endpoints fixed, by performing a field redefinition such as $\Phi_4 \to e^{\frac{2\pi i}{N}} \Phi_4$. Alternatively, we can redefine the transition functions to shrink the disk. In the presence of a Wilson line, these manipulations will lead to a phase. This action is consistent with the open Wilson lines, since the bulk of the disk acts in a compensating fashion on the endpoint (see Figure~\ref{fig:BGTransf}). 

\begin{figure}[t!]
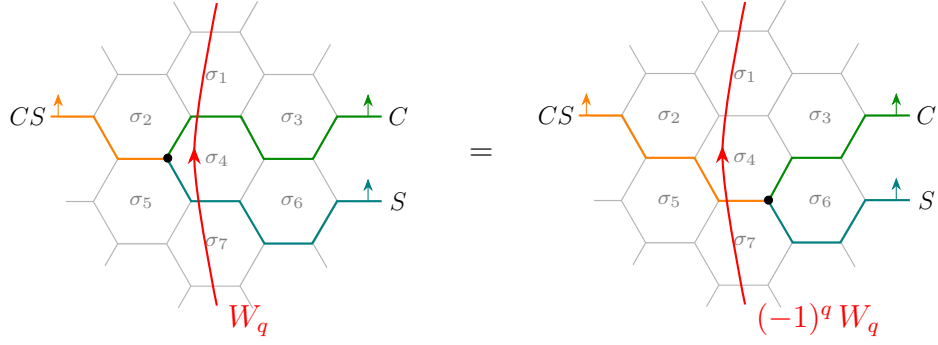

    \centering
    \begin{tikzpicture}
        \draw[draw=white,fill=white,
            path picture={
                \node[] at (0,0.5)
                {
                    \includegraphics[scale=0.135]{honeycomb.png}
                };
            }
            ] (0,0) circle (2.1);  
        \node[anchor = south west, opacity=.5] at (-0.3,0.8) {$\scriptstyle \s_1$};
        \node[anchor = south west, opacity=.5] at (-1.3,0.25) {$\scriptstyle \s_2$};
        \node[anchor = south west, opacity=.5] at (0.7,0.25) {$\scriptstyle \s_3$};
        \node[anchor = south west, opacity=.5] at (-0.3,-0.3) {$\scriptstyle \s_4$};
        \node[anchor = south west, opacity=.5] at (-1.3,-0.85) {$\scriptstyle \s_5$};
        \node[anchor = south west, opacity=.5] at (0.7,-0.85) {$\scriptstyle \s_6$};
        \node[anchor = south west, opacity=.5] at (-0.3,-1.4) {$\scriptstyle \s_7$};

        \draw[color = teal, thick] 
%        (-2.2, 0.53) --
%        (-1.62, 0.53) --
%        (-1.3,-0.02) --
        (-0.68,-0.02) --
        (-0.34,-0.62) -- 
        (0.3,-0.62) --
        (0.63,-1.18) --
        (1.26,-1.18) --
        (1.59,-0.62) --
        (2.18,-0.62);

         \draw[color = dgreen, thick] 
%        (-2.2, 0.5) --
%        (-1.63, 0.5) --
%        (-1.31,-0.05) --
        (-0.67,-0.05) --
        (-0.34, 0.50) --
        (0.32, 0.50) --
        (0.62, -0.06) --
        (1.27, -0.06) --
        (1.59, 0.50) --
        (2.18, 0.50);
        
        \draw[color = orange, thick] 
        (-2.2, 0.5) --
        (-1.63, 0.5) --
        (-1.31,-0.06) --
        (-0.67,-0.06);

        \draw[fill] (-0.65,-0.05) circle (1.5pt);
        \draw[red, thick]
        plot [smooth] coordinates {(-0.,-2) (-0.3, 0)(-0.,2)};
        \draw[red, thick, -Stealth] (-0.3, 0) -- (-0.3, 0.1);
        \node[red, anchor = west] at (0,-2.2) {$W_q$};
        \node[scale = 0.8] at (-2.5, 0.5) {$CS$};
        \node[scale = 0.8] at (2.4, 0.50) {$C$};
        \node[scale = 0.8] at (2.4,-0.62) {$S$};
        \draw[color = dgreen, -Stealth] (2, 0.5) -- (2, 0.8);
        \draw[color = teal, -Stealth] (2, -0.62) -- (2, -0.32);
        \draw[color = orange, -Stealth] (-2.1,0.5) -- (-2.1, 0.8);

        \node at (3.5,0) {$=$};
    \end{tikzpicture}
    \hskip .2cm
    \begin{tikzpicture}
        \draw[draw=white,fill=white,
            path picture={
                \node[] at (0,0.5)
                {
                    \includegraphics[scale=0.135]{honeycomb.png}
                };
            }
            ] (0,0) circle (2.1); 

        \node[anchor = south west, opacity=.5] at (-0.3,0.8) {$\scriptstyle \s_1$};
        \node[anchor = south west, opacity=.5] at (-1.3,0.25) {$\scriptstyle \s_2$};
        \node[anchor = south west, opacity=.5] at (0.7,0.25) {$\scriptstyle \s_3$};
        \node[anchor = south west, opacity=.5] at (-0.3,-0.3) {$\scriptstyle \s_4$};
        \node[anchor = south west, opacity=.5] at (-1.3,-0.85) {$\scriptstyle \s_5$};
        \node[anchor = south west, opacity=.5] at (0.7,-0.85) {$\scriptstyle \s_6$};
        \node[anchor = south west, opacity=.5] at (-0.3,-1.4) {$\scriptstyle \s_7$};
        
        \draw[color = teal, thick] 
%        (-2.2, 0.53) --
%        (-1.62, 0.53) --
%        (-1.3,-0.02) --
%        (-0.66,-0.02) --
%        (-0.34,-0.62) -- 
        (0.3,-0.62) --
        (0.63,-1.18) --
        (1.26,-1.18) --
        (1.59,-0.62) --
        (2.18,-0.62);

         \draw[color = dgreen, thick] 
%        (-2.2, 0.5) --
%        (-1.63, 0.5) --
%        (-1.31,-0.05) --
%        (-0.67,-0.05) --
%        (-0.35,-0.65) -- 
        (0.3,-0.65) --
        (0.62, -0.06) --
        (1.27, -0.06) --
        (1.59, 0.50) --
        (2.18, 0.50);
        
        \draw[color = orange, thick] 
        (-2.2, 0.5) --
        (-1.63, 0.5) --
        (-1.31,-0.06) --
        (-0.67,-0.06) --  
        (-0.35,-0.63) -- 
        (0.3,-0.63);

        \draw[fill] (0.31,-0.62) circle (1.5pt); 
        \draw[red, thick]
        plot [smooth] coordinates {(-0.,-2) (-0.3, 0)(-0.,2)};
        \draw[red, thick, -Stealth] (-0.3, 0) -- (-0.3, 0.1);
        \node[red, anchor = west] at (0,-2.2) {$(-1)^q \, W_q$};
        \node[scale = 0.8] at (-2.5, 0.5) {$CS$};
        \node[scale = 0.8] at (2.4, 0.50) {$C$};
        \node[scale = 0.8] at (2.4,-0.62) {$S$};
       \draw[color = dgreen, -Stealth] (2, 0.5) -- (2, 0.8);
        \draw[color = teal, -Stealth] (2, -0.62) -- (2, -0.32);
        \draw[color = orange, -Stealth] (-2.1,0.5) -- (-2.1, 0.8);

    \end{tikzpicture}
    \caption{Realizing the projective $SO(3)$ action on Wilson lines in the $N = 2$ abelian Higgs model. Here we labeled the $SO(3)$ symmetry operators using their lifts to $SU(2)$. The black dot indicates a non-trivial $\Lambda_{ijk}$. To get from the left to right, we performed the field redefinitions $\Phi_4 \to C\, \Phi_4$ and $\phi_{45} \to \phi_{45} + \pi$, $\phi_{47} \to \phi_{47} + \pi$. The fundamental Wilson line includes a factor of $e^{-i\phi_{47}}$, so it flips sign under this transformation.}
    \label{fig:patches_action}
\end{figure}

To further illustrate the connection to the projective phases from Section~\ref{sec:no1formsymfrac}, let us take $N = 2$ and consider the $\ZZ_2 \times \ZZ_2 \subset SO(3)$ subgroup that descends from the quotient of $\mathbb{Q}_8 \subset SU(2)$ generated by the clock and shift matrices $C = i\sigma_3$ and $S = i \sigma_1$, which obey
\begin{equation}
\mathbb{Q}_8: \quad S^2 = C^2 = -\mathbbm{1}\,,\quad  SC = -CS\,. 
\end{equation}
If we denote by $A^\1, B^\1$ the background fields for the faithfully acting $\ZZ_2\times\ZZ_2$ symmetry, the symmetry fractionalization \eqref{eq:fractional_flux} on orientable manifolds reduces to 
\begin{equation} \label{eq:reduced_fractional_flux} 
    \int_\Sigma \frac{da}{2\pi} = \frac{1}{2}\int_\Sigma A^\1 \cup A^\1 + A^\1 \cup B^\1 + B^\1 \cup B^\1 =  \frac{1}{2}\int_\Sigma A^\1 \cup B^\1  \text{ mod } 1\,,
\end{equation}
where we used the fact that $A \cup A = \frac{1}{2}\delta A$ for a $\ZZ_2$ gauge field. Let us denote the elements of $\ZZ_2\times\ZZ_2$ as $(a,b)$ with $a,b =0,1$. We choose the lift
\begin{equation}
    (a,b) \to C^b  S^a\,,
\end{equation}
which is characterized by the group 2-cocycle (see Appendix~\ref{app:clockandshift})
\begin{equation}
    c_{(a_1,b_1), (a_2,b_2)} = (-1)^{a_1a_2 + b_1b_2 + a_1 b_2}\,.
\end{equation}
Now consider, for instance, a trivalent junction with $g = (1,0)$ and $h = (0,1)$ fusing into $gh = (1,1)$. The set of background fields dual to this configuration consists of a set of transition functions with a non-trivial triple overlap on $\sigma_{ijk}$ with $\hatG_{ij} = S$, $\hatG_{jk} = C$, $\hatG_{ik} = CS$. Since $c_{(1,0),(0,1)} = (-1)$, this junction has $\Lambda_{ijk} = 1$ mod $2$. We show a 2d slice of this configuration in Figure~\ref{fig:patches_action}, where we see that moving the junction across a charge-$q$ Wilson line $W_q$ results in the phase $(-1)^q$. Moving a general $\ZZ_2 \times \ZZ_2$ junction across $W_q$ results in the phase 
\begin{equation}
W_q \to \omega_{q; (a_1,b_1),(a_2,b_2)}\, W_q = (-1)^{q(a_1a_2 + b_1b_2 + a_1 b_2)}\, W_q\,.
\end{equation}
Now let us adjust counterterms on the Wilson line, i.e. redefine any correlation function involving the two operators by
\begin{equation}
\langle\, \cdots\, U_{(a,b)}(\Sigma) \, W_q(\gamma) \,\cdots\, \rangle \  \longrightarrow \  e^{\frac{iq\pi}{2}(a^2+b^2)\text{Int}(\Sigma,\gamma) }\, \langle \, \cdots \, U_{(a,b)}(\Sigma) \, W_q(\gamma)  \, \cdots \, \rangle\,,
\end{equation}
we can change the representative of the group cohomology class to $\omega_{q;(a_1,b_1),(a_2,b_2)} =  (-1)^{q\, a_1 b_2}$ which nicely matches Eq.~\eqref{eq:reduced_fractional_flux}.

%%%%%%%%%%%%%%%%%%%%%%%%%%%%%%%%
\section{Selection Rules and Exact Vacuum Degeneracy }
\label{sec:selectionrulesanddegeneracies}
%%%%%%%%%%%%%%%%%%%%%%%%%%%%%%%%

We now move on to describe some basic implications of symmetry fractionalization for line operator correlation functions and vacuum structure. The constraints are kinematic in nature and can be derived by using the topological junctions of 0-form symmetry operators, or equivalently with the disk operator we constructed in the previous section. 

%%%%%%%%%%%%%%%%%%%%%%%%%%%%%%%%
\subsection{Selection Rules in Twisted Sectors}
\label{sec:selectionrules}
%%%%%%%%%%%%%%%%%%%%%%%%%%%%%%%%

One of the most basic implications of global symmetries in quantum systems is the existence of selection rules for correlation functions of charged operators in symmetric states. For instance, given a 0-form symmetry described by a group $G$, finite-volume correlation functions involving $G$-charged operators vanish in any $G$-symmetric state unless there is non-vanishing overlap with the trivial representation of $G$.

Selection rules for extended operators charged under higher-form symmetries are a bit more subtle. Strictly speaking, exact selection rules (i.e. exactly vanishing correlation functions) only hold for extended operators supported on non-trivial cycles in spacetime, while symmetries can at best constrain the expectation value of a contractible extended operator in a scaling limit where its size goes to infinity.\footnote{In this way, contractible extended operators are analogous to the two-point functions of charged local operators, which vanish only in the limit that the separation goes to infinity. } For line operators charged under 1-form symmetry, this corresponds to the observation that in confining theories Polyakov loops (Wilson loops wrapping non-contractible cycles) have vanishing expectation values while contractible Wilson loops vanish faster than perimeter law. 

Let us first review how a 1-form global symmetry implies selection rules for non-contractible line operators on e.g. the torus. Consider a $d$-dimensional theory with $\IZ_N^\1$ global symmetry on $T^d$. In this theory we have a collection of line operators defined by some data (such as a representation of a Lie group) and a collection of codimension-2 operators $V_n(\Sigma)$ where $n\in \IZ_N$. For the current discussion, only the 1-form charge of the lines matter (for instance given by the $N$-ality of a representation for Wilson lines in $SU(N)$ gauge theory), and we denote by $\scrL_q(\gamma)$ a line operator with charge $q$ under $\IZ_N^\1$. The $\IZ_N^\1$ global symmetry acts on the $\scrL_q(\gamma)$ by a linking action:
\eq{
\langle V_n(\Sigma)~\scrL_q(\gamma)\rangle=e^{\frac{2\pi i}{N}nq \,{\rm Link}(\Sigma,\gamma)}\,\langle \scrL_q(\gamma)\rangle~,
}
where we have assumed that $\Sigma$ is contractible to a point.

\begin{figure}[t!]
\centering
  \begin{tikzpicture}[scale=.5]

  	\coordinate (a) at (0,0);
	\coordinate (b) at (6.5,0);
	\coordinate (c) at (13,0);
	\coordinate (d) at (19.5,0);
	\coordinate (e) at (26,0);

  %%%%%%%%%%% (a) %%%%%%%%%%%%%%%%%%%%%%%	
 		
		\draw[black] ($(a) + (4,-1.685)$) -- ($(a) + (4,2.315)$) -- ($(a) + (8,2.315)$) -- ($(a) + (8,-1.685)$) -- ($(a) + (4,-1.685)$);		
		\draw[black] ($(a) + (3,-2.4)$) -- ($(a) + (4,-1.685)$);
		\draw[black] ($(a) + (7,-2.4)$) -- ($(a) + (8,-1.685)$);
		\draw[black] ($(a) + (3,1.6)$) -- ($(a) + (4,2.315)$);
		\draw[black] ($(a) + (7,1.6)$) -- ($(a) + (8,2.315)$);

        \draw[black] ($(a) + (3,-2.4)$) -- ($(a) + (3,1.6)$) -- ($(a) + (7,1.6)$) -- ($(a) + (7,-2.4)$) -- ($(a) + (3,-2.4)$);
               
        \draw [thick, red] ($(a) + (5.5,-2)$) -- ($(a) + (5.5,2)$);
        \draw [thick, red, -Stealth] ($(a) + (5.5,0.1)$) -- ($(a) + (5.5,0.2)$);
        \node[red] at ($(a) + (5.6,2.8)$) {$\scrL_q$};
         \node at ($(a) + (8.75, 0.2)$) {$=$};
         	
  %%%%%%%%%%% (b) %%%%%%%%%%%%%%%%%%%%%%%
           
        \draw[black] ($(b) + (4,-1.685)$) -- ($(b) + (4,2.315)$) -- ($(b) + (8,2.315)$) -- ($(b) + (8,-1.685)$) -- ($(b) + (4,-1.685)$);		
		\draw[black] ($(b) + (3,-2.4)$) -- ($(b) + (4,-1.685)$);
		\draw[black] ($(b) + (7,-2.4)$) -- ($(b) + (8,-1.685)$);
		\draw[black] ($(b) + (3,1.6)$) -- ($(b) + (4,2.315)$);
		\draw[black] ($(b) + (7,1.6)$) -- ($(b) + (8,2.315)$);

        \draw[black] ($(b) + (3,-2.4)$) -- ($(b) + (3,1.6)$) -- ($(b) + (7,1.6)$) -- ($(b) + (7,-2.4)$) -- ($(b) + (3,-2.4)$);
               
        \draw [thick, red] ($(b) + (5.5,-2)$) -- ($(b) + (5.5,2)$);
        \draw [thick, red, -Stealth] ($(b) + (5.5,.5)$) -- ($(b) + (5.5,.6)$);
        \node[red] at ($(b) + (5.6,2.8)$) {$\scrL_q$};
         \node at ($(b) + (8.75, 0.2)$) {$=$};
        \node[black] at ($(b) + (4.5,0.6)$) {$V_n$};
        
             \draw[black,thick] ($(b) + (4.3,0)$) to[out=20, in=20, looseness=1.5] ($(b) + (5,-0.15)$) to[out=200, in=200, looseness=1.5] ($(b) + (4.3,0)$); 
         \draw [black, -Stealth] ($(b) + (4.6,-0.21)$) -- ($(b) + (4.5,-0.22)$);
	
  %%%%%%%%%%% (c) %%%%%%%%%%%%%%%%%%%%%%%

         		\draw[black] ($(c) + (4,-1.685)$) -- ($(c) + (4,2.315)$) -- ($(c) + (8,2.315)$) -- ($(c) + (8,-1.685)$) -- ($(c) + (4,-1.685)$);		
		\draw[black] ($(c) + (3,-2.4)$) -- ($(c) + (4,-1.685)$);
		\draw[black] ($(c) + (7,-2.4)$) -- ($(c) + (8,-1.685)$);
		\draw[black] ($(c) + (3,1.6)$) -- ($(c) + (4,2.315)$);
		\draw[black] ($(c) + (7,1.6)$) -- ($(c) + (8,2.315)$);	
      	\draw[black] ($(c) + (3,-2.4)$) -- ($(c) + (3,1.6)$) -- ($(c) + (7,1.6)$) -- ($(c) + (7,-2.4)$) -- ($(c) + (3,-2.4)$);
                      
           \draw[thick, red] ($(c) + (5.5, -0.44)$) -- ($(c) + (5.5, -2)$); 
           \draw [thick, red, -Stealth] ($(c) + (5.5,.7)$) -- ($(c) + (5.5,.8)$);      
        \node[red] at ($(c) + (5.6,3)$) {$e^{\frac{2\pi i}{N}nq} \scrL_q$};
        \node at ($(c) + (8.75, 0.2)$) {$=$};
        \node[black] at ($(c) + (4.5,0.8)$) {$V_n$};
        
         \draw[black,thick] ($(c) + (4.65,0.1)$) to[out=20, in=20, looseness=1.5] ($(c) + (6.35,-0.25)$) to[out=200, in=200, looseness=1.5] ($(c) + (4.65,0.1)$); 
    		\draw [thick, white] ($(c) + (5.48,0.2)$) -- ($(c) + (5.48,0.3)$);
		\draw [thick, white] ($(c) + (5.52,0.2)$) -- ($(c) + (5.52,0.3)$); 	
		     \draw [thick, red] ($(c) + (5.5,-0.35)$) -- ($(c) + (5.5,2)$);
		     
		       \draw [black, -Stealth] ($(c) + (5.8,-0.37)$) -- ($(c) + (5.7,-0.38)$);
	
  %%%%%%%%%%% (d) %%%%%%%%%%%%%%%%%%%%%%%

        \draw[black] ($(d) + (4,-1.685)$) -- ($(d) + (4,2.315)$) -- ($(d) + (8,2.315)$) -- ($(d) + (8,-1.685)$) -- ($(d) + (4,-1.685)$);		
		\draw[black] ($(d) + (3,-2.4)$) -- ($(d) + (4,-1.685)$);
		\draw[black] ($(d) + (7,-2.4)$) -- ($(d) + (8,-1.685)$);
		\draw[black] ($(d) + (3,1.6)$) -- ($(d) + (4,2.315)$);
		\draw[black] ($(d) + (7,1.6)$) -- ($(d) + (8,2.315)$);

        \draw[black] ($(d) + (3,-2.4)$) -- ($(d) + (3,1.6)$) -- ($(d) + (7,1.6)$) -- ($(d) + (7,-2.4)$) -- ($(d) + (3,-2.4)$);
               
        \draw [thick, red] ($(d) + (5.5,-2)$) -- ($(d) + (5.5,2)$);
        \draw [thick, red, -Stealth] ($(d) + (5.5,.5)$) -- ($(d) + (5.5,.6)$);
        \node[red] at ($(d) + (5.6,3)$) {$e^{\frac{2\pi i}{N}nq} \scrL_q$};
         \node at ($(d) + (8.75, 0.2)$) {$=$};
         \node[black] at ($(d) + (4.5,0.8)$) {$V_n$};
        
             \draw[black,thick] ($(d) + (3.3,-0.28)$) to[out=0, in=25, looseness=1.4] ($(d) + (4.4,-0.43)$);
             \draw[black,thick] ($(d) + (7.3,-0.28)$) to[out=180, in=40, looseness=1] ($(d) + (6,-0.43)$);   
              \draw[black,thick] ($(d) + (7.8,0.12)$) to[out=180, in=-150, looseness=1.5] ($(d) + (6.7,0.32)$);     
             \draw[black,thick] ($(d) + (3.8,0.12)$) to[out=0, in=-150, looseness=1.5] ($(d) + (5.2,0.32)$);         
                  
                    \draw [black, -Stealth] ($(d) + (3.6,-0.27)$) -- ($(d) + (3.5,-0.27)$);
                       \draw [black, -Stealth] ($(d) + (4.3,0.09)$) -- ($(d) + (4.4,0.08)$);
                         \draw [black, -Stealth] ($(d) + (6.7,-0.235)$) -- ($(d) + (6.6,-0.23)$);
                       \draw [black, -Stealth] ($(d) + (7.4,0.11)$) -- ($(d) + (7.5,0.11)$);
                          	
  %%%%%%%%%%% (e) %%%%%%%%%%%%%%%%%%%%%%%
  
 			\draw[black] ($(e) + (4,-1.685)$) -- ($(e) + (4,2.315)$) -- ($(e) + (8,2.315)$) -- ($(e) + (8,-1.685)$) -- ($(e) + (4,-1.685)$);		
		\draw[black] ($(e) + (3,-2.4)$) -- ($(e) + (4,-1.685)$);
		\draw[black] ($(e) + (7,-2.4)$) -- ($(e) + (8,-1.685)$);
		\draw[black] ($(e) + (3,1.6)$) -- ($(e) + (4,2.315)$);
		\draw[black] ($(e) + (7,1.6)$) -- ($(e) + (8,2.315)$);

        \draw[black] ($(e) + (3,-2.4)$) -- ($(e) + (3,1.6)$) -- ($(e) + (7,1.6)$) -- ($(e) + (7,-2.4)$) -- ($(e) + (3,-2.4)$);
               
        \draw [thick, red] ($(e) + (5.5,-2)$) -- ($(e) + (5.5,2)$);
        \draw [thick, red, -Stealth] ($(e) + (5.5,0.1)$) -- ($(e) + (5.5,0.2)$);
        \node[red] at ($(e) + (5.6,3)$) {$e^{\frac{2\pi i}{N}nq} \scrL_q$};

	    \end{tikzpicture}
	    \\ \vspace{.5cm}
	      \begin{tikzpicture}[scale=.5]

  	\coordinate (a) at (0,0);
	\coordinate (b) at (6.5,0);
	\coordinate (c) at (13,0);
	\coordinate (d) at (19.5,0);
	\coordinate (e) at (26,0);

  %%%%%%%%%%% (a) %%%%%%%%%%%%%%%%%%%%%%%	
 		
		\draw[black] ($(a) + (4,-1.685)$) -- ($(a) + (4,2.315)$) -- ($(a) + (8,2.315)$) -- ($(a) + (8,-1.685)$) -- ($(a) + (4,-1.685)$);		
		\draw[black] ($(a) + (3,-2.4)$) -- ($(a) + (4,-1.685)$);
		\draw[black] ($(a) + (7,-2.4)$) -- ($(a) + (8,-1.685)$);
		\draw[black] ($(a) + (3,1.6)$) -- ($(a) + (4,2.315)$);
		\draw[black] ($(a) + (7,1.6)$) -- ($(a) + (8,2.315)$);

        \draw[black] ($(a) + (3,-2.4)$) -- ($(a) + (3,1.6)$) -- ($(a) + (7,1.6)$) -- ($(a) + (7,-2.4)$) -- ($(a) + (3,-2.4)$);
               
        \draw [thick, red] ($(a) + (5.5,-2)$) -- ($(a) + (5.5,2)$);
        \draw [thick, red, -Stealth] ($(a) + (5.5,0.1)$) -- ($(a) + (5.5,0.2)$);
        \node[red] at ($(a) + (5.6,2.8)$) {$\scrL$};
         \node at ($(a) + (8.75, 0.2)$) {$=$};
         	
  %%%%%%%%%%% (b) %%%%%%%%%%%%%%%%%%%%%%%
           
        \draw[black] ($(b) + (4,-1.685)$) -- ($(b) + (4,2.315)$) -- ($(b) + (8,2.315)$) -- ($(b) + (8,-1.685)$) -- ($(b) + (4,-1.685)$);		
		\draw[black] ($(b) + (3,-2.4)$) -- ($(b) + (4,-1.685)$);
		\draw[black] ($(b) + (7,-2.4)$) -- ($(b) + (8,-1.685)$);
		\draw[black] ($(b) + (3,1.6)$) -- ($(b) + (4,2.315)$);
		\draw[black] ($(b) + (7,1.6)$) -- ($(b) + (8,2.315)$);

        \draw[black] ($(b) + (3,-2.4)$) -- ($(b) + (3,1.6)$) -- ($(b) + (7,1.6)$) -- ($(b) + (7,-2.4)$) -- ($(b) + (3,-2.4)$);
               
        \draw [thick, red] ($(b) + (5.5,-2)$) -- ($(b) + (5.5,2)$);
        \draw [thick, red, -Stealth] ($(b) + (5.5,.5)$) -- ($(b) + (5.5,.6)$);
        \node[red] at ($(b) + (5.6,2.8)$) {$\scrL$};
         \node at ($(b) + (8.75, 0.2)$) {$=$};
         
         \node[olive] at ($(b) + (4.5,0.6)$) {$\scrD_c$};
        
             \draw[black] ($(b) + (4.3,0)$) to[out=20, in=20, looseness=1.5] ($(b) + (5,-0.15)$) to[out=200, in=200, looseness=1.5] ($(b) + (4.3,0)$); 
             \draw[fill=olive,opacity=0.5] ($(b) + (4.3,0)$) to[out=20, in=20, looseness=1.5] ($(b) + (5,-0.15)$) to[out=200, in=200, looseness=1.5] ($(b) + (4.3,0)$); 
         \draw [black, -Stealth] ($(b) + (4.6,-0.21)$) -- ($(b) + (4.5,-0.22)$);
	
  %%%%%%%%%%% (c) %%%%%%%%%%%%%%%%%%%%%%%

         		\draw[black] ($(c) + (4,-1.685)$) -- ($(c) + (4,2.315)$) -- ($(c) + (8,2.315)$) -- ($(c) + (8,-1.685)$) -- ($(c) + (4,-1.685)$);		
		\draw[black] ($(c) + (3,-2.4)$) -- ($(c) + (4,-1.685)$);
		\draw[black] ($(c) + (7,-2.4)$) -- ($(c) + (8,-1.685)$);
		\draw[black] ($(c) + (3,1.6)$) -- ($(c) + (4,2.315)$);
		\draw[black] ($(c) + (7,1.6)$) -- ($(c) + (8,2.315)$);	
      	\draw[black] ($(c) + (3,-2.4)$) -- ($(c) + (3,1.6)$) -- ($(c) + (7,1.6)$) -- ($(c) + (7,-2.4)$) -- ($(c) + (3,-2.4)$);
                      
           \draw[thick, red] ($(c) + (5.5, -0.42)$) -- ($(c) + (5.5, -2)$); 
           \draw [thick, red, -Stealth] ($(c) + (5.5,.7)$) -- ($(c) + (5.5,.8)$);      
        \node[red] at ($(c) + (5.6,3)$) {$\rho_c\, \scrL$};
        \node at ($(c) + (8.75, 0.2)$) {$=$};
        \node[olive] at ($(c) + (4.5,0.8)$) {$\scrD_c$};

         \draw[black] ($(c) + (4.65,0.1)$) to[out=20, in=20, looseness=1.5] ($(c) + (6.35,-0.25)$) to[out=200, in=200, looseness=1.5] ($(c) + (4.65,0.1)$); 
         \draw [thick, red,opacity=0.5] ($(c) + (5.5,-0.37)$) -- ($(c) + (5.5,-0.08)$);
         \draw[fill=olive,opacity=0.5] ($(c) + (4.65,0.1)$) to[out=20, in=20, looseness=1.5] ($(c) + (6.35,-0.25)$) to[out=200, in=200, looseness=1.5] ($(c) + (4.65,0.1)$); 
    %		\draw [thick, white] ($(c) + (5.48,0.2)$) -- ($(c) + (5.48,0.3)$);
	%	\draw [thick, white] ($(c) + (5.52,0.2)$) -- ($(c) + (5.52,0.3)$); 	
		     \draw [thick, red] ($(c) + (5.5,-0.08)$) -- ($(c) + (5.5,2)$);
		     \draw [fill,red] ($(c) + (5.5,-0.08)$) circle (1pt);
		     
		       \draw [black, -Stealth] ($(c) + (5.8,-0.37)$) -- ($(c) + (5.7,-0.38)$);
	
  %%%%%%%%%%% (d) %%%%%%%%%%%%%%%%%%%%%%%

        \draw[black] ($(d) + (4,-1.685)$) -- ($(d) + (4,2.315)$) -- ($(d) + (8,2.315)$) -- ($(d) + (8,-1.685)$) -- ($(d) + (4,-1.685)$);		
		\draw[black] ($(d) + (3,-2.4)$) -- ($(d) + (4,-1.685)$);
		\draw[black] ($(d) + (7,-2.4)$) -- ($(d) + (8,-1.685)$);
		\draw[black] ($(d) + (3,1.6)$) -- ($(d) + (4,2.315)$);
		\draw[black] ($(d) + (7,1.6)$) -- ($(d) + (8,2.315)$);

        \draw[black] ($(d) + (3,-2.4)$) -- ($(d) + (3,1.6)$) -- ($(d) + (7,1.6)$) -- ($(d) + (7,-2.4)$) -- ($(d) + (3,-2.4)$);
               
            \draw [thick, red, opacity=0.5] ($(d) + (5.5,-0.08)$) -- ($(d) + (5.5,-0.43)$);
        \node[red] at ($(d) + (5.6,3)$) {$\rho_c\, \scrL$};
         \node at ($(d) + (8.75, 0.2)$) {$=$};
         \node[olive] at ($(d) + (4.5,0.8)$) {$\scrD_c$};
        
                        \filldraw[olive,opacity=0.5]  ($(d) + (3.3,-0.28)$) to[out=0, in=25, looseness=1.4] ($(d) + (4.4,-0.43)$) --  ($(d) + (6,-0.43)$)  to[out=40, in=180, looseness=1] ($(d) + (7.3,-0.28)$) -- ($(d) + (7.8,0.12)$) to[out=180, in=-150, looseness=1.5] ($(d) + (6.7,0.32)$) -- ($(d) + (5.2,0.32)$) to[out=-150, in=0, looseness=1.5] ($(d) + (3.8,0.12)$);
		
        \draw[black] ($(d) + (3.3,-0.28)$) to[out=0, in=25, looseness=1.4] ($(d) + (4.4,-0.43)$);
             \draw[black] ($(d) + (7.3,-0.28)$) to[out=180, in=40, looseness=1] ($(d) + (6,-0.43)$);   
              \draw[black] ($(d) + (7.8,0.12)$) to[out=180, in=-150, looseness=1.5] ($(d) + (6.7,0.32)$);     
             \draw[black] ($(d) + (3.8,0.12)$) to[out=0, in=-150, looseness=1.5] ($(d) + (5.2,0.32)$);         
                  
                   \draw [black, -Stealth] ($(d) + (3.6,-0.27)$) -- ($(d) + (3.5,-0.27)$);
                       \draw [black, -Stealth] ($(d) + (4.3,0.09)$) -- ($(d) + (4.4,0.08)$);
                         \draw [black, -Stealth] ($(d) + (6.7,-0.235)$) -- ($(d) + (6.6,-0.23)$);
                       \draw [black, -Stealth] ($(d) + (7.4,0.11)$) -- ($(d) + (7.5,0.11)$);
                                
     \draw [thick, red] ($(d) + (5.5,-0.08)$) -- ($(d) + (5.5,2)$);
		     \draw [fill,red] ($(d) + (5.5,-0.08)$) circle (1pt);
		      \draw [thick, red, -Stealth] ($(d) + (5.5,0.8)$) -- ($(d) + (5.5,0.9)$);
		         \draw[thick, red] ($(d) + (5.5, -0.44)$) -- ($(d) + (5.5, -2)$);

  %%%%%%%%%%% (e) %%%%%%%%%%%%%%%%%%%%%%%
  
        \draw[black] ($(e) + (4,-1.685)$) -- ($(e) + (4,2.315)$) -- ($(e) + (8,2.315)$) -- ($(e) + (8,-1.685)$) -- ($(e) + (4,-1.685)$);		
		\draw[black] ($(e) + (3,-2.4)$) -- ($(e) + (4,-1.685)$);
		\draw[black] ($(e) + (7,-2.4)$) -- ($(e) + (8,-1.685)$);
		\draw[black] ($(e) + (3,1.6)$) -- ($(e) + (4,2.315)$);
		\draw[black] ($(e) + (7,1.6)$) -- ($(e) + (8,2.315)$);

        \draw[black] ($(e) + (3,-2.4)$) -- ($(e) + (3,1.6)$) -- ($(e) + (7,1.6)$) -- ($(e) + (7,-2.4)$) -- ($(e) + (3,-2.4)$);
               
            \draw [thick, red] ($(e) + (5.5,-0.44)$) -- ($(e) + (5.5,-2)$);
        \node[red] at ($(e) + (5.6,3)$) {$\rho_c\, \scrL$};

 \draw [thick, red, opacity=0.5] ($(e) + (5.5,-0.08)$) -- ($(e) + (5.5,-0.43)$);
 
        \node[olive] at ($(e) + (4.5,0.8)$) {$\scrD_c$};
                        \filldraw[olive,opacity=0.5]  ($(e) + (3,-0.48)$) -- ($(e) + (4,0.235)$) -- ($(e) + (8,0.235)$) -- ($(e) + (7,-0.48)$) -- ($(e) + (3,-0.48)$);

     \draw [thick, red] ($(e) + (5.5,-0.08)$) -- ($(e) + (5.5,2)$);
		     \draw [fill,red] ($(e) + (5.5,-0.08)$) circle (1pt);
		      \draw [thick, red, -Stealth] ($(e) + (5.5,0.8)$) -- ($(e) + (5.5,0.9)$);

	    \end{tikzpicture}
\caption{In the top line we derive a selection rule on $T^3$ for a line $\scrL_q$ which has charge $q$ under a $\ZZ_N$ 1-form symmetry. In the second line, we demonstrate how the same argument fails if one repeats the same manipulations with a disk operator $\scrD$.  \label{fig:1formselectionruletorus}}
\end{figure}
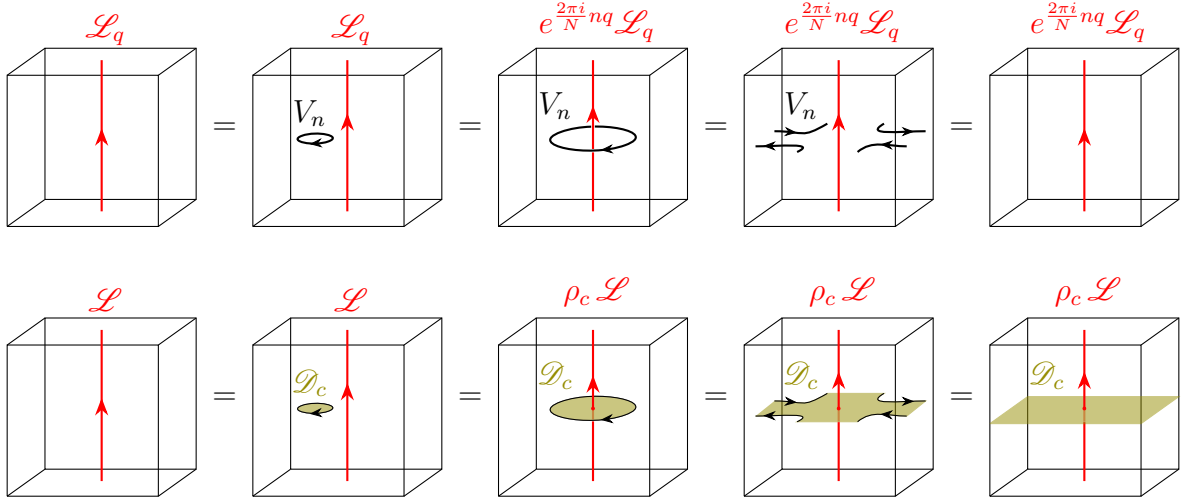

Now consider the theory on $T^d$ where $\scrL_q$ wraps a $S^1$ generator of $H_1(T^d,\IZ)$. We can nucleate a 1-form symmetry operator $V_n(S^{d-2})$ as in Figure~\ref{fig:1formselectionruletorus}. Because of the topology of the torus, we can pull the nucleated $V_n$ across the entire $T^{d-1}$ that is orthogonal to the $\scrL_q(S^1)$, giving the relation 
\eq{
\langle \scrL_q(S^1)\rangle =e^{\frac{2\pi i}{N}nq}\, \langle \scrL_q(S^1)\rangle \,,
}
so that $\scrL_q(S^1)$ has vanishing expectation value when $q \not=0$ mod $N$.

On general grounds one expects such selection rules to be violated if a 1-form symmetry is broken. To gain some intuition, let us consider a pure gauge theory (with 1-form symmetry) perturbed by charged matter fields which break the 1-form symmetry explicitly. The 1-loop functional determinant describing the fluctuations of charged matter fields can be captured by formally summing over Wilson lines, with some weight, in pure gauge theory~\cite{PhysRev.80.440,PhysRev.84.108,PhysRev.82.664,PhysRevLett.66.1669,BERN1992451,STRASSLER1992145} (see Appendix~\ref{app:lattice} for a demonstration of this within the hopping parameter expansion in lattice gauge theory). Such dynamical charged particle worldlines give rise to new contributions to contractible and non-contractible Wilson loops alike, and are responsible for the violation of the selection rules enjoyed by the theory with 1-form symmetry. Perturbing a theory with unbroken 1-form symmetry and tensionful confining strings by heavy charged matter fields dramatically changes the long-distance behavior of Wilson lines from area to perimeter law. Similarly, the expectation value of a single Polyakov loop on a large circle of length $L$ becomes nonzero, scaling like $N_f \, e^{-M L}$, where $N_f$ is the number of charged matter fields of mass $M$.

Now let us consider the situation where a gauge theory with an exact 1-form symmetry is perturbed by heavy charged fields transforming under a flavor symmetry $G^\0$, but with non-trivial symmetry fractionalization. In this case, Wilson lines transform in projective representations of the flavor symmetry, and there is a sense in which they are `charged' under disk operators. One might be tempted to conclude that Wilson loops, since they carry charges under a global symmetry, obey the same selection rules as in the pure gauge theory without matter. However, this is not the case, and symmetry fractionalization alone does not prevent the charged matter fields from screening the Wilson loop, turning area to perimeter law. 

Symmetry fractionalization can, however, imply selection rules for line operators on $T^d$ in the presence of certain $G^\0$ backgrounds.\footnote{This derivation can also be extended to more general oriented manifolds $S^1\times M_{d-1}$.} Such selection rules were studied in~\cite{Cherman:2017tey} in the context of $SU(N_c)$ QCD with $N_f$ flavors of fundamental quarks (with $\text{gcd}(N_f,N_c) \not =1$), and were dubbed `color-flavor-center' symmetries. Here we derive the same selection rules using topological operators, in a way which generalizes to any theory with symmetry fractionalization.\footnote{We are grateful to Nabil Iqbal for suggesting to one of the authors that one may be able to understand these selection rules from the viewpoint of generalized symmetries.  } 

Consider the case where a 0-form symmetry $G^\0$ is fractionalized on a line $\scrL$ which is neutral under any 1-form symmetry. To start, we could attempt to repeat the procedure described above, but replacing the 1-form symmetry operator with the disk operator $\scrD_c$ labeled by $c \in \Gamma$. This leads to the relation (see Figure~\ref{fig:1formselectionruletorus})
\eq{
\langle \scrL(S^1)\rangle = \rho_c \langle \scrD_c(T^{d-1})\,\scrL(S^1)\rangle~.
}
Rather than a selection rule for a single expectation value, we simply get a relation between correlation functions computed with or without the insertion of the disk operator.

To find a genuine selection rule, we \emph{start} in a $G^\0$-twisted sector by performing the path integral with a symmetry operator $U_g$ inserted on a non-contractible codimension-1 cycle. Recall that such twisted sectors are labeled by conjugacy classes $[g]$ --- in the absence of other operator insertions, nucleating a contractible symmetry operator $U_k$ just has the effect of conjugating $g \to k^{-1}gk$. 

\begin{figure}[t!]
\centering
  \begin{tikzpicture}[scale=.75]
  	\coordinate (a) at (0,0);
	\coordinate (b) at (7,0);
	\coordinate (c) at (14,0);
	\coordinate (d) at (3.5,-7);
	\coordinate (e) at (10.5,-7);
	
  %%%%%%%%%%% (a) %%%%%%%%%%%%%%%%%%%%%%%	
 		
		\draw[black] ($(a) + (4,-1.685)$) -- ($(a) + (4,2.315)$) -- ($(a) + (8,2.315)$) -- ($(a) + (8,-1.685)$) -- ($(a) + (4,-1.685)$);		
		\draw[black] ($(a) + (3,-2.4)$) -- ($(a) + (4,-1.685)$);
		\draw[black] ($(a) + (7,-2.4)$) -- ($(a) + (8,-1.685)$);
		\draw[black] ($(a) + (3,1.6)$) -- ($(a) + (4,2.315)$);
		\draw[black] ($(a) + (7,1.6)$) -- ($(a) + (8,2.315)$);	

              \draw [thick, red, opacity=0.5] ($(a) + (5.5,0.03)$) -- ($(a) + (5.5,-.4)$);
		 
        \filldraw[teal ,opacity=0.5] ($(a) + (7,-.4)$) -- ($(a) + (8,0.315)$) -- ($(a) + (4,0.315)$) -- ($(a) + (3,-.4)$) -- ($(a) + (7,-.4)$);
        \draw[thick, color = teal, -Stealth] ($(a) + (3.5,-0.3)$) -- ($(a) + (3.5, 0.1)$);
        \node[color = teal] at ($(a) + (8.35,0.3)$) {$g$};

        \draw[black] ($(a) + (3,-2.4)$) -- ($(a) + (3,1.6)$) -- ($(a) + (7,1.6)$) -- ($(a) + (7,-2.4)$) -- ($(a) + (3,-2.4)$);
               
        \draw [thick, red] ($(a) + (5.5,0.03)$) -- ($(a) + (5.5,2)$);
        \draw[fill, red] ($(a) + (5.5,0.03)$) circle (1pt); 

        \draw[thick, red] ($(a) + (5.5, -.41)$) -- ($(a) + (5.5, -2)$);       
        \draw [thick, red, -Stealth] ($(a) + (5.5,.5)$) -- ($(a) + (5.5,.6)$);
        \node[red] at ($(a) + (5.6,2.8)$) {$\scrL$};
         \node at ($(a) + (9.25, 0.2)$) {$=$};
         	
  %%%%%%%%%%% (b) %%%%%%%%%%%%%%%%%%%%%%%
           
         		\draw[black] ($(b) + (4,-1.685)$) -- ($(b) + (4,2.315)$) -- ($(b) + (8,2.315)$) -- ($(b) + (8,-1.685)$) -- ($(b) + (4,-1.685)$);		
		\draw[black] ($(b) + (3,-2.4)$) -- ($(b) + (4,-1.685)$);
		\draw[black] ($(b) + (7,-2.4)$) -- ($(b) + (8,-1.685)$);
		\draw[black] ($(b) + (3,1.6)$) -- ($(b) + (4,2.315)$);
		\draw[black] ($(b) + (7,1.6)$) -- ($(b) + (8,2.315)$);	
        
         \draw [thick, red, opacity=0.5] ($(b) + (5.5,0.03)$) -- ($(b) + (5.5,-.4)$);
		 
        \filldraw[teal ,opacity=0.5] ($(b) + (7,-.4)$) -- ($(b) + (8,0.315)$) -- ($(b) + (4,0.315)$) -- ($(b) + (3,-.4)$) -- ($(b) + (7,-.4)$);
        \draw[thick, color = teal, -Stealth] ($(b) + (3.5,-0.3)$) -- ($(b) + (3.5, 0.1)$);
        \node[color = teal] at ($(b) + (8.35,0.3)$) {$g$};

        \draw[black] ($(b) + (3,-2.4)$) -- ($(b) + (3,1.6)$) -- ($(b) + (7,1.6)$) -- ($(b) + (7,-2.4)$) -- ($(b) + (3,-2.4)$);
               
        \draw [thick, red] ($(b) + (5.5,0.03)$) -- ($(b) + (5.5,2)$);
        \draw[fill, red] ($(b) + (5.5,0.03)$) circle (1pt); 
     
        \draw[thick, red] ($(b) + (5.5, -.41)$) -- ($(b) + (5.5, -2)$);       
        \draw [thick, red, -Stealth] ($(b) + (5.5,.5)$) -- ($(b) + (5.5,.6)$);
        \node[red] at ($(b) + (5.6,2.8)$) {$\scrL$};
         \node at ($(b) + (9.25, 0.2)$) {$=$};

            \filldraw[dgreen,opacity=0.5] ($(b) + (4.7,-0.9)$) [partial ellipse=0:360:.6];
  		\draw[dgreen,opacity=0.5] ($(b) + (4.7,-0.9)$) [partial ellipse=180:360:0.6 and 0.18];
   		\draw[dgreen,dotted] ($(b) + (4.7,-0.9)$) [partial ellipse=0:180:0.6 and 0.18];
	   	\draw[thick, color = dgreen, -Stealth] ($(b) + (4.8,-1.25)$) -- ($(b) + (4.9,-1.55)$);
		 \node[color = dgreen] at ($(b) + (3.8,-1)$) {$h$};

  %%%%%%%%%%% (c) %%%%%%%%%%%%%%%%%%%%%%%

         		\draw[black] ($(c) + (4,-1.685)$) -- ($(c) + (4,2.315)$) -- ($(c) + (8,2.315)$) -- ($(c) + (8,-1.685)$) -- ($(c) + (4,-1.685)$);		
		\draw[black] ($(c) + (3,-2.4)$) -- ($(c) + (4,-1.685)$);
		\draw[black] ($(c) + (7,-2.4)$) -- ($(c) + (8,-1.685)$);
		\draw[black] ($(c) + (3,1.6)$) -- ($(c) + (4,2.315)$);
		\draw[black] ($(c) + (7,1.6)$) -- ($(c) + (8,2.315)$);

        \draw [thick, red, opacity=0.5] ($(c) + (5.5,0.03)$) -- ($(c) + (5.5,-.4)$);
		 
        \filldraw[teal ,opacity=0.5] ($(c) + (7,-.4)$) -- ($(c) + (8,0.315)$) -- ($(c) + (4,0.315)$) -- ($(c) + (3,-.4)$) -- ($(c) + (7,-.4)$);
        \draw[thick, color = teal, -Stealth] ($(c) + (3.5,-0.3)$) -- ($(c) + (3.5, 0.1)$);
        \node[color = teal] at ($(c) + (8.35,0.3)$) {$g$};

        \draw[black] ($(c) + (3,-2.4)$) -- ($(c) + (3,1.6)$) -- ($(c) + (7,1.6)$) -- ($(c) + (7,-2.4)$) -- ($(c) + (3,-2.4)$);

            \draw [thick, red, opacity=0.5] ($(c) + (5.5,-1.1)$) -- ($(c) + (5.5,-1.55)$);             
 		
	     \filldraw[dgreen ,opacity=0.5] ($(c) + (7,-1.5)$) -- ($(c) + (8,-0.785)$) -- ($(c) + (4,-0.785)$) -- ($(c) + (3,-1.5)$) -- ($(c) + (7,-1.5)$);
         \draw[thick, color = dgreen, -Stealth] ($(c) + (3.5, -1.4)$) -- ($(c) + (3.5, -1)$);
        \node[color = dgreen] at ($(c) + (8.35,-0.8)$) {$h$};

              \draw [thick, red, opacity=0.5] ($(c) + (5.5,0.76)$) -- ($(c) + (5.5,1.15)$);
        
         \filldraw[dgreen ,opacity=0.5] ($(c) + (7,0.75)$) -- ($(c) + (8,1.465)$) -- ($(c) + (4,1.465)$) -- ($(c) + (3,0.75)$) -- ($(c) + (7,0.75)$);
         \draw[thick, color = dgreen, -Stealth] ($(c) + (3.5, 0.85)$) -- ($(c) + (3.5,0.45)$);
        \node[color = dgreen] at ($(c) + (8.35,1.5)$) {$h$};
        
        \draw [thick, red] ($(c) + (5.5,1.15)$) -- ($(c) + (5.5,2)$);
        \draw[fill, red] ($(c) + (5.5,1.15)$) circle (1pt); 
  
        \draw [thick, red] ($(c) + (5.5,0.03)$) -- ($(c) + (5.5,0.735)$);
        \draw [thick, red, -Stealth] ($(c) + (5.5,.5)$) -- ($(c) + (5.5,.6)$);
        \draw[fill, red] ($(c) + (5.5,0.03)$) circle (1pt); 
      
         \draw [thick, red] ($(c) + (5.5,-1.1)$) -- ($(c) + (5.5,-0.42)$);
        \draw[fill, red] ($(c) + (5.5,-1.1)$) circle (1pt); 
     
        \draw[thick, red] ($(c) + (5.5, -1.5)$) -- ($(c) + (5.5, -2)$);       
        \node[red] at ($(c) + (5.6,2.8)$) {$\scrL$};

  %%%%%%%%%%% (d) %%%%%%%%%%%%%%%%%%%%%%%

         		\draw[black] ($(d) + (4,-1.685)$) -- ($(d) + (4,2.315)$) -- ($(d) + (8,2.315)$) -- ($(d) + (8,-1.685)$) -- ($(d) + (4,-1.685)$);		
		\draw[black] ($(d) + (3,-2.4)$) -- ($(d) + (4,-1.685)$);
		\draw[black] ($(d) + (7,-2.4)$) -- ($(d) + (8,-1.685)$);
		\draw[black] ($(d) + (3,1.6)$) -- ($(d) + (4,2.315)$);
		\draw[black] ($(d) + (7,1.6)$) -- ($(d) + (8,2.315)$);

        \draw[black] ($(d) + (3,-2.4)$) -- ($(d) + (3,1.6)$) -- ($(d) + (7,1.6)$) -- ($(d) + (7,-2.4)$) -- ($(d) + (3,-2.4)$);  

          \draw [thick, red, opacity=0.5] ($(d) + (5.5,-1)$) -- ($(d) + (5.5,-1.4)$);

		\node[color = teal] at ($(d) + (8.35,-0.7)$) {$g$};
        \node[color = dgreen] at ($(d) + (8.35,0.4)$) {$h$};
        \node[color = dgreen] at ($(d) + (8.35,1.55)$) {$h$};

  \filldraw[teal ,opacity=0.5] ($(d) + (7,-1.4)$) -- ($(d) + (8,-0.685)$) -- ($(d) + (4,-0.685)$) -- ($(d) + (3,-1.4)$) -- ($(d) + (7,-1.4)$);
         \draw[thick, color = teal, -Stealth] ($(d) + (3.5, -1.3)$) -- ($(d) + (3.5, -0.9)$);

          \draw [thick, red, opacity=0.5] ($(d) + (5.5,0.7)$) -- ($(d) + (5.5,1.2)$);
             \draw [thick, red, opacity=0.5] ($(d) + (5.5,-0.3)$) -- ($(d) + (5.5,0.25)$);
             
        \filldraw[dgreen ,opacity=0.5] ($(d) + (7,0.75)$) -- ($(d) + (8,1.465)$) -- ($(d) + (4,1.465)$) -- ($(d) + (3,0.75)$) -- ($(d) + (7,0.75)$);
        \draw[thick, color =dgreen, -Stealth] ($(d) + (3.5,0.85)$) -- ($(d) + (3.5,0.45)$);
        \filldraw[dgreen ,opacity=0.5] ($(d) + (7,-0.3)$) -- ($(d) + (8,0.415)$) -- ($(d) + (4,0.415)$) -- ($(d) + (3,-0.3)$) -- ($(d) + (7,-0.3)$);
        \draw[thick, color =dgreen, -Stealth] ($(d) + (3.5,-0.2)$) -- ($(d) + (3.5,0.2)$);

 		\draw [thick, red] ($(d) + (5.5,1.1)$) -- ($(d) + (5.5,2)$);
		\draw[fill, red] ($(d) + (5.5,1.1)$) circle (1pt);
       
		\draw [thick, red] ($(d) + (5.5,0.05)$) -- ($(d) + (5.5,0.74)$);
		\draw [thick, red, -Stealth] ($(d) + (5.5,0.5)$) -- ($(d) + (5.5,0.6)$);
        \draw[fill, red] ($(d) + (5.5,0.05)$) circle (1pt);

        \draw [thick, red] ($(d) + (5.5,-1)$) -- ($(d) + (5.5, -0.32)$);
        \draw[fill, red] ($(d) + (5.5, -1)$) circle (1pt);                
      
        \draw[thick, red] ($(d) + (5.5, -1.41)$) -- ($(d) + (5.5, -2)$);     
        \node[red] at ($(d) + (5.6,2.8)$) {$\chi_{h,g}\, \scrL$};
         \node at ($(d) + (9.25, 0.2)$) {$=$};
		\node at ($(d) + (2, 0.2)$) {$=$};
	
  %%%%%%%%%%% (e) %%%%%%%%%%%%%%%%%%%%%%%

		\draw[black] ($(e) + (4,-1.685)$) -- ($(e) + (4,2.315)$) -- ($(e) + (8,2.315)$) -- ($(e) + (8,-1.685)$) -- ($(e) + (4,-1.685)$);		
		\draw[black] ($(e) + (3,-2.4)$) -- ($(e) + (4,-1.685)$);
		\draw[black] ($(e) + (7,-2.4)$) -- ($(e) + (8,-1.685)$);
		\draw[black] ($(e) + (3,1.6)$) -- ($(e) + (4,2.315)$);
		\draw[black] ($(e) + (7,1.6)$) -- ($(e) + (8,2.315)$);

          \draw [thick, red, opacity=0.5] ($(e) + (5.5,0.03)$) -- ($(e) + (5.5,-.4)$);
          
        \filldraw[teal ,opacity=0.5] ($(e) + (7,-.4)$) -- ($(e) + (8,0.315)$) -- ($(e) + (4,0.315)$) -- ($(e) + (3,-.4)$) -- ($(e) + (7,-.4)$);
        \draw[thick, color = teal, -Stealth] ($(e) + (3.5,-0.3)$) -- ($(e) + (3.5, 0.1)$);
        \node[color = teal] at ($(e) + (8.35,0.3)$) {$g$};

        \draw[black] ($(e) + (3,-2.4)$) -- ($(e) + (3,1.6)$) -- ($(e) + (7,1.6)$) -- ($(e) + (7,-2.4)$) -- ($(e) + (3,-2.4)$);
               
        \draw [thick, red] ($(e) + (5.5,0.03)$) -- ($(e) + (5.5,2)$);
        \draw[fill, red] ($(e) + (5.5,0.03)$) circle (1pt); 
       
        \draw[thick, red] ($(e) + (5.5, -.41)$) -- ($(e) + (5.5, -2)$);       
        \draw [thick, red, -Stealth] ($(e) + (5.5,.5)$) -- ($(e) + (5.5,.6)$);
        \node[red] at ($(e) + (5.6, 2.8)$) {$\chi_{h,g}\, \scrL$};
    \end{tikzpicture}
\caption{This Figure illustrates how to derive selection rules for a line operator $\scrL$ (red) wrapped along a non-contractible $S^1$ in the $g$-twisted  sector (teal). First we nucleate a $U_h$ defect (green) with $h$ in the centralizer of $g$. Then, we can use Eq.~\eqref{eq:commutator} (depicted in Figure~\ref{fig:commutator} and Figure~\ref{fig:commutator_disk} using disk operators) to commute $U_h$ and $U_g$, returning to the original configuration but with a net phase.
\label{fig:torusselectionrule} }
\end{figure}
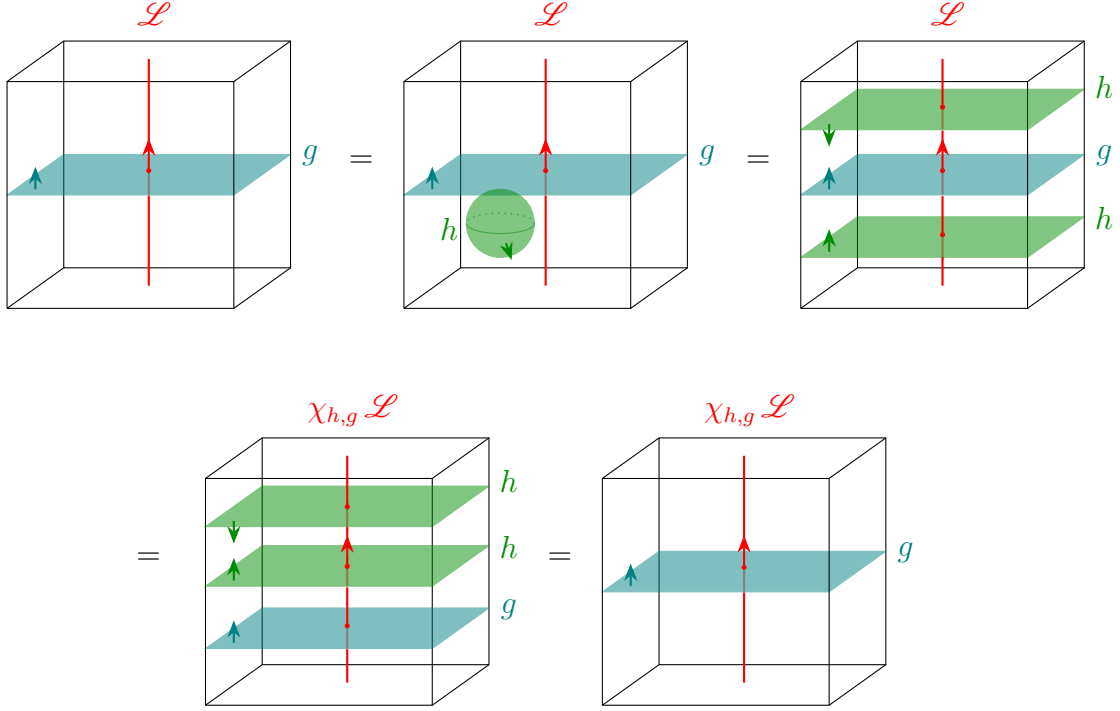

In the current context, we insert $U_g$ on $T^{d-1}$ and consider the line operator $\scrL$ on the orthogonal $S^1$. Now we nucleate a contractible symmetry operator $U_h$ on $T^d$. We take $h$ to be in the centralizer $C(g)$ of $g$, in other words $gh = hg$, and assume that the associated symmetry operators wrapped on $T^{d-1}$ commute in the absence of the line $\scrL$. Then following the steps in Figure~\ref{fig:torusselectionrule}, we find that the line operator expectation value in the $g$-twisted sector is equal to itself times a phase,
\begin{equation}
    \langle \scrL(S^1) \rangle_{g} =  \chi_{h,g}\, \langle  \scrL(S^1) \rangle_{g}\,, 
\end{equation}
where the phase factor $\chi_{h,g}$ captures the failure of $U_g$ and $U_h$ to commute in the presence of the line $\scrL$, as explained near Eq.~\eqref{eq:commutator}. If $\chi_{h,g}$ is non-trivial, the above twisted-sector expectation value identically vanishes. In other words, 
\begin{equation}\label{torusselectionrule}
    \langle \scrL(S^1) \rangle_g = 0 \quad \text{ if } \exists \ h \in C(g) \text{ such that } \chi_{h,g} \not =1\,.  
\end{equation}
If we think in terms of the lift to $\hatG^\0$, we can satisfy the above condition by finding pairs of group elements whose lifts only commute up to elements of $\Gamma$. An immediate consequence is that if $\hatG^\0$ is abelian then we cannot derive any selection rules in twisted sectors.

%%%%%%%%%%%%%%%%%%%%%%%%%%%%%%%%%%%%%%%%%%%%%%%%%%%%%%
\subsubsection{Gauge Theory Examples}
%%%%%%%%%%%%%%%%%%%%%%%%%%%%%%%%%%%%%%%%%%%%%%%%%%%%%%

In the following we consider various examples of the above selection rule which arise due to symmetry fractionalization in gauge theories. 

\bigskip
%%%%%%%%%%%%%%%%%%%%%%%%%%%%%%%%
\noindent
\underline{\emph{$G^\0 = SU(N)/\ZZ_k$}:}
%%%%%%%%%%%%%%%%%%%%%%%%%%%%%%%%

\smallskip
\noindent
Perhaps the most well-known examples of symmetry fractionalization involve $PSU(N)$ or more generally $G^\0 = SU(N)/\ZZ_k$ flavor symmetries. For instance in Section~\ref{sec:qed_junction}, we discussed the case of QED with $N$ complex scalars of charge one. Another example is the $\mathbb{CP}^{N-1}$ non-linear sigma model, which can be realized as the above abelian Higgs model with a constraint $\sum_i |\phi_i|^2 = 1$. Both theories have faithfully acting symmetry $PSU(N)$ \cite{Gaiotto:2017yup}. We can consider the theory on the torus $T^d$ twisted by a defect corresponding to some element $g \in PSU(N)$. Wilson lines obey the selection rule \eqref{torusselectionrule} if we can find $g,h \in PSU(N)$ such that $\chi_{h,g}\neq 1$. As we show in Appendix~\ref{app:clockandshift}, when $k=N$ there is a single conjugacy class $[g]$ where $\chi_{h,g} = e^{\frac{2\pi i}{k}}$ for some $h$. This pair can be taken to lift to the clock and shift matrices in $SU(N)$. In such a twisted sector, we have
\begin{equation}
    \left< W_q(S^1) \right>_{g} = 0, \quad  q=1,2, \dots, N-1~.
    \label{PSU(N)}
\end{equation}
These selection rules are visible in Monte Carlo lattice simulations of the $\ICP^{N-1}$ model with clock-twisted boundary conditions~\cite{Fujimori:2019skd,Misumi:2019upg,Fujimori:2020zka}. 

Similarly, one can consider $SU(N_c)$ QCD coupled to $N_f$ color-fundamental scalars (or fermions) of equal mass transforming under a $SU(N_f)$ flavor symmetry~\cite{Cherman:2017tey}. In this case, the global symmetry is $G^\0 = SU(N_f)/\ZZ_{\gcd(N_c, N_f)}$. As in the previous example, Wilson lines in representations whose $N$-ality mod $k=\gcd(N_c,N_f)$ is non-trivial have vanishing one-point functions in certain twisted sectors. In Appendix~\ref{app:clockandshift} we show how for $1<k< N$ there is a continuous family of conjugacy classes whose twisted sectors host selection rules for endable Wilson lines. When $N_f =1$ there is only a $U(1)$ baryon symmetry. In this case, the fractionalization phases can be made trivial by choosing counterterms appropriately, all $\chi_{h,g}  = 1$, and there are no selection rules on $T^d$.

\bigskip
\bigskip
%%%%%%%%%%%%%%%%%%%%%%%%%%%%%%%%
\noindent
\underline{\emph{$G^\0 = O(2)$}:}
%%%%%%%%%%%%%%%%%%%%%%%%%%%%%%%%

\smallskip
\noindent
Consider the $U(1)$ gauge theory studied in Refs.~\cite{Cherman:2020hbe,Cherman:2024exo}, consisting of two complex scalar fields $\phi_\pm$ with charges $\pm 1$ coupled to a neutral scalar $\phi_0$ via the cubic coupling
\begin{equation}
\epsilon \, \phi_0 \, \phi_+ \, \phi_- + \text{c.c.}   
\end{equation}
The theory has a $U(1)$ global symmetry under which $\phi_0$ has charge 2, while the gauge-non-invariant fields $\phi_\pm$ have charge $-1$,
\begin{equation}
\widehat{U(1)}: \ \phi_0 \to e^{2i\alpha}\, \phi_0\,,\  \phi_\pm \to e^{-i\alpha} \, \phi_\pm\,.      
\end{equation}
All gauge-invariant operators such as $\phi_+\phi_-$ have even charges under this $\widehat{U(1)}$. In addition, there is a $\ZZ_2$ global symmetry that leaves the gauge field alone and acts as\footnote{This is the diagonal combination of the charge conjugation and `$(\ZZ_2)_F$' symmetries discussed in~\cite{Cherman:2020hbe,Cherman:2024exo}. One can also consider combining the above action with $\phi_\pm \to \pm \phi_\pm$ and $\phi_0 \to - \phi_0$ to obtain the group $Pin^-(2)$.}
\begin{equation}
 \ZZ_2^\CC:\    \phi_\pm \to  \phi_\mp^* \,, \quad \phi_0 \to \phi_0^*\,.
\end{equation}
Together these symmetries form the group $Pin^+(2)$, which is a double cover of $O(2)$ (see Appendix~\ref{app:O(2)} for more details about the projective representations of $O(2)$). The  center $Z( Pin^+(2)) = \ZZ_2$ acts by multiplying $\phi_\pm \to - \phi_\pm$ and coincides with a gauge transformation. The faithfully acting flavor symmetry is thus 
\begin{equation}
    O(2) =\frac{Pin^+(2)}{\ZZ_2} =  \frac{\widehat{U(1)}}{\ZZ_2}\rtimes \ZZ_2^\CC\,,
\end{equation}
As a result, $O(2)$ is fractionalized on the fundamental Wilson line, which has the worldline anomaly inflow
\begin{equation}
    \mathcal A = i\pi \int \frac{F}{2\pi}\,,
\end{equation}
where $F$ is the $U(1)$ field strength. This is the same anomaly as the particle on a circle at $\theta = \pi$~\cite{Gaiotto:2017yup}. 

Now let $y$ denote the generator of the $\ZZ_2$ `reflection' and $\widehat r_{\pi/2} = e^{i \pi/2}$ the generator of the $\widehat{\ZZ_4} \subset \widehat{U(1)}$. The latter generates the $\ZZ_2$ center of the faithfully acting $O(2)$ symmetry. However, in the extended group $Pin^+(2)$ these two generators only commute up to the center,
\begin{equation}
    \widehat r_{\pi/2}\, y = y \,\widehat r_{-\pi/2}  = \widehat{r}_\pi\,  y\, \widehat r_{\pi/2} \,.
\end{equation}
This immediately implies that the expectation value of the charge-1 Polyakov loop in the $\ZZ_2 \subset U(1) \subset O(2)$-twisted sector vanishes, 
\begin{equation}
    \langle W(S^1) \rangle_{(-1)} = 0 \,. 
\end{equation}

This example generalizes to $SO(2N)$ gauge theory with a complex scalar field $\Phi$ in the vector representation. This theory also has a $U(1)$ global symmetry which is conjugated by the action of charge conjugation so that $\Phi$ transforms faithfully under 
\eq{
\CG=\frac{SO(2N)_g\times (U(1)\rtimes \IZ_2^\CC)}{\IZ_2}~. 
}
Any potential for $\Phi$ that depends only on $|\Phi|^2$ is $\CG$-preserving. In this theory, the $O(2)$ global symmetry is fractionalized on the Wilson line in the vector representation,  and the above computation implies that its expectation value in the $(-1)\in O(2)$-twisted sector vanishes identically: 
\eq{
\langle W_{\mathbf{2N}}(S^1)\rangle_{(-1)}=0\,. 
}

A slight variation occurs when there is a shared $\IZ_N$ between the gauge group and the non-faithfully acting flavor symmetry, 
\begin{equation}
\CG=\frac{G_g\times G^\0}{\IZ_N}\rtimes \IZ_2^\CC\,,
\end{equation}
where now the $\IZ_2^\CC$ necessarily acts non-trivially on the quotient in $G_g$. In this case, we get selection rules for Wilson lines in real representations which can either by irreducible, or direct sums of complex irreducible reps. 

For instance, consider $SU(N)$ gauge theory with a (potentially massive) Dirac fermion in the fundamental representation. In this theory, the fermions transform faithfully under 
\eq{
\CG=\frac{SU(N)_g\times U(1)}{\IZ_{N}}\rtimes \IZ_2^\CC~,
}
where here $U(1)$ is the vector-like symmetry that acts as $\psi\mapsto e^{i \alpha}\psi$. The Wilson lines 
$W_{R\oplus \bar R}$ are invariant under $\CC$ since it acts non-trivially on the representations $\CC:R\mapsto {\bar{R}}$. Let $R$ have $N$-ality $n$. To obtain a selection rule in the sector twisted by $g \in U(1)/\ZZ_N$, it's lift $\widehat{g}$ must satisfy 
\begin{equation}
\CC \, \widehat{g}^{\,n} =\widehat{g}^{\,-2n}\, \widehat{g}^{\,n} \CC \ \Longrightarrow \ \widehat{g}^{\,-2n} \in \ZZ_N\,.   
\end{equation}
On the other hand, the representations $R$ and $\bar{R}$ must transform by the same phase, which requires $\widehat{g}^{\,2n} \in \ZZ_2$ and hence $N \in 2\ZZ$. Finally, we require this phase to be the same regardless of what lift we choose. This only happens when $n =N/2$, and $\widehat g = e^{\frac{2\pi i}{2N}}$ (modulo $\ZZ_N$) is a lift of $g = (-1) \in U(1)/\ZZ_N$. Hence, we have the selection rule
\begin{equation}
    \langle W_{R \oplus \bar R}(S^1) \rangle_{(-1)} = 0\,,
\end{equation}
for representations $R$ with $N$-ality $N/2$. Since we can decompose $W_{R\oplus \bar{R}}=W_R+W_{\bar{R}}$, this is equivalent to the relation 
\begin{equation}
    \langle W_R(S^1) \rangle_{(-1)} = - \langle W_{\bar{R}}(S^1) \rangle_{(-1)}\,.
\end{equation}

%%%%%%%%%%%%%%%%%%%%%%%%%%%%%%%%
\subsection{Exact Vacuum Degeneracy}
\label{sec:Degeneracy}
%%%%%%%%%%%%%%%%%%%%%%%%%%%%%%%%

Selection rules can typically be reinterpreted as the orthogonality of states that are generated by path integrals computed over open manifolds with insertions of charged operators. When said charged operators are themselves topological, these orthogonal states are guaranteed to be exactly degenerate. Put differently, when topological operators are charged under global symmetries (indicating a specific kind of 't Hooft anomaly), the entire Hilbert space (including its ground state subspace) will have some topology-dependent degeneracy.\footnote{The degeneracy depends on the topology of the spatial manifold, because this dictates the number of distinct charged states one can construct by placing the symmetry operators on non-contractible cycles.} 

The selection rules we derived in the previous section for lines with symmetry fractionalization can also be viewed as the orthogonality of states generated by the lines (in certain $G$-twisted sectors). When these lines are topological, this further implies that these twisted Hilbert spaces enjoy some exact degeneracy which depends on the spatial topology. 

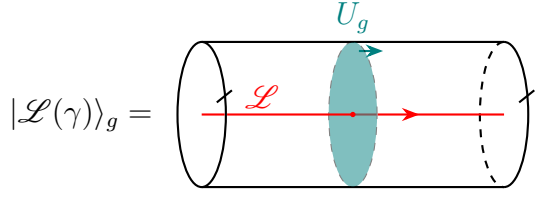
\begin{figure}[t!]
\centering
\begin{tikzpicture}[scale=0.8]

% % % % %   panel coordinates

  	\coordinate (a) at (0,0);
	\coordinate (b) at (8,0);
	\coordinate (c) at (16,0);
	\coordinate (d) at (3.5,-7);
	\coordinate (e) at (10.5,-7);
	
% % % % % (a) % % % % % 		

% cylinder 
	
	\draw[black,thick] ($(a) + (-2.5,0)$) [partial ellipse=0:360:0.4 and 1.2];
	\draw[black,thick] ($(a) + (2.5,0)$) [partial ellipse=-90:90:0.4 and 1.2];
	\draw[black,thick,dashed] ($(a) + (2.5,0)$) [partial ellipse=90:270:0.4 and 1.2];
	\draw[black,thick] ($(a) + (-2.5,1.2)$) -- ($(a) + (2.5,1.2)$);
	\draw[black,thick] ($(a) + (-2.5,-1.2)$) -- ($(a) + (2.5,-1.2)$);
	\draw[black,thick] ($(a) + (-2.25,0.2)$) -- ($(a) + (-2,0.4)$);
	\draw[black,thick] ($(a) + (2.75,0.2)$) -- ($(a) + (3,0.4)$);

% defect	

	\draw[thick, red, opacity=0.8] ($(a) + (0,0)$) -- ($(a) + (0.39,0)$);

	\draw[thick, color = teal, -Stealth] ($(a) + (0.1,1.05)$) -- ($(a) + (0.5, 1.05)$);
	\draw[black, dashed,fill = teal,opacity=0.5] ($(a) + (0,0)$) [partial ellipse=0:360:0.4 and 1.2];
    \node[color = teal] at ($(a) + (0,1.6)$) {$U_g$};
    
% line 
	
	\draw[thick, red] ($(a) + (-2.5,0)$) -- ($(a) + (-2.12,0)$);
	\draw[thick, red] ($(a) + (-2.08,0)$) -- ($(a) + (0,0)$);
	\draw[thick, red, -Stealth] ($(a) + (1,0)$) -- ($(a) + (1.1,0)$);
	\draw[fill, red] ($(a) + (0,0)$) circle (1pt);
%	\draw[thick, red, dotted] ($(a) + (0.08,0)$) -- ($(a) + (0.4,0)$);
	\draw[thick, red] ($(a) + (0.41,0)$) -- ($(a) + (2.5,0)$);
	\node[red] at ($(a) + (-1.5,0.3)$) {$\scrL$};

	\node at ($(a) + (-4.5,0)$) {$|\scrL(\gamma)\rangle_{g} = $};

\end{tikzpicture}
\caption{A state in the $g$-twisted Hilbert space on $S^1 \times S^1$ created by inserting a line $\scrL$ on a closed non-contractible loop $\gamma$.  \label{fig:linestate}}
\end{figure}

To see how this arises, consider a (generically non-topological) line $\scrL$ which carries a worldline anomaly for $G$, characterized by a symmetry fractionalization class $\omega \in H^2(BG,U(1))$. We place the theory on a spacetime $S^1 \times M_{d-1}$, where the $S^1$ factor is time, with a Hilbert space $\CH[M_{d-1}]$. For simplicity, we take $M_{d-1} = S^1 \times S^{d-2}$ so that there is (generically) just one non-contractible 1-cycle. Let us study the $g$-twisted Hilbert space $\CH[S^1_g\times S^{d-2}]$ where we apply $g$-twisted boundary conditions along the $S^1$ (i.e. insert $U_g$ on the orthogonal $S^{d-2}$, extended in time). States in this twisted Hilbert space can be generated by evaluating the path integral (with possible operator insertions) on $S^1\times D^{d-1}$ with a $U_g$ operator wrapping the $D^{d-1}$. We will denote the state generated by the path integral with no other operator insertions as $|\mathbbm{1}\rangle_g$:
\eq{
|\mathbbm{1}\rangle_g&=\mathcal{N}\, \dlangle U_g( D^{d-1})\drangle_{S^1\times D^{d-1}}\,,
}
where the double braces denote unnormalized expectation values. The normalization factor is $\mathcal{N} = \dlangle \mathbbm{1}\drangle_{S^1 \times S^{d-1}}^{-1/2}$ where the $S^{d-1}$ is obtained by gluing $D^{d-1} \cup \overline{D}^{d-1}$. Since we're in a non-topological theory, this state depends on the geometry of the $S^1 \times D^{d-1}$. Now consider the states $|\scrL(\gamma)\rangle_g$ that are created by the insertion of the fractionalized line operator along some contour $\gamma$ homologous to the $S^1$ cycle, as depicted in Figure~\ref{fig:linestate}:
\eq{
|\scrL(\gamma) \rangle_g &= \mathcal{N}\, \dlangle  \scrL(\gamma) \, U_g( D^{d-1})\drangle_{S^1\times D^{d-1}}\,.
}
In general, when the line operator $\scrL$ is not topological, the above states also depend on $\gamma$ and are not degenerate with $|\mathbbm{1}\rangle_g$ (the splitting is set by the free energy of the line).

\begin{figure}[t!]
\centering
\begin{tikzpicture}[scale=0.75]

% % % % %   panel coordinates

	\coordinate (b) at (0,0);
	\coordinate (c) at (7.5,0);
	\coordinate (d) at (15,0);
	
% % % % % (b) % % % % % 		

% cylinder 
	
	\draw[black,thick] ($(b) + (-2.5,0)$) [partial ellipse=0:360:0.4 and 1.2];
	\draw[black,thick] ($(b) + (2.5,0)$) [partial ellipse=-90:90:0.4 and 1.2];
	\draw[black,thick,dashed] ($(b) + (2.5,0)$) [partial ellipse=90:270:0.4 and 1.2];
	\draw[black,thick] ($(b) + (-2.5,1.2)$) -- ($(b) + (2.5,1.2)$);
	\draw[black,thick] ($(b) + (-2.5,-1.2)$) -- ($(b) + (2.5,-1.2)$);
	\draw[black,thick] ($(b) + (-2.25,0.2)$) -- ($(b) + (-2,0.4)$);
	\draw[black,thick] ($(b) + (2.75,0.2)$) -- ($(b) + (3,0.4)$);

% defect	

	\draw[thick, red,opacity=0.8] ($(b) + (0,0)$) -- ($(b) + (0.39,0)$);

	\draw[thick, color = teal, -Stealth] ($(b) + (0.1,1.05)$) -- ($(b) + (0.5, 1.05)$);
	\draw[black, dashed,fill = teal,opacity=0.5] ($(b) + (0,0)$) [partial ellipse=0:360:0.4 and 1.2];
	
    \node[color = teal] at ($(b) + (0,1.6)$) {$U_g$};
    
% line 

	\draw[thick, red] ($(b) + (-2.08,0)$) -- ($(b) + (0,0)$);
	\draw[thick, red, -Stealth] ($(b) + (1,0)$) -- ($(b) + (1.1,0)$);
	\draw[fill, red] ($(b) + (0,0)$) circle (1pt);
%	\draw[thick, red, dotted] ($(b) + (0.08,0)$) -- ($(b) + (0.4,0)$);
	\draw[thick, red] ($(b) + (0.41,0)$) -- ($(b) + (2.5,0)$);

% U_h cylinder 
	
	%\draw[dgreen, fill = dgreen, opacity = 0.5] ($(b) + (-2.5,0)$) [partial ellipse=0:360:0.45 and 1.35];
	 \filldraw[dgreen,opacity=0.3] ($(b) + (-2.5,1.25)$) 
	 to [out=-10, in=10, looseness=.65] ($(b) + (-2.5,-1.25)$) -- ($(b) + (2.5,-1.25)$)
	 to [out=10, in=-10, looseness=.65] ($(b) + (2.5,1.25)$);
       
     \filldraw[dgreen,opacity=0.3] ($(b) + (-2.5,1.25)$) 
	 to [out=190, in=170, looseness=.65] ($(b) + (-2.5,-1.25)$) -- ($(b) + (2.5,-1.25)$)
	 to [out=170, in=190, looseness=.65] ($(b) + (2.5,1.25)$);  
	 
	 \draw[thick, color = dgreen, -Stealth] ($(b) + (2.5,1.05)$) -- ($(b) + (2.7, 1.5)$);
	 \node[color = dgreen] at ($(b) + (3.1,1.25)$) {$U_h$};
	
	\draw[thick, red] ($(b) + (-2.5,0)$) -- ($(b) + (-2.12,0)$);
	\node[red] at ($(b) + (-1.5,0.3)$) {$\scrL$};
	
	\node at ($(b) + (3.75,0)$) {$=$};

% % % % % (c) % % % % % 		

% cylinder 
	
	\draw[black,thick] ($(c) + (-2.5,0)$) [partial ellipse=0:360:0.4 and 1.2];
	\draw[black,thick] ($(c) + (2.5,0)$) [partial ellipse=-90:90:0.4 and 1.2];
	\draw[black,thick,dashed] ($(c) + (2.5,0)$) [partial ellipse=90:270:0.4 and 1.2];
	\draw[black,thick] ($(c) + (-2.5,1.2)$) -- ($(c) + (2.5,1.2)$);
	\draw[black,thick] ($(c) + (-2.5,-1.2)$) -- ($(c) + (2.5,-1.2)$);
	\draw[black,thick] ($(c) + (-2.25,0.2)$) -- ($(c) + (-2,0.4)$);
	\draw[black,thick] ($(c) + (2.75,0.2)$) -- ($(c) + (3,0.4)$);

% defect	

	\draw[thick, red,opacity=0.8] ($(c) + (0,0)$) -- ($(c) + (0.39,0)$);

	\draw[thick, color = teal, -Stealth] ($(c) + (0.1,1.05)$) -- ($(c) + (0.5, 1.05)$);
	\draw[black, dashed,fill = teal,opacity=0.5] ($(c) + (0,0)$) [partial ellipse=0:360:0.4 and 1.2];
	
    \node[color = teal] at ($(c) + (0,1.6)$) {$U_g$};
    
% line 

	\draw[thick, red] ($(c) + (-2.08,0)$) -- ($(c) + (0,0)$);
	\draw[thick, red, -Stealth] ($(c) + (1.3,0)$) -- ($(c) + (1.4,0)$);
	\draw[fill, red] ($(c) + (0,0)$) circle (1pt);
%	\draw[thick, red, dotted] ($(c) + (0.08,0)$) -- ($(c) + (0.4,0)$);
	\draw[thick, red] ($(c) + (0.41,0)$) -- ($(c) + (2.5,0)$);

% U_h sphere 
	
  \filldraw[dgreen,opacity=0.5] ($(c) + (0,0)$) [partial ellipse=0:360:1 and 0.6];
  	%	\draw[dgreen] ($(c) + (0,0)$) [partial ellipse=180:360:1 and 0.15];
		\draw[dgreen] ($(c) + (0,0)$) [partial ellipse=-90:90:0.17 and 0.6];
		\draw[dgreen,dotted] ($(c) + (0,0)$) [partial ellipse=90:270:0.17 and 0.6];
   	%	\draw[dgreen,dotted] ($(c) + (0,0)$) [partial ellipse=0:180:0.6 and 0.18];
	   	\draw[thick, color = dgreen, -Stealth] ($(c) + (0.9,.2)$) -- ($(c) + (1.2,.3)$);
		 \node[color = dgreen] at ($(c) + (0.9,-0.8)$) {$U_h$};

	\draw[thick, red] ($(c) + (-2.5,0)$) -- ($(c) + (-2.12,0)$);
	\node[red] at ($(c) + (-1.5,0.3)$) {$\scrL$};
	\draw[fill, red] ($(c) + (1,0)$) circle (1pt);
	\draw[fill, red] ($(c) + (-1,0)$) circle (1pt);
	
	\node at ($(c) + (3.75,0)$) {$=$};
	
	% % % % % (d) % % % % % 		

% cylinder 
	
	\draw[black,thick] ($(d) + (-2.5,0)$) [partial ellipse=0:360:0.4 and 1.2];
	\draw[black,thick] ($(d) + (2.5,0)$) [partial ellipse=-90:90:0.4 and 1.2];
	\draw[black,thick,dashed] ($(d) + (2.5,0)$) [partial ellipse=90:270:0.4 and 1.2];
	\draw[black,thick] ($(d) + (-2.5,1.2)$) -- ($(d) + (2.5,1.2)$);
	\draw[black,thick] ($(d) + (-2.5,-1.2)$) -- ($(d) + (2.5,-1.2)$);
	\draw[black,thick] ($(d) + (-2.25,0.2)$) -- ($(d) + (-2,0.4)$);
	\draw[black,thick] ($(d) + (2.75,0.2)$) -- ($(d) + (3,0.4)$);

% defect	

	\draw[thick, red, opacity=0.8] ($(d) + (0,0)$) -- ($(d) + (0.39,0)$);

	\draw[thick, color = teal, -Stealth] ($(d) + (0.1,1.05)$) -- ($(d) + (0.5, 1.05)$);
	\draw[black, dashed,fill = teal,opacity=0.5] ($(d) + (0,0)$) [partial ellipse=0:360:0.4 and 1.2];
    \node[color = teal] at ($(d) + (0,1.6)$) {$U_g$};
    
% line 
	
	\draw[thick, red] ($(d) + (-2.5,0)$) -- ($(d) + (-2.12,0)$);
	\draw[thick, red] ($(d) + (-2.08,0)$) -- ($(d) + (0,0)$);
	\draw[thick, red, -Stealth] ($(d) + (1,0)$) -- ($(d) + (1.1,0)$);
	\draw[fill, red] ($(d) + (0,0)$) circle (1pt);
%	\draw[thick, red, dotted] ($(d) + (0.08,0)$) -- ($(d) + (0.4,0)$);
	\draw[thick, red] ($(d) + (0.41,0)$) -- ($(d) + (2.5,0)$);
	\node[red] at ($(d) + (-1.2,0.35)$) {$\chi_{h,g}\, \scrL$};

\end{tikzpicture}
\caption{Action of $U_h$ on the state $|\scrL(\gamma)\rangle_g$ in the $g$-twisted sector. \label{fig:stateaction}}
\end{figure}
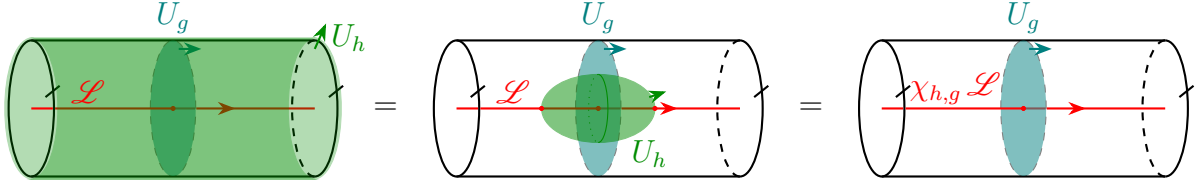

 Since this line carries a projective representation of $G$, the state constructed above might transform in a non-trivial representation of the centralizer $C(g)$ of $g$. Assuming $U_h$ acts trivially on $|\mathbbm{1}\rangle_g$, the action of $U_h$ (with $h \in C(g)$) on the state with the line is
\begin{equation}
U_h(S^1\times S^{d-2})|\scrL(\gamma) \rangle_g=\chi_{\scrL;h,g}|\scrL(\gamma)\rangle_g\,,
\end{equation}
as illustrated in Figure~\ref{fig:stateaction}. In other words, the state $|\scrL(\gamma)\rangle_g$ furnishes a one-dimensional linear representation of $C(g)$.\footnote{Repeated use of the cocycle condition $\omega_{g,h}\,,\omega_{gh,k} = \omega_{h,k}\, \omega_{g,hk}$ shows that $\chi_{h_1,g}\,\chi_{h_2,g} = \chi_{h_1h_2,g}$ for $h_1,h_2 \in C(g)$, so the representation here is indeed linear, not projective.}

In general, different lines will give rise to distinct representations of $C(g)$, and we have the selection rule: 
\begin{equation} \label{eq:fractionalizedlineorthogonality} 
\,_g\langle \scrL'(\gamma')|\scrL(\gamma)\rangle_g=0 \ \text{ if } \exists\, h \in C(g) \text{ such that }\, \chi_{\scrL;h,g} \not = \chi_{\scrL';h,g}\,,
\end{equation}
where $\gamma$ and $\gamma'$ are homologous to the $S^1$ factor of the spatial manifold. A special case is when $\scrL' = \mathbbm{1}$ is the identity operator, in which case the above selection rule can be written as 
\eq{ \label{eq:selectionrule} 
\,_g\langle \mathbbm{1}|\scrL\rangle_g = \langle U_g(S^{d-1})\, \scrL(S^1)\rangle_{S^1\times S^{d-1}} = 0 \text{ if } \chi_{\scrL;h,g} \not= 1\,,
}
which is analogous to Eq.~\eqref{torusselectionrule}. In the above we only considered states in the same twisted sector. Twisted sectors corresponding to elements in the same conjugacy class are isomorphic and related by a unitary operator. 

Now we specialize to the case where the lines $\scrL$ are topological and commute with the Hamiltonian. This further ensures that the orthogonal states in Eq.~\eqref{eq:fractionalizedlineorthogonality} are \emph{exactly} degenerate. The extent of the degeneracy depends on the number of distinct representations of $C(g)$ furnished by the set of lines. 

In this paper, we are especially interested in the scenario where an IR 1-form global symmetry is broken in the UV, but in such a way that there is symmetry fractionalization for the symmetry $G^\0$. In these scenarios, it is often the case that anomalies in the IR that involve the (emergent) 1-form symmetry descend from exact anomalies of the UV theory that exist due to symmetry fractionalization~\cite{Brennan:2023ynm,Brennan:2023vsa,Brennan:2022tyl,Wang:2018qoy,Anber:2020gig,Anber:2021iip,Tanizaki:2017mtm}. In particular, the above discussion is relevant when the theory flows at long distances to a topological quantum field theory (TQFT), with topological ground state degeneracies protected by the emergent 1-form symmetry. Since the 1-form symmetry is not exact, one expects exponentially small splittings of said ground states, at least in the ordinary, untwisted Hilbert space. Our analysis shows that if one instead works in a $G^\0$-twisted Hilbert space, one can enforce (at least some amount of) \emph{exact} degeneracy. Intuitively, if the local operators that are charged under $G^\0$ are very heavy (which is indeed the case in this putative flow to a TQFT), the low-energy dynamics of local observables is insensitive to the twisting by $G^\0$, and it is merely a tool to preserve some IR features (the ground state degeneracy) along the entire RG flow.

%%%%%%%%%%%%%%%%%%%%%%%%%%%%%%%%
\subsubsection{Example: Flows to $\ZZ_N$ Gauge Theory}
\label{sec:ZNdeformations}
%%%%%%%%%%%%%%%%%%%%%%%%%%%%%%%%

To illustrate the idea presented above, we consider various deformations of topological $\ZZ_N$ gauge theory on a spacetime $S^1 \times M_{d-1}$. The pure gauge theory is a TQFT with an exact vacuum degeneracy which depends on the topology of $M_{d-1}$ and which is protected by the spontaneously broken 1-form symmetry (or alternatively through its  spontaneously broken $(d-2)$-form global symmetry). After reviewing the exact degeneracy of the pure gauge theory we consider various UV completions where the higher-form symmetries are explicitly broken, and compare and contrast scenarios with and without symmetry fractionalization. 

In the pure $\ZZ_N$ gauge theory, we have the topological operators $V_k(\Sigma)$ and $\scrL_m(\gamma)$ that generate the $\ZZ_N^\1$ and $\ZZ_N^{(d-2)}$ respectively. These operators obey the correlation function 
\eq{\label{linkZNgaugetheory}
\langle V_k(\Sigma) \, \scrL_m(\gamma)\rangle=e^{\frac{2\pi i k m }{N}{\rm Link}(\Sigma,\gamma)}~,
}
when both $\gamma$ and $\Sigma$ are contractible. Let us consider the Hilbert space $\CH[M_{d-1}]$ on the spatial manifold $M_{d-1}=S^1\times S^{d-2}$. For generic $d$, the theory has $N$ degenerate ground states on such a topology (when $d=3$ this enhances to $N^2$). The states in the Hilbert space can be generated for example by computing the path integral over the manifold $S^1\times D^{d-1}$ with various operators inserted in the bulk. 

Let us begin by choosing a basis of $\CH[M_{d-1}]$ which diagonalizes $V_k(S^{d-2})$, 
\begin{equation} \label{eq:Veigenstates} 
V_k(S^{d-2})|\Psi_m\rangle = e^{\frac{2\pi i m}{N} k} |\Psi_m \rangle\,, \quad m = 0,\ldots,N-1\,. 
\end{equation}
These state can be constructed by computing the path integral on $S^1 \times D^{d-1}$ with the $\ZZ_N$ Wilson line $\scrL_m$ inserted along the $S^1$,
\begin{equation}
|\Psi_m\rangle = \mathcal{N}\, \dlangle \scrL_m(S^1)  \drangle_{S^1 \times D^{d-1}}\,.
\end{equation}
The linking relation between $V_k$ and $\scrL_m$ gives rise to the action in Eq.~\eqref{eq:Veigenstates}. Similarly, we have $\scrL_n(S^1)|\Psi_m \rangle = |\Psi_{m+n} \rangle$, giving rise to the familiar algebra
\begin{equation}
V_k(S^{d-2}) \, \scrL_n(S^1)  = e^{\frac{2\pi i}{N}kn} \, \scrL_n(S^1) \, V_k(S^{d-2})\,. 
\end{equation}
A simple but important observation is that the states $|\Psi_m\rangle$ furnish distinct representations of the $\ZZ_N^\1$ symmetry, and hence they are orthogonal, 
\begin{equation}
\langle \Psi_m |  \Psi_n \rangle = \delta_{m,n}\,.
\end{equation}
These states span the Hilbert space $\CH[S^1 \times S^{d-2}]$. On the other hand, these states are (obviously) degenerate, since $\scrL_m$ is a topological line which commutes with the Hamiltonian (which in this example vanishes). Together, these two facts (the states are orthogonal, and they are related by the action of a line which commutes with the Hamiltonian) immediately tell us that there is an $N$-fold degeneracy in the entire Hilbert space (which in this case is the space of ground states). 

The expectation value of $\scrL_m(S^1)$ on $S^1\times M_{d-1}$ is given by its trace, 
\begin{equation}
\langle \scrL_m(S^1) \rangle_{S^1 \times(S^1 \times S^{d-2})} = \frac{1}{N} \sum_{n=0}^{N-1} \langle \Psi_n|\scrL_m(S^1) | \Psi_n \rangle = \frac{1}{N}\sum_{n=0}^{N-1} \langle \Psi_n|\Psi_{m+n}\rangle = \delta_{m,0} \,,
\end{equation}
so that the selection rule for the charged line $\scrL$ comes from the orthogonality of $|\Psi_m\rangle$. Alternatively, we can work in the basis $|\psi_k\rangle$ that diagonalizes $\scrL_m(S^1)$, constructed by computing the path integral on $D^2 \times S^{d-2}$ with an insertion of the 1-form symmetry generator $V_k(S^{d-2})$. In this basis
\begin{equation}
\scrL_m(S^1)|\psi_k\rangle = e^{\frac{2\pi ik}{N}m}|\psi_k\rangle\,, \quad V_\ell(S^{d-2})|\psi_k\rangle = |\psi_{k+\ell}\rangle \,, \quad \langle \psi_k|\psi_\ell\rangle = \delta_{k,\ell}\,,
\end{equation}
and the selection rule for $\scrL$ comes from the interference between these states. 

At a practical level, the ground state degeneracy in this theory arises because the algebra of topological operators can only be represented on a Hilbert space of non-trivial dimension. The algebra of symmetry operators, and the resulting degeneracy, simply reflect the mixed anomaly between the $\ZZ_N^\1\times \ZZ_N^{(d-2)}$ global symmetries, described by the inflow action
\begin{equation} \label{eq:ZNanomaly} 
\CA = \frac{2\pi i}{N} \int_{M_{d+1}} B_e^{(2)} \cup B_m^{(d-1)}\, . 
\end{equation}  
It is therefore reasonable to expect that the vacuum degeneracy is lifted in UV completions where either (or both) of the symmetries are explicitly broken and the anomaly is trivialized. We now consider different microscopic theories that flow to $\ZZ_N$ gauge theory, and discuss the fate of the topological ground state degeneracy.   

\bigskip
%%%%%%%%%%%%%%%%%%%%%%%%%%%%%%%%
\noindent
\underline{\emph{Breaking $\ZZ_N^{(d-2)}$:}}
%%%%%%%%%%%%%%%%%%%%%%%%%%%%%%%%

\smallskip
\noindent
To start, we UV complete the topological $\ZZ_N$ gauge theory in $d=4$ by the charge-$N$ abelian Higgs model --- $U(1)$ gauge theory coupled to a charge-$N$ complex scalar field $\phi$:
\eq{
S=\int d^4x \, \frac{1}{2g^2}F^2+|D\phi|^2+\lambda(|\phi|^2-v^2)^2~.
}
The potential causes $\phi$ to condense, Higgsing the gauge group $U(1)\to \ZZ_N$. In this UV completion, the $\ZZ_N^\1$ symmetry of the IR $\ZZ_N$ gauge theory is preserved while the $\ZZ_N^{(2)}$ symmetry is explicitly broken.\footnote{See~\cite{Nguyen:2024ikq} for an analysis of other $\ZZ_N^\1$-preserving deformations of $\ZZ_N$ gauge theory and connections to confinement. } Instead, the above theory has a $U(1)^\1$ 1-form magnetic (or `topological') global symmetry with conserved current $\ast J_{\text{top}} = \frac{F}{2\pi}$, which has a mixed anomaly with $\ZZ_N^\1$. If we restrict to a $\ZZ_N^{(1)} \subset U(1)^{(1)}$ subgroup of the topological symmetry, we can write a simple anomaly inflow action 
\eq{\label{MaxwellAnomaly}
\CA=\frac{2\pi i}{N}\int B_e^{(2)} \cup \frac{1}{N}\delta B_{\text{top}}^{(2)} \,. 
}
Despite the fact that we have \emph{some} anomaly in the UV theory, this anomaly is not powerful enough to give rise to any exact vacuum degeneracy on a generic spatial manifold such as the $S^1 \times S^2$ we considered above. States in the deformed theory can still be labeled by their quantum numbers under the $\ZZ_N^{(1)}$ symmetry (like the $|\Psi_m\rangle$ states constructed above). On the other hand, since $\ZZ_N^{(2)}$ is explicitly broken, the states $|\psi_k\rangle$ of the pure $\ZZ_N$ gauge theory are no longer orthogonal, and can mix. Dynamically, this occurs via a process by which an ANO vortex of finite tension $T\sim v^2$ spontaneously appears on the spatial $S^2$. In finite volume these nucleating vortices are simply instantons that tunnel between the $|\psi_k\rangle$ states. As in the case of the double well potential, these dynamical vortices lead to exponentially suppressed splitting between the approximately degenerate vacua. In this case, the the tunneling is suppressed by the factor $\sim e^{-T\text{Area}(S^2)}$ which vanishes in the infinite volume limit. This is also consistent with the fact that the Wilson line is not topological and therefore there is no reason for the $|\Psi_m\rangle$ states to be degenerate.

Similar considerations hold for $4d$ $SU(N)$ QCD with massive quarks and a collection of adjoint scalar fields which condense so that the theory flows to $\ZZ_N$ gauge theory in the IR. In this theory, the IR $\ZZ_N^{(2)}$ is again broken in the UV by dynamical vortices, but unlike the abelian example there is no topological symmetry to take its place.

\bigskip
%%%%%%%%%%%%%%%%%%%%%%%%%%%%%%%%
\noindent
\underline{\emph{Breaking $\ZZ_N^{\1}$ without fractionalization:}}
%%%%%%%%%%%%%%%%%%%%%%%%%%%%%%%%

\smallskip
\noindent
Now consider a different UV completion where we simply deform the $4d$ $\ZZ_N$-gauge theory by coupling it to a massive charge-1 scalar. The $\ZZ_N^{(2)}$ magnetic symmetry remains intact (so that $\scrL$ remains topological, i.e. it commutes with the Hamiltonian). On the other hand, this `UV' theory has no exact 1-form symmetry, but it emerges in the infrared below the mass scale $M$ of the charged matter. Let us study the Hilbert space on a spatial $S^1\times \IR^2$ where the circle has length $L$. Since we have broken $\ZZ_N^\1$ but $\ZZ_N^{(2)}$ is intact, it is convenient to use eigenstates of the topological $\ZZ_N$ Wilson lines $\scrL_m$.

The $N$ vacua in the IR are given by the flat $\ZZ_N$ connections on the $S^1$. This vacuum degeneracy is affected by the fact that the massive particle can run in loops, leading to an effective potential for the holonomy $V_{\rm eff}(\theta)$ where $\theta\in \frac{2\pi}{N}\ZZ$ parameterizes the $\ZZ_N$ holonomies on the circle (see Appendix \ref{app:holonomy} for the derivation) 
\eq{ 
V_{\rm eff}(\theta)=-\frac{2}{L^4}\left(\frac{ML}{2\pi}\right)^{3/2} \cos(\theta)\, e^{-ML}+\mathcal{O}(e^{-2ML})~,
}
From this, we see that the massive charged particle induces an exponentially suppressed splitting of the $N$ formerly degenerate ground states. The unique vacuum corresponds to vanishing holonomy, $\theta = 0$.

\bigskip
%%%%%%%%%%%%%%%%%%%%%%%%%%%%%%%%
\noindent
\underline{\emph{Breaking $\ZZ_N^{\1}$ with fractionalization:}}
%%%%%%%%%%%%%%%%%%%%%%%%%%%%%%%%

\smallskip
\noindent
We now consider a mild generalization of the previous case, where we couple $N_f$ scalars to the $\ZZ_N$ gauge field. We give the scalars a degenerate mass $M$ such that we preserve a $U(N_f)/\ZZ_N$ global symmetry. In what follows, we will focus on the non-abelian flavor symmetry
\begin{equation} \label{eq:fermion_flavor_symmetry}
G^\0 = SU(N_f)/\ZZ_k\,, \quad k \equiv \text{gcd}(N,N_f)\,. 
\end{equation}
The gauge-non-invariant scalars transform in the simply-connected cover $SU(N_f)$, so that the Wilson lines $\scrL_m$ experience symmetry fractionalization. This indicates a worldline anomaly, and (since the lines $\scrL_m$ are topological) a bulk 't Hooft anomaly between $G^\0$ and $\ZZ_N^{(d-2)}$, 
\begin{equation} \label{eq:full_anomaly}
\mathcal{A}_{\scrL_m} = \frac{2\pi i m}{k}\int_\sigma w_2(A)\, \quad \Longrightarrow \quad \mathcal{A}_{\text{bulk}} = \frac{2\pi i}{k}\int_{M_{d+1}} w_2(A) \cup B^{(d-1)}\,. 
\end{equation}
As discussed in Section~\ref{sec:topological_lines}, this induces a mixed anomaly between the $\ZZ_N^{(2)}$ symmetry generated by the lines $\scrL_m$ and the $G^\0$ flavor symmetry. As is well known, the above anomaly can be activated by looking at the subgroup of generalized clock and shift matrices in $SU(N_f)$ (discussed in Appendix~\ref{app:clockandshift}), which descends to a $\ZZ_k^\0 \times \ZZ_k^\0$ subgroup of $G^\0$. From Eq.~\eqref{eq:clockshift_cocycle}, we can read off the relevant phase 
\begin{equation}
\chi_{m; (a_1,b_1),(a_2,b_2) } = e^{\frac{2\pi i}{k}m (a_1 b_2 - a_2 b_1)}  ~.
\end{equation}
We can now apply the general analysis from above to find that the states
\begin{equation}
\,_{(1,b)} \langle \scrL_{m'}|\scrL_m \rangle_{(1,b)}  =  \,_{(a,1)} \langle \scrL_{m'}|\scrL_m \rangle_{(a,1)}  = 0 \text{ if } m' \not= m \text{ mod } k\,,
\end{equation}
for any $a,b$. Since the $\scrL_m$ operators are topological, the above orthogonal states are also degenerate. This implies a $k$-fold degeneracy in the $\ZZ_k^\0\times\ZZ_k^\0$-twisted Hilbert space $\CH[S_{(a,b)}^1\times S^{d-2}]$ with either $a,b=1$.\footnote{We can repeat the analysis on a spatial manifold with a $T^2$ factor to get an $N^2$-fold degeneracy provided we twist the two cycles by the two $\ZZ_k$ factors, for instance by considering $\CH[S_{(1,0)}^1 \times S_{(0,1)}^1 \times S^{d-3}]$. } To reiterate, this degeneracy comes from the fact that in the UV, the full anomaly in Eq.~\eqref{eq:full_anomaly} reduces to a mixed $\ZZ_N^{(d-2)} \times (\ZZ_k^\0 \times \ZZ_k^\0)$ anomaly 
\eq{
\CA_{\text{UV}}= \frac{2\pi i}{k}\int_{M_{d+1}} A_1^{(1)} \cup A_2^{(1)} \cup B^{(d-1)}\,.
}
In the IR, the $G^\0$ flavor symmetry acts trivially and a $\ZZ_N^{(1)}$ 1-form symmetry emerges with the anomaly in Eq.~\eqref{eq:ZNanomaly}. In the IR, we can match the UV anomaly by setting $B_e^{(2)} = A_1^{(1)} \cup A_2^{(1)}$, as discussed in Section~\ref{sec:1formsymfrac}. 

To see an explicit demonstration of the degeneracy, we look again to the holonomy effective potential on $S^1_L \times \IR^3$, but this time with $N_f$ charged scalars with equal mass $M$ and (for concreteness) clock-twisted boundary conditions as in Eq.~\eqref{eq:gen_clock}, 
\begin{equation}
V_{\rm eff}(\theta) = - \frac{2}{L^4}\left(\frac{M L}{2\pi}\right)^{3/2} \sum_{n=1}^\infty \frac{e^{-nML}}{n^{5/2}} \sum_{j=1}^{\ell} \sum_{w=0}^{k-1}   \cos\left( n \theta +n\lambda_j +  \frac{2\pi n w}{k} \right) \,.
\end{equation}
Here $\ell = N_f/k$ and $\lambda_j$ are phases satisfying $\sum_{j=1}^{\ell} \lambda_j = - \frac{\pi\ell(k-1)}{k}$. Performing the sum over $w$ causes destructive interference such that only the terms with $n = rk$ survive, 
\begin{equation}
V_{\rm eff}(\theta)  = - \frac{2k}{L^4}\left(\frac{ML}{2\pi}\right)^{3/2} \sum_{r=1}^\infty \frac{e^{-r k ML}}{(rk)^{5/2}} \sum_{j=1}^\ell \cos(r k\theta + rk\lambda_j)\,.
\end{equation}
This clearly demonstrates the $k$-fold degeneracy, as one can easily verify that $V_{\rm eff}(\theta +\frac{2\pi}{k}) = V_{\rm eff}(\theta)$. Moreover, one can verify that the degeneracy remains exact regardless of how light the charged matter field is.

\subsubsection{Example: Symmetry Fractionalization for a Non-Genuine Line Operator}

In addition to non-topological, but genuine lines,  line-like junctions of higher-dimensional topological operators can also carry fractional quantum numbers under a 0-form $G^\0$ symmetry. An example occurs in $4d$ theories with a $\IZ_N^\1$ 1-form symmetry and $G^\0=\IZ_N^\0 \times G_f^\0$ symmetry where the $G_f^\0$ has $\IZ_N$-valued obstruction classes, which participates in a mixed anomaly with the $\IZ_N^\0\times \IZ_N^\1$ symmetry: 
\eq{ \label{eq:4danomaly} 
\CA=\frac{2\pi i }{N}\int A^{(1)} \cup B^{(2)} \cup w_2(G_f^\0)~,
}
where $A^{(1)},\, B^{(2)}$ are the background gauge fields for $\IZ_N^\0$ and $\IZ_N^\1$ global symmetries respectively and $w_2(G_f^\0)$ is the $\IZ_N$-classified $G_f^\0$-symmetry fractionalization class. Here we will denote the symmetry operators for $G_f^\0$, $\IZ_N^{(0)}$, and $\IZ_N^{(1)}$ as $U_g$, $Y_k$, and $V_m$ respectively (with $m,k = 0,\ldots,N-1$). This anomaly implies that the (line-like) intersection of $Y_k(\Omega_3)$ and $V_m(\Sigma_2)$ operators has a `worldline anomaly' characterized by the class $km\, w_2(G_f^\0)$ --- i.e. $G_f^\0$ is fractionalized on the line-like intersection of $Y_k$ and $V_m$.

We are interested in whether or not the anomaly \eqref{eq:4danomaly} implies an exact vacuum degeneracy. As we will now show, there is an exact degeneracy in a twisted, defect Hilbert space, as in our previous examples. 

Consider the Hilbert space on $T^{3,xyz} = S^{1,x} \times S^{1,y} \times S^{1,z}$, $\CH[T^{3,xyz}]$, whose states can be constructed by evaluating the path integral over $T^{2,xy}\times D^{2,z}$ (where the $T^{2,xy} =S^{1,x} \times S^{1,y}$ and $\partial D^{2,z} = S^{1,z}$) with operator insertions in the bulk. Here we will specialize to states in the twisted Hilbert space $\CH[S^{1,x}_k\times S^{1,y}_g\times S^{1,z}]$ which are generated by evaluating the partition function on $T^{2,xy}\times D^{2,z}$ where $U_g$ is inserted along $S^{1,x}\times D^{2,z}$ and $Y_k$ is inserted along $S^{1,y}\times D^{2,z}$.  
We will denote state which is generated by evaluating the partition function with only $U_g,Y_k$ inserted in this way by $|0\rangle_{k,g}$:
\eq{
|0\rangle_{k,g}:=\dlangle U_g(S^{1,x}\times D^{2,z})\, Y_k(S^{1,y}\times D^{2,z})\drangle_{T^{2,xy}\times D^{2,z}}~. 
}
Because of the anomaly, the state $|0\rangle_{k,g}$ will lead to the relation 
\eq{
U_h(T^{3,xyz})\, V_m(T^{2,xy})|0\rangle_{k,g}=e^{\frac{2\pi i k m}{N}}V_m(T^{2,xy})\, U_h(T^{3,xyz})|0\rangle_{k,g}~,
}
for $U_h,V_m$ which form a rational torus algebra where $h\in G_f^\0$ such that  $\omega_{h,g}^{-1}\,\omega_{g,h} = e^{2\pi i/N}$ where $\omega_{g,h}$ is a representative of $w_2(G_f)\in H^2(BG_f,\IZ_N)$. A basis of the representations of this algebra are given by the states 
\eq{
|\psi_n\rangle=V_n(T^{2,xy})|0\rangle_{k,g}~, 
}
which diagonalize the  $U_h$ symmetry operators:
\eq{
U_h(T^{3,xyz})|\psi_n\rangle=e^{\frac{2\pi i kn}{N}}|\psi_n\rangle~. 
}
The states $|\psi_n\rangle$ form independent irreducible representations of $\IZ_N^{(0)}$ and  are therefore orthogonal and the action of $V_m(T^{2,xy})$:
\eq{
V_m(T^{2,xy})|\psi_n\rangle=|\psi_{n+m}\rangle~,
}
ensures that the $|\psi_n\rangle$ states are degenerate in $\CH[S^{1,x}_k\times S^{1,y}_g\times S^{1,z}]$.\footnote{Vacuum degeneracy can also be seen in the $g,h$-twisted Hilbert space which is generated by the operators $Y_k,V_m$ acting on the state  $|0\rangle_{g,h}=\dlangle U_g(S^{1,y}\times D^{2,z})\,U_h(S^{1,x}\times D^{2,z})\drangle_{T^{2,xy}\times D^{2,z}}$. In this case, the $Y_k,V_m$ operators form a rational torus algebra which also implies vacuum degeneracy. }

\bigskip
%%%%%%%%%%%%%%%%%%%%%%%%%%%%%%%%
\noindent
\underline{\emph{Example: $Spin(8N)$ QCD:}}
%%%%%%%%%%%%%%%%%%%%%%%%%%%%%%%%

\smallskip
\noindent
This anomaly naturally occurs for example in  $Spin(8N)$ Yang-Mills with two Weyl fermions in the vector representation. The classical $U(1)_\chi$ chiral symmetry 
 \eq{
\psi^a\longmapsto e^{i \alpha}\, \psi^a\quad \Longrightarrow\quad \Delta S=4 i \alpha \int \frac{\Tr [F_g\wedge F_g]}{8\pi^2}~,
 }
is broken by an ABJ anomaly $U(1)_\chi\mapsto \IZ_{4,\chi}$. Therefore, the fields in this theory all transform faithfully under the symmetry structure $\CG=\left(\frac{Spin(8N)}{\IZ_2}\times \frac{SU(2)\times \IZ_{4,\chi}}{\IZ_2^{\prime\prime}}\right)\big{\slash}\IZ_2^\prime$ with 0-form global symmetry $G^\0=\frac{SU(2)\times \IZ_{4,\chi}}{\IZ_2^\prime\times\IZ_2^{\prime\prime}}\cong \IZ_2^\0 \times SO(3)^\0$ and 1-form global symmetry $G^\1 = \ZZ_2^\1$. In $\CG$, the three $\IZ_2$ quotients identify 
\eq{
\IZ_2~:~-\mathds{1}_L\sim -\mathds{1}_R\quad, \quad \IZ_2^{\prime\prime}~:~ -\mathds{1}_{SU(2)}\sim (-1)_\chi \quad, \quad \IZ_2^\prime~:~-\mathds{1}_L\sim -\mathds{1}_{SU(2)}~,
}
where $Z(Spin(8N))=\IZ_2[-\mathds{1}_L]\times \IZ_2[-\mathds{1}_R]$ and the $\IZ_2$ quotient is what gives rise to the $\IZ_2^{(1)}$ global symmetry. 

This theory has an anomaly given by 
 \eq{\label{Spin8nanomaly}
\CA=i\pi \int A^{(1)}\cup B^{(2)} \cup w_2(SO(3))~,
 }
where $A^{(1)},B^{(2)}$ are the $\IZ_2^{(0)},\IZ_2^{(1)}$ background gauge fields respectively, and we are restricting ourselves to spin manifolds. Because of the quotient structure above, $B^{(2)},w_2(SO(3))$ are related to the generators of $H^2(BPSO(8N),\IZ_2)=\IZ_2[w_2^{(L)}]\times \IZ_2[w_2^{(R)}]$ as\footnote{Note that the identification of $w_2(SO(3))$ in \eqref{w2spin8n} is only determined up to a choice of parametrization of the $\IZ_2$ quotient in $\CG$. In other words, we could have made a different choice e.g. $w_2(SO(3))=w_2^{(R)}$, but the anomalies involving $w_2(SO(3))$ will be independent of this choice.}
 \eq{\label{w2spin8n}
B^{(2)}=w_2^{(L)}\,, \quad w_2(SO(3))=w_2^{(L)}+w_2^{(R)}~.
}
The anomaly comes from the fact that the chiral $U(1)$ is broken to $\IZ_4$ by an ABJ anomaly whereas the quantization of a $Spin(8N)$ instanton is modified when activating discrete $w_2^{(L)},w_2^{(R)}$ flux to~\cite{Aharony:2013hda,Witten:2000nv}
 \eq{
\int \frac{\Tr[F_g\wedge F_g]}{8\pi^2}=\frac{1}{2}\int w_2^{(L)}\cup w_2^{(R)} \, \in \frac{1}{2} \ZZ\,.
 } 
Acting with the generator of the unbroken chiral symmetry then produces an anomalous phase of the partition function coupled to background fields  
\eq{
\IZ_2^\0 ~:~Z[B_2,w_2(SO(3))]\longmapsto Z[B_2,w_2(SO(3))]\times e^{\pi i \int B^{(2)}\cup w_2(SO(3))}~,
}
which implies the anomaly in \eqref{Spin8nanomaly}. 
 Note that the $\IZ_2^\0$ has no mixed anomaly with just the $SO(3)^\0$ global symmetry and no self-anomaly due to the dimension of the gauge representations.

Our results imply that this theory has an exact vacuum degeneracy in a defect Hilbert space twisted by an element in the conjugacy class of the clock matrix in $G^\0 = SO(3) \times \ZZ_2$ (see Appendix \ref{app:clockandshift}). The theory can be deformed by coupling to adjoint scalars with a potential which causes them to condense in orthogonal directions, breaking the gauge group to its center $Spin(8N)\to \IZ_2\times \IZ_2$. In this phase, the IR theory is described by $\IZ_2\times \IZ_2$ gauge theory coupled to two Weyl fermions with charges (1,1) and carries the anomaly in Eq.~\eqref{Spin8nanomaly}.

Without deforming by adjoint scalars, the $Spin(8N)$ gauge theory is thought to confine, spontaneously breaking $SO(3)\times \IZ_2 \to SO(2)$ as discussed in \cite{Witten:1983tx}. The IR theory is believed to be described by a pair of $\ICP^1$ NLSMs which are related by the spontaneously broken $\IZ_2$. In this scenario, the anomaly \eqref{Spin8nanomaly} can be matched by a worldvolume anomaly on the $\IZ_2$-domain walls:
\eq{
\CA_{w.v.}=\pi i \int B_2\cup w_2(SO(3))~.
}

%%%%%%%%%%%%%%%%%%%%%%%%%%%%%%%%%%%%%%%%%%%%%%%%
\section{Worldline Anomaly Matching and RG Flows}
\label{sec:matching}
%%%%%%%%%%%%%%%%%%%%%%%%%%%%%%%%%%%%%%%%%%%%%%%%

An important feature of generalized symmetries is that the symmetry operators and their junctions are topological and hence trackable along RG flows. In other words, the choice of how we couple the UV partition function to background gauge fields for the global symmetries fixes the way the IR partition function couples to background gauge fields for the symmetries preserved along the RG flow. The same is true for symmetry fractionalization --- it is trackable along RG flows and therefore the projective representations carried by line operators (measured by the non-trivial correlation functions with disk operators/junctions of $G^{(0)}$ symmetry operators) are also trackable. This implies that a line operator with $G^\0$ symmetry fractionalization cannot flow to the completely trivial line along any symmetry-preserving RG flow.\footnote{Here we mean trivial in the strongest sense --- i.e. we consider a decoupled quantum mechanical theory on the line to be non-trivial. } The interpretation of symmetry fractionalization in terms of worldline anomalies was emphasized in~\cite{Brennan:2022tyl,Antinucci:2024izg} and utilized to constrain RG flows on line operators in~\cite{Aharony:2023amq}.

In this section, we discuss different scenarios that can match the non-trivial symmetry fractionalization on 
a line, namely: 
\begin{itemize}
    \item In the long-distance effective theory the line cannot end, and is charged under an emergent 1-form symmetry.
    \item The fractionalization is matched by a worldline theory which (other than the coupling to $G^\0$ symmetry) is decoupled from the bulk.
    \item The line remains non-trivial and endable, without a decoupled worldline quantum mechanics.  
\end{itemize}
Before considering these possibilities in more detail below, we pause to emphasize that the scenario which is realized by a given line operator does not necessarily have implications for the bulk phases of a theory (for instance, the second option can always be realized, even in a trivially gapped phase). With that said, some options can be ruled out in certain phases (e.g. the first and third options cannot occur in trivially gapped phases), while certain options may appear to be more natural in others (e.g. matching by emergent 1-form symmetry in a topologically ordered phase).

%%%%%%%%%%%%%%%%%%%%%%%%%%%%%%%%%%%%%%%%%%%%%%%%
\subsection{Matching by Emergent 1-Form Symmetry}
\label{sec:emergent1form}
%%%%%%%%%%%%%%%%%%%%%%%%%%%%%%%%%%%%%%%%%%%%%%%%

One way that the symmetry fractionalization can be matched along an RG flow is if the IR theory has an emergent 1-form global symmetry $G^\1_{\rm IR}$. The statement that there is emergent 1-form symmetry is equivalent to the inability for (certain) line operators to end on local operators in the low-energy effective theory. In this case, the symmetry fractionalization can be matched by activating a 1-form background gauge field which is fixed by the appropriate obstruction class in $H^2(BG^\0,G^\1_{\rm IR})$ as discussed in~\cite{Delmastro:2022pfo,Brennan:2022tyl,Barkeshli:2014cna}. This is equivalent to the dressing of the $G^\0$ junctions by 1-form symmetry defects as described in Section~\ref{sec:1formsymfrac}. In this scenario, the (topological) disk operator flows to the (topological) 1-form symmetry defect operator in the IR, and the 2-form background gauge field $B^{(2)}$ for $G^\1_{\rm IR}$ is fixed by the topological class of the $G^\0$ background gauge field in the UV. 

It is common to think about emergent 1-form symmetries as being associated with codimension-2 operators which at long distances become topological~\cite{Cordova:2022rer} and act non-trivially \cite{Cherman:2023xok}. In the special case where an IR-emergent 1-form symmetry descends from symmetry fractionalization in the UV, the disk operator provides an alternative that `descends' to the 1-form symmetry operator in the IR, but remains topological along the entire RG flow. The price we pay is that in the UV the operator is \emph{non-genuine}, i.e. it is the boundary of a codimension-1 topological surface.

\begin{figure}[t!]
\centering
    \begin{tikzpicture}[scale=0.8]
        \draw[black,thick, fill = olive!50] (3,0) [partial ellipse=0:360:0.4 and 1.2];
        \draw[-Stealth] (3.4,0) -- (3.4,0.1);
        \node[] at (4,0) {$\rightarrow$};
        \draw[black,thick] (5,0) [partial ellipse=0:360:0.4 and 1.2];
        \draw[-Stealth] (5.4,0) -- (5.4,0.1);
  
    \end{tikzpicture}
    \caption{In the IR, the bulk of the disc operator can become trivial giving rise to an emergent 1-form symmetry. 
    \label{fig:DiscRG}}
\end{figure}
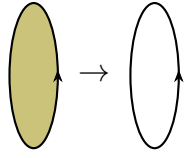
Let us give a sketch of how a non-genuine line operator in the UV can flow to a genuine line operator in the IR. We start in the UV with a theory without 1-form symmetry but with symmetry fractionalization for $G^\0$ as detected by codimension-2 junctions of $G^\0$ operators, or equivalently, the disk operator. Now suppose that all local operators charged under $G^\0$ decouple at a high scale, so that $G^\0$
acts trivially in the IR. Then the corresponding $G^\0$ symmetry operators become transparent in the IR, leaving behind a genuine, topological codimension-2 operator that formerly lived at the junction of codimension-1 operators. If the resulting operator acts non-trivially in the IR, it generates an emergent 1-form symmetry. Similarly, in this situation the interior of the disk operator (which already acts trivially on all local operators) becomes completely transparent, leaving only its edge as in Figure \ref{fig:DiscRG}. 

On the other hand, when the symmetry fractionalization is matched by an emergent 1-form symmetry, the $G^\0$ symmetry in general need not be trivialized. It is consistent that there is an emergent 1-form global symmetry and that $G^\0$ acts on operators that transform under linear representations of $G^\0$ as long as there are no non-genuine open lines that transform projectively under $G^\0$. When there is an emergent 1-form symmetry and $G^\0$ is not trivialized, the $G^\0$ symmetry still fractionalizes through the 1-form global symmetry in a way fixed by the UV physics as described in Section~\ref{sec:1formsymfrac}. 

Let us consider the emergence of 1-form symmetries in the context of gauge theories with heavy charged matter fields. One may wonder how it is consistent to have an emergent 1-form global symmetry when the line operators can be cut in the full theory. Consider the case where a Wilson line $W(\gamma)$ can end on a charged field $\phi(x)$. The correlation function that measures the possibility for the Wilson line to be cut is then given by 
\eq{
\langle \phi(x)W(\gamma_{x,y})\phi^\dagger(y)\rangle~,
}
where $\gamma_{x,y}$ is a path from $x\to y$. Intuitively, such an open Wilson line creates the dynamical, but heavy, $\phi$ particle which propagates from $x$ to $y$ and interacts with the static probe charge represented by $W$. 

This correlation function is scheme-dependent, and its value can be altered by adjusting a local counterterm which integrates to the length of the string 
\begin{equation}
    W(\gamma_{x,y}) \to W(\gamma_{x,y})\, e^{\mu\, \int_{\gamma_{x,y}} ds}\,.
\end{equation}
Let us give $\phi$ a mass $M$ which is parametrically larger than any other scale in the problem. Taking $\gamma_{x,y}$ to be a straight line, the above string operator decays at long distances as
\eq{
\langle \phi(x)W(\gamma_{x,y})\phi^\dagger(y)\rangle \sim \, e^{-(M-\mu)|x-y|}\,. 
}
We can cancel the exponential decay by taking $\mu = M$. Crucially, however, the value of this counterterm must be identical to those accompanying \emph{closed} Wilson loops $W(\gamma)$. 

Since we are assuming that microscopically there is no 1-form symmetry, asymptotically large circular closed Wilson loops decay with perimeter law, $W(\gamma) \sim e^{-\Lambda L(\gamma)}$. This sets an effective upper bound on the coefficient of the counterterm on $W(\gamma)$ --- in particular, $\mu \le \Lambda$ (otherwise, asymptotically large contractible loops will not be finite). This further implies that 
\begin{equation}
\lim_{|x-y| \to \infty} \langle \phi(x)W(\gamma_{x,y})\phi^\dagger(y)\rangle  = 0 ~,
\end{equation}
in \emph{any} scheme if $M > \Lambda$. Only if $M = \Lambda$ can we choose a scheme where the long-distance limit of $\langle \phi(x)W(\gamma_{x,y})\phi^\dagger(y)\rangle$ remains finite and the operator survives in the infrared. Such an equality is expected precisely when the pure gauge theory without matter confines with a non-vanishing string tension $T_{\rm string}$. Then at distances larger than the string-breaking scale $\sim M^2/T_{\rm string}$, the Wilson loop becomes unstable to the pair production of heavy charged fields, and the dominant long-distance fall-off turns from area- to perimeter-law $\sim e^{-M  L(\gamma)}$. In this case, the expectation values of asymptotically large Wilson loops are not controlled by any emergent 1-form symmetry~\cite{Cherman:2023xok}, and correspondingly, one can choose counterterms such that open Wilson lines remain finite at long distances.  

On the other hand, if the emergent 1-form global symmetry is spontaneously broken in a gapped phase, the gauge sector will be  topological in the IR, and the perimeter scaling of the Wilson lines can be canceled by a fixed local counterterm.\footnote{For spontaneously broken continuous 1-form symmetries, the expectation value of the line operators may have additional geometric dependence on the curve. However, for large circular loops, there similarly exist local counterterms that can cancel the perimeter scaling as the radius $r\to \infty$.} In this case, we expect $\Lambda$ to be set by some IR scale $\ll M$, so that $\mu\ll M$ and the open Wilson lines vanish at long distances in agreement with the emergent 1-form symmetry.

The expectation value of a line on large non-contractible cycles also serves as an indicator of emergent 1-form symmetry. Let $\langle \scrL(S^1)\rangle_{S^1\times S^{d-1}}$ denote the expectation value of a line $\scrL$ wrapping the non-contractible cycle of $S^1_\beta\times S^{d-1}$ with radius $\beta$. If the fractionalized line operator is protected by an emergent 1-form symmetry, then this expectation value will vanish up to exponentially small corrections. More precisely, in this scenario there are no local perimeter counterterms which can make the $\beta \to \infty$ limit of $\langle \scrL(S^1)\rangle_{S^1\times S^{d-1}}$ non-vanishing while keeping the analogous expectation value for asymptotically large contractible loops finite.

In cases without an emergent 1-form symmetry, this expectation value can be made $O(1)$ and equal to the dimension of a projective representation of $G^\0$. The reason is that we can interpret the expectation value $\langle \scrL(S^1)\rangle_{S^1\times S^{d-1}}$ as the trace over the $\scrL$-defect Hilbert space generated by the partition function over the $d$-ball containing the end point of the fractionalized line. Since the endpoint necessarily transforms projectively under $G^\0$, so does the entire defect Hilbert space. However, when there is an emergent 1-form symmetry the open line operator necessarily flows to the zero operator (i.e. the ground state of the defect Hilbert space has non-vanishing energy) and consequently in the IR limit the trace over the $\scrL$-defect Hilbert space/expectation value of $\scrL$ vanishes.

%%%%%%%%%%%%%%%%%%%%%%%%%%%%%%
\subsubsection{2-Groups and Extension of Fractionalization by 1-Form Symmetry}
%%%%%%%%%%%%%%%%%%%%%%%%%%%%%%

\begin{figure}[t!]
\centering
    \begin{tikzpicture}[scale=0.8]
        \draw[black,thick, fill = olive,fill opacity=0.5] (-0.2,0) [partial ellipse=0:360:0.4 and 1.2];
        \draw[-Stealth] (0.2,0) -- (0.2,0.1);
        \draw[black,thick, fill = olive,fill opacity=0.5] (0.8,0) [partial ellipse=0:360:0.4 and 1.2];
        \draw[-Stealth] (1.2,0) -- (1.2,0.1);
        \node[] at (2,0) {\Large $ \dots$};
        \draw[black,thick, fill = olive,fill opacity=0.5] (3,0) [partial ellipse=0:360:0.4 and 1.2];
        \draw[-Stealth] (3.4,0) -- (3.4,0.1);
        \node[] at (4,0) {$=$};
        \draw[black,thick] (5,0) [partial ellipse=0:360:0.4 and 1.2];
        \draw[-Stealth] (5.4,0) -- (5.4,0.1);

        \draw [decorate,decoration={brace,amplitude=5pt, mirror}]
  (-0.5,-1.5) -- (3.5,-1.5) ;
  \node[color = olive] at (1.5,-2) {$\scrD \times \scrD \times \dots \times \scrD$};
  
    \end{tikzpicture}
    \caption{Extension of symmetry fractionalization by the 1-form symmetry --- the codimension-2 1-form symmetry operator appears in the fusion of disk operators. 
    \label{fig:extension}}
\end{figure}
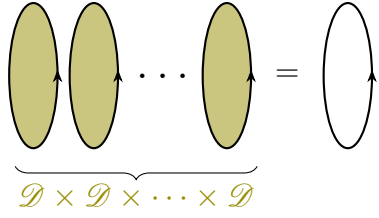

In the case where the UV theory already has a 1-form symmetry $G^\1_{\rm UV}$, the symmetry fractionalization can mix with the 1-form global symmetry to combine into a 2-group. Roughly speaking, in this scenario some power of a 1-form symmetry-charged line is endable and has symmetry fractionalization. This means that fusing some number of disk operators produces a 1-form symmetry generator, as in Figure~\ref{fig:extension}. This structure is described by the short exact sequence
\eq{
1\longrightarrow G^\1_{\rm UV}\longrightarrow \IG \longrightarrow \Gamma \longrightarrow 1~,
}
where $\Gamma$ is again a subgroup of the center of $\hatG^\0$ such that $G^\0 = \hatG^\0/\Gamma$.  When the above extension is non-trivial, the UV theory has a 2-group global symmetry as described in e.g.~\cite{Benini:2018reh,Hsin:2020nts,Bhardwaj:2021wif,Bhardwaj:2022dyt}. This 2-group structure can be matched in the IR by a theory with an emergent 1-form global symmetry $G^\1_{\rm IR} = \IG$ which is an extension of $\Gamma$ by the UV 1-form symmetry. 

For example, in the case when $\Gamma = \ZZ_N$ and $G^\1_{\rm UV}=\IZ_M$, the possible Postnikov classes characterizing the 2-group are given by multiples of the Bockstein of the fractionalization class $w_3(G^\0) = \beta(w_2(G^\0))\in H^3(BG^\0,\IZ_M)$, associated to the sequence
\eq{
1\longrightarrow G^\1_{\rm UV} = \ZZ_M \longrightarrow \ZZ_{MN} \longrightarrow \Gamma = \ZZ_N \longrightarrow 1~,
}
which can only be non-trivial when the above sequence does not split --- i.e. when $M,N$ are not coprime (otherwise $\ZZ_{MN} = \ZZ_M \times \ZZ_N$). 

In this example, the $N$-fold fusion of the disk operator yields the generator of the 1-form symmetry: $\scrD^N=V$, and $V^M = \mathbbm{1}$. In terms of background fields, we have the 2-group relation
\begin{equation} \label{eq:2group}
    \delta B^{(2)}_{\rm UV} +  w_3(G^\0) =  \delta B^{(2)}_{\rm UV} + \beta(w_2(G^\0)) = 0 \text{ mod } M\,.
\end{equation}
Now consider an RG flow where the bulk of the disk operators become transparent, leading to an emergent $\ZZ_{MN}$ 1-form symmetry. The IR background field can be decomposed in terms of the backgrounds of the exact UV global symmetries as 
\begin{equation}
    B^{(2)}_{\rm IR} = w_2(G^\0) + N B^{(2)}_{\rm UV} \ (\text{mod } MN)\,,
\end{equation}
where the flatness condition of $B_{\rm IR}^{(2)}$ is guaranteed by the UV 2-group relation \eqref{eq:2group}. 

A simple example of the above structure is furnished by the $N$-flavor abelian Higgs model with scalars of charge $M$~\cite{Freed:2017rlk}. Repeating the steps in Section~\ref{sec:qed_junction}, we find a cocycle condition for the gauge field transition functions
\begin{equation} 
    (\delta\phi)_{ijk} = \frac{2\pi}{MN}\left( \Lambda_{ijk} + N B_{ijk}  + MN m_{ijk}\right)\,,
\end{equation}
with $\Lambda_{ijk}$ as defined before, and $B_{ijk}$ the background field for the $\ZZ_M$ 1-form symmetry. Taking the coboundary of the above equation, we find 
\begin{equation}
(\delta B)_{ijk\ell} + \frac{1}{N}(\delta \Lambda)_{ijk\ell} =(\delta B)_{ijk\ell} + \beta(\Lambda)_{ijk\ell} = 0 \text{ mod } M\,.
\end{equation}
This modified cocycle condition for the background 1-form gauge field indicates a 2-group with Postnikov class $\beta(\Lambda)$~\cite{Benini:2018reh}. One can imagine RG flows where the charged scalars are heavy and decouple, leaving behind an emergent $\ZZ_{MN}^\1$ 1-form symmetry. 

As another example, consider a 3-dimensional $U(1)$ gauge theory with a Chern-Simons term at level $N$, coupled to $N$ massive scalars $\Phi_i$ of charge $q$ 
\begin{equation}
    S = \frac{iN}{4\pi} \int a da + \int d^3x\, \frac{1}{4e^2}f^2 + |(\partial - i q a)\Phi|^2 + m^2 |\Phi|^2 \,,
\end{equation}
where $\gcd(N,q)\neq 1$. 
This model has a $\ZZ_{\gcd(N,q)}^\1$ 1-form symmetry. In addition, there is a 0-form $PSU(N)$ symmetry acting on the scalars.\footnote{The model also has a $U(1)$ magnetic 0-form symmetry which also fractionalizes.} Because of the Chern-Simons term, monopole operators $\mathcal M$ are not gauge invariant~\cite{Pisarski:1986gr,Affleck:1989qf}. In fact, gauge transformations act as
\begin{equation}
    a \rightarrow a + d\l, 
    \quad \Phi_i \rightarrow e^{iq \l }\,\Phi_i, \quad \mathcal M \rightarrow e^{-iN \l}\,\mathcal M~.
\end{equation}
To make monopole operators gauge invariant, we need to dress them with $\Phi$'s. The minimal gauge-invariant monopole operator is 
\begin{equation}
    \Phi_{i_1}\dots\Phi_{i_{N'}} \mathcal{M}^{q'}~,
\end{equation}
where $N'= \frac{N}{\gcd(N,q)} = \frac{\text{lcm}(N,q)}{q}$ and $q' = \frac{q}{\gcd(N,q)} = \frac{\text{lcm}(N,q)}{N}$. In addition to genuine local operators there are open Wilson lines 
\begin{equation}
    \Phi(x)_{i_1}\dots\Phi(x)_{i_k}\, \mathcal{M}(x)^\ell \; e^{i(q k  -\ell N)\int_x a}~,
\end{equation}
which are not charged under the 1-form symmetry but are charged under disk operators $\scrD_c$ where $c \in \ZZ_{N'}$. 

If the scalar $\Phi$ is heavy, we can integrate it out and in the deep IR the theory flows to $U(1)_N$ with a $\ZZ_N$ 1-form symmetry. Hence, this is a model in which the IR emergent 1-form symmetry is a non-trivial extension of the UV symmetry fractionalization by the UV 1-form symmetry
\begin{equation}
    1 \longrightarrow G^\1_{\rm UV} = \ZZ_{\gcd(N,q)}^\1 \longrightarrow G^\1_{\rm IR} = \ZZ_N^\1 \longrightarrow \ZZ_{N'}^\1 \rightarrow 1~.
\end{equation} 
%
%
%
%
%
%
%
%
%
%%%%%%%%%%%%%%%%
\subsection{Matching by Decoupled Quantum Mechanics}
%%%%%%%%%%%%%%%%

\smallskip

Line operators with $G$ symmetry fractionalization can in general be matched by a line defect that supports a decoupled quantum mechanics with a worldline anomaly for the $G^\0$ symmetry  (i.e. $G^\0$ is realized projectively on the worldline Hilbert space). Here by decoupled we mean that in all correlation functions, the expectation value of the fractionalized line $\scrL$ is given by the expectation value of a (possibly trivial) line $L$ times the partition function of a quantum mechanics up to  exponentially small corrections.  
A distinguished possibility is when the IR line $L$ is the trivial line $L=\mathds{1}$ so that the UV fractionalized line is entirely matched by a completely decoupled quantum mechanics (QM) $\scrL\to Z_{\rm QM}\mathds{1}$.  
In particular, symmetry fractionalization can and must be matched in this way in  trivially gapped phases (i.e. in gapped IR phases with a unique ground state and no topological order).

Even in trivially gapped phases where the fractionalized line is matched by a decoupled worldline quantum mechanics, the worldline anomaly alone does not uniquely specify the worldline theory. Some examples of quantum mechanical theories with worldline anomalies are a charged particle in a magnetic field~\cite{Seiberg:2024yig} for $G^\0 = \ZZ_k \times \ZZ_k$, a charged particle on $S^2$ with odd magnetic flux~\cite{RABINOVICI1984523,Gaiotto:2017yup} or free qubit for $G^\0 =SO(3)$, a particle on $\mathbb{CP}^{N-1}$ with a WZW term~\cite{Tong:2014yla,Nguyen:2022lie} for $G^\0 =PSU(N)$, and a particle on a circle at $\theta = \pi$ ~\cite{Gaiotto:2017yup} for $G^\0 =O(2)$. In particular, since these worldline anomalies are global in nature, they do not constrain whether the QM theory is gapped or gapless.

In the IR effective theory where the fractionalized line is matched by a worldline quantum mechanics, integrating out the degrees of freedom on the line is equivalent to computing the worldline quantum mechanical partition function. When the line operator is placed on a closed curve, tracing out the worldline degrees of freedom turns the line operator into some operator (in the trivially gapped theory, the identity operator) multiplied by its partition function. This partition function can match the selection rules from Section~\ref{sec:selectionrules} because it vanishes identically in the appropriately twisted sector due to the trace over  states which transform projectively under $G$. In cases where the worldline quantum mechanics is gapped, the line operator becomes topological at long distances.\footnote{Or rather, there exists a choice of local counterterms that makes the line operator topological. This local counterterm shifts the ground state energy in the worldline quantum mechanics to zero. } 

A prototypical example of such a quantum mechanical partition function coupled to background fields is 
\begin{equation} \label{eq:bgd_wilson_line}
 Z_{\text{QM}}[A] =   \text{Tr}_{\mathfrak{R}}\, \mathcal P \, e^{i \oint_\gamma  \widehat A} \,,
\end{equation}
where $\mathfrak{R}$ is a projective representation of $G$ and $\widehat A$ is an appropriately lifted background gauge field, so that $\mathfrak{R}(\hatg)\, \mathfrak{R}(\hath) = \omega_{g,h}\, \mathfrak{R}(\hatgh)$. Indeed, evaluating the partition function in the $g$-twisted sector is equivalent to setting $\CP \, e^{i \oint_\gamma\widehat A} = \hatg$. Because of our choice of $g$, the trace of $\hatg$ satisfies
\begin{equation}
    \text{Tr}_{\mathfrak{R}}(\hatg) =  \text{Tr}_{\mathfrak{R}}(\,\hath^{-1}\,\hatg\, \hath\,) =  \text{Tr}_{\mathfrak{R}} (\chi_{g,h}\,\hatg) = \chi_{g,h}\,   \text{Tr}_{\mathfrak{R}}(\hatg)\,,
\end{equation}
which vanishes if $\chi_{g,h} \not = 1$. This implies that the quantum mechanical partition function also vanishes in the $g$-twisted sector, consistent with the selection rule in Eq.~\eqref{eq:selectionrule}. Essentially, inserting the $g$-twist causes destructive interference in the QM Hilbert space, such that the above trace over $g$ in the projective representation causes the partition function to  vanish identically.

We can sharply distinguish between the cases where a fractionalized line operator flows to a decoupled worldline quantum mechanics and the other possibilities. % a topological $(d-2)$-form symmetry operator. 
In both cases, the fractionalized line becomes topological at long distances, and (as we discussed in Section \ref{sec:selectionrules}) obeys selection rules on $S^1\times T^{d-1}$ in certain $G^\0$-twisted sectors. Let us instead consider the two-point function of fractionalized lines $\scrL$ wrapped on a non-trivial cycle, in the $g$-twisted sector:
\eq{\label{2pointconstraints}
\langle \scrL^\dagger(\gamma) \scrL(\gamma_r)\rangle_{g} = 
\langle \scrL^\dagger(\gamma) \scrL(\gamma_r)\,U_g(T^{d-1})\rangle_{S^1\times T^{d-1}}\,,
}
Here we take $\gamma$ and $\gamma_r$ to be wrapped along the $S^1$ with parallel separation $r$ along the $T^{d-1}$, and $U_g$ is a $G^\0$ symmetry operator which is chosen such that $\langle \scrL(\gamma)\rangle_g = 0$. However the two-point function is not constrained to vanish, since $\scrL^\dagger$ and $\scrL$ have opposite fractionalization. 

On the other hand, if the line operator flows to a decoupled worldline quantum mechanics, then in the limit of large separation the two-point function must approach its factorized value, namely zero:
\eq{ \label{eq:vanishing2pt} 
\lim_{r\to\infty} \langle \scrL^\dagger(\gamma) \scrL(\gamma_r)\rangle_{g} = 0\,.
}
This implies that if the two-point function approaches a non-zero value for $r\to \infty$, then the fractionalized line cannot be matched by a decoupled QM. 

On the other hand, if the line $\scrL$ is exactly topological, then the above two-point function must be independent of $r$, which would be inconsistent with the above behavior (since $\scrL$ is not the zero operator). We conclude that \emph{a topological line with symmetry fractionalization cannot flow to a decoupled worldline quantum mechanics.} This resonates with the discussion in Section~\ref{sec:topological_lines}, where we explained how topological lines with symmetry fractionalization are indicators of bulk anomalies. Our analysis here confirms that any topological line participating in such a mixed anomaly cannot flow to a decoupled worldline quantum mechanics. Additionally, since symmetry fractionalization can only be matched in a trivially gapped phase by a decoupled worldline quantum mechanics, we also see that if the above correlation function does not vanish in the IR limit, then the IR phase cannot be trivially gapped. 

Additionally, we can use Eq.~\eqref{eq:vanishing2pt} to distinguish the scenario with a decoupled quantum mechanics from the case when a non-topological line in the UV flows to a topological line in the IR. For such a line, the large separation limit of the above two-point function should approach a constant different from zero.

When the UV fractionalized line is matched in the IR by a gapped decoupled QM, there exists a scheme where the expectation value of the line on a non-contractible cycle of length $\beta$ can be used to tell us the ground state degeneracy of the worldline quantum mechanics: 
\eq{
\CI_{\rm QM}:=\lim_{\beta\to \infty}Z_{\rm QM}(\beta)=\lim_{\beta\to \infty} \Tr_{\CH_{\rm QM}} ~e^{-\beta H}=\dim[\CH_{\rm QM}]\neq 0~.
}
In this case, the symmetry fractionalization implies that $\CI_{\rm QM}$ is necessarily the dimension of a projective representation of $G^\0$ associated to the symmetry fractionalization class. We can entertain the possibility that the integer $\CI_{\rm QM}$ can jump discontinuously. This is similar to the wall crossing of framed BPS states that bind to bare, UV line operators in supersymmetric gauge theories~\cite{Gaiotto:2010be,Moore:2015szp,Brennan:2016znk,Brennan:2018ura,Chuang:2013wt,Cordova:2013bza,Brennan:2018rcn,Lee:2011ph,Cordova:2016uwk,Brennan:2019hzm}. 
As we alluded to in the beginning of this section, such jumps in $\CI_{\rm QM}$ are not necessarily tied to bulk phase transitions as they can be matched by level crossing in the worldline QM. However, in specific examples it may nonetheless be interesting to investigate how this quantity changes as a function of bulk parameters, as it serves as an indicator of the microscopic physics of how the line is screened.

\subsubsection{Example: Wilson Lines in  $4d$ Confining Gauge Theories}

An important   class of theories where a fractionalized line is believed to flow to a decoupled quantum mechanics are $4d$ confining QCD-like theories with massive matter. Take for example $SU(N_c)$ gauge theory coupled to $N_c$ Dirac fermions in the fundamental representation with a 0-form $G^\0=SU(N_c)/\IZ_{N_c}$-preserving mass. In this case, the Vafa-Witten theorem \cite{Vafa:1983tf} implies that the matter bound states are massive and the theory is believed to flow to a trivially gapped theory.  

In this theory the fundamental Wilson line experiences $G^\0$-symmetry fractionalization because it is endable on  gauge-non-invariant fermion operators. The same is true for any Wilson line in a representation in which the center acts non-trivially. Our previous discussion then implies that the Wilson lines/large Wilson loops must be matched by a decoupled worldline quantum mechanics with a $G^\0$ anomaly. Intuitively, this anomalous quantum mechanics can be viewed as the worldline theory of the particle that screens the Wilson line. For the fundamental Wilson line, the natural candidate is the quantum mechanics with a $N_c$-dimensional ground state that transforms in the projective representation of $PSU(N_c)$. 

Another class of theories where the Wilson lines are expected to be matched by a decoupled worldline quantum mechanics are a subset of $4d$ gauge theories that exhibit symmetric mass generation. These theories are chiral gauge theories (i.e. theories without gauge-invariant mass deformations) whose global symmetries have no 't Hooft anomalies, and are believed to flow to a trivially gapped phase. See for example~\cite{Tong:2021phe,Bars:1981se,Wang:2022ucy,Shirman:2023hhk,Creutz:1996xc,Poppitz:2010at,Wen:2013ppa,DeMarco:2017gcb,Kikukawa:2017ngf,Wang:2018ugf,Butt:2018nkn,Haldane:1995xgi}. 

A standard example is the $4d$ $SU(N)$ gauge theory with fermions in the anti-fundamental and 2-index symmetric representation as well as neutral fermions which have an additional interaction that preserves a $SU(N+4)\times U(1)$ 0-form global symmetry without any anomalies~\cite{Tong:2021phe,Bars:1981se}. 
However, when $N$ is even a $\IZ_2$ subgroup of this 0-form global symmetry is identified with part of the center of the gauge group and results in the global symmetry $G^\0=\frac{SU(N+4)}{\IZ_2}\times U(1)$. In this case, the fundamental Wilson line experiences $SU(N+4)/\IZ_2$ symmetry fractionalization due to the fact that it can end on local fermion fields. If the conjecture that this theory flows to the trivially gapped phase is correct, then the Wilson line must also be matched by a decoupled quantum mechanics with a $SU(N+4)/\IZ_2$ worldline anomaly. This conjecture can also be analyzed by studying the Wilson line 2-point function as in \eqref{eq:vanishing2pt}.

\subsection{Matching in Nonlinear Sigma Models }

In this last section, we analyze the situation where the theory with symmetry fractionalization in the UV flows to a nonlinear sigma model (NLSM) in the IR. This case can logically realize all three possibilities listed at the beginning of this section, depending on the symmetry breaking pattern.  We assume that that IR is described by a $\CC=G/H$ NLSM where a continuous group $G$ is spontaneously broken down to a subgroup $H$.

The NLSM still has the 0-form global symmetry group $G$ although it might be realized non-linearly. In addition, there might be emergent symmetries associated to stable winding configurations. These emergent global symmetries are encoded in the homotopy structure of the coset space $\CC$. In general, the homotopy group $\pi_k(\CC)$ classifies topologically protected operators of codimension-$(k+1)$ on which the global symmetries of the theory acts. The homotopy groups of $\CC$ can be determined by the induced long exact sequence in homotopy:
\eq{
\label{homotopy sequence}
\hdots \to \pi_{k+1}(\CC)\to \pi_k(H)\to \pi_k(G)\to \pi_k(\CC)\to \pi_{k-1}(H)\to \hdots
 }
However, the  symmetry of the $\CC$ NLSM in general is not simply the product of the Pontryagin dual groups $\pi_k(\CC)^\vee$, but may include categorical symmetries which can only be determined by the full homotopy category of $\CC$~\cite{Chen:2022cyw,Chen:2023czk,Pace:2023kyi,Pace:2023mdo,Hsin:2025ria}. Note however, that the lowest non-trivial homotopy group always has a corresponding  higher-form global symmetry (which may however participate in a larger categorical symmetry structure) \cite{Sheckler:2025rlk}.

One scenario that can match the symmetry fractionalization is when the NLSM has an emergent 1-form global symmetry $G_{\rm IR}^\1$ in the IR.  In this case, the symmetry fractionalization can be matched in the way discussed above in Section~\ref{sec:1formsymfrac}. However, this scenario is also restricted by the fact that the global symmetry $G^\0\times G_{\rm IR}^\1$ always has a mixed anomaly in the IR NLSM when $\pi_1(G)\neq 0$ due to the effects of coupling a $G^\0$ background gauge field on the $\pi_{d-2}$-winding density \cite{Sheckler:2025rlk}. If this mechanism is used to match the symmetry fractionalization, then this necessarily leads to an additional $G^\0$ self anomaly which must be matched along the entire RG flow.\footnote{Similar considerations were used to match the $w_2w_3$ type anomalies of $4d$ 2-flavor adjoint QCD as in \cite{Cordova:2018acb,Brennan:2022tyl,DHoker:2024vii,Brennan:2023vsa}.}

Another scenario is that the defect lines remain non-trivial and endable in the IR. When they further become topological there is a natural candidate that matches the symmetry fractionlization in the UV. This scenario can occur iff $\pi_1(\CC)\neq 0$ and depends on the embedding $\pi_1(G)\hookrightarrow \pi_1(\CC)$. This is because as we showed in Section~\ref{sec:diskoperator}, symmetry fractionalization in the UV is associated to $\Gamma$ and $\pi_1(G) = \Gamma$. Due to the structure of the long exact sequence of homotopy groups, $\pi_1(G)$ at most shows up in the homotopy group $\pi_1(\CC)$ which corresponds to a $(d-2)$-form symmetry $G_{\rm IR}^{(d-2)}=\pi_1(\CC)^\vee$ that is generated by a topological line. However, the way that $\pi_1(G)$ maps into $\pi_1(\CC)$  is determined by the breaking pattern $G\to H$ and the way $\pi_1(H)$ maps into  $\pi_1(G)$. Let us consider the case where  $\pi_1(\CC)=\pi_1(G)$. In this case, the natural candidate for the IR fractionalized line is the line that generates the corresponding $G_{\rm IR}^{(d-2)}$ symmetry. It exhibits $G$ symmetry fractionalization because of the mixed anomaly with the $G$ global symmetry~\cite{Brennan:2023vsa,Cordova:2018acb,Sheckler:2025rlk,Lee:2020ojw}:
\eq{
\CA=\int w_2(G)\cup B^{(d-1)}~,
}
where $B^{(d-1)}$ is the background gauge field for $G_{\rm IR}^{(d-2)}$. This anomaly  implies that the topological line operator has a non-trivial linking with the junction of $G$ symmetry defects reproducing the $G$ symmetry fractionalization. In the case where $\pi_1(G)=\pi_1(\CC)$, the topological line operators are also endable. 
One way to see this is to rewrite the $\CC=G/H$ NLSM as a $\widehat \CC$ NLSM coupled to a $\Gamma=\pi_1(G)$ gauge theory where here $\widehat{\CC}$ is the simply connected cover of $\CC$.  In this formulation, the topological $\pi_1(\CC)$ lines corresponds to the $\Gamma$ Wilson lines and they can end on gauge-non-invariant local operators. Upon integrating out the gauge field, the gauge-non-invariant local operators corresponds to local (defect) operators that the $\pi_1(\CC)$ lines can end on. 

However, in the cases where $\pi_1(\CC)=\emptyset$ and $\pi_{d-2}(\CC) = \emptyset$ there is generically no candidate line defect in the NLSM that matches the symmetry fractionalization and we expect it to flow to a line which hosts a worldline quantum mechanics.  
Similarly, even if $\pi_1(G)= \pi_1(\CC)$, it does not necessarily imply that the UV line operator with symmetry fractionalization flows to the $\pi_1(\CC)$ topological operator as the symmetry fractionalization may also be matched by a distinct defect operator with a worldline quantum mechanics.

\bigskip
%%%%%%%%%%%%%%%%%%%%%%%%%%%%%%%%
\noindent
\underline{\emph{Example: $\pi_1(\CC)=\pi_1(G)$}}
%%%%%%%%%%%%%%%%%%%%%%%%%%%%%%%%

\smallskip
\noindent
A class of examples with $\pi_1(\CC) = \pi_1(G)$ are cases when $G\to \emptyset$ so that $\CC=G$. This can happen for example in $4d$  $\IZ_2$ gauge theory coupled to two complex scalar doublets $z_a,\tildez_a$ with $a=1,2$ of charge 1 with potential 
\eq{
V(z,\tildez)=\lambda (|z|^2-v^2)^2+\tilde\lambda (|\tildez|^2-\tilde v^2)^2+m^2 \epsilon^{ab}z_a\tildez_b+\tilde{m}^2z_a(\tildez_a)^\dagger+c.c.
}
This theory, has a $SO(3)=SU(2)/\IZ_2$ global symmetry where the $\IZ_2$ is identified with the gauge group. The potential  causes $z,\tildez$ to condense in the IR along orthogonal directions in $\mathfrak{so}(3)$ so that the $SO(3)$ global symmetry is broken $SO(3)\to\emptyset$ and the IR theory is described by a $SO(3)$ NLSM. Here, the local operators $z_a,\tildez_a$ are gauge-non-invariant and transform projectively under $SO(3)$ so that the $\IZ_2$ Wilson line has $SO(3)$ symmetry fractionalization. 

The IR $SO(3)$ NLSM has a natural description in terms of a $SU(2)$ NLSM coupled to the topological $\IZ_2$ gauge theory. In this case, the topological $\IZ_2$ Wilson line is identified with the generator of the $(d-2)$-form $\pi_1(SO(3))$ global symmetry which has $SO(3)$ symmetry fractionalization. Therefore, in this example the $SO(3)$-fractionalized $\IZ_2$ Wilson line is directly matched by the topological $\pi_1(SO(3))$ line. 

\bigskip
%%%%%%%%%%%%%%%%%%%%%%%%%%%%%%%%
\noindent
\underline{\emph{Example: $\pi_1(\CC)=\emptyset$}}
%%%%%%%%%%%%%%%%%%%%%%%%%%%%%%%%

\smallskip
\noindent
A class of theories with $\pi_1(\CC)=\emptyset$ are 
$SU(N_c)$ QCD theories.  
Consider a $4d$ $SU(N_c)$ gauge theory with $N_c$ massless Dirac fermions in the fundamental representation. This theory has a 
\eq{
G=\frac{SU(N_c)_L\times SU(N_c)_R\times U(1)_B}{\IZ_{N_c}\times \IZ_{N_c}^\prime}~,
}
global symmetry where the $SU(N_c)_{L/R}$ acts on the fermions in the fundamental/anti-fundamental representation. Here, the quotients relate 
\eq{
z_L\circ z_R\sim 1\quad, \quad z_L\sim e^{2\pi i /N_c}\in U(1)_B~,
}
where $Z(SU(N_c)_{L/R})=\langle z_{L/R}\rangle$ and the first relation identifies the combined center of $SU(N_c)_L\times SU(N_c)_R$ with the center of the gauge group. Since $H^2(BG,U(1))\supset \IZ_{N_c}$, we see that the fundamental Wilson lines experience $G$-symmetry fractionalization. 

It is believed that in the IR this theory exhibits chiral symmetry breaking by fermion bilinear condensation which spontaneously breaks $G\to H=\frac{SU(N_c)\times U(1)_B}{\IZ_{N_c}\times \IZ_{N_c}^\prime}$. In this case, the IR is described by a $\CC=G/H=SU(N_c)$ NLSM which has $\pi_1(\CC)=\emptyset$. Thus, we expect that the fundamental Wilson line, which is trackable along the RG flow due to its symmetry fractionalization, flows to an operator that hosts a decoupled worldline quantum mechanics.

\section*{Acknowledgements}

We thank Ken Intriligator, Thomas Dumitrescu, Pierluigi Niro, Andrea Grigoletto, John McGreevey, Aleksey Cherman, Tudor Dimofte, Aiden Sheckler, and Zhengdi Sun for their helpful discussions, and are grateful to Aleksey Cherman for comments on a draft. TDB is supported by Simons Foundation award 568420 (Simons Investigator) and award 888994 (The Simons Collaboration on Global Categorical Symmetries). TJ is supported by a Schwinger Fellowship at the Mani L. Bhaumik Institute at UCLA. KR is supported in part by the Simons Collaboration on Global Categorical Symmetries and also by the NSF grant PHY-2412361.

\appendix

%%%%%%%%%%%%%%%%%%%%%%%%%%%%%%%%%%%%%%%%%%%%%%%%%%%%%%%%%%%%%%%%%
\section{Generalized Clock and Shift Matrices}
\label{app:clockandshift} 
%%%%%%%%%%%%%%%%%%%%%%%%%%%%%%%%%%%%%%%%%%%%%%%%%%%%%%%%%%%%%%%%%

In Section \ref{sec:selectionrules} we showed that if there exist two group elements $g,h \in G^\0$ which commute in $G^\0$ but their lifts in $\hatG^\0$ do not, the line operators charged under the disk operator obey selection rules in the sectors twisted by either $g$ or $h$. Here, we determine which elements have this property for $\hatG = SU(N)/\ZZ_k$ for some $k|N$. 

Consider the lifts $\hatg$ and $\hath$ of $g$ and $h$ respectively in $\hatG^\0 = SU(N)$. They need to satisfy
\begin{equation} \label{congugation}
    \hath^{-1}\, \hatg\, \hath = \omega\, \hatg, \quad \omega\in \ZZ_k~,
\end{equation}
where the $\ZZ_k$ is in the center of $SU(N)$. Every element $\hatg$ can be conjugated to the maximal torus of $SU(N)$, so we can assume that $\hatg$ is a diagonal matrix. Then, Eq.~\eqref{congugation} requires that $\hatg$ and $\omega\hatg$ have the same eigenvalues. Let's consider the fundamental representation and take $\omega = e^{\frac{2\pi i }{k}}\mathbbm{1}_{N\times N}$. If $\hatg$ has an eigenvalue $e^{i\lambda}$, then it also has the eigenvalue $e^{i\lambda + \frac{2\pi i }{k}}$ since $\hatg$ and $\omega \hatg$ have the same eigenvalues. Then, we can proceed recursively and conclude that $\hatg$ has all the eigenvalues $e^{i\lambda + \frac{2\pi i n }{k}}$ for $n = 0,1,\dots, k-1$. Therefore, the eigenvalues of $\hatg$ consist of the above sets, perhaps with different $\lambda$:
\begin{equation}
    e^{i\lambda_j + \frac{2\pi i n }{k}},\quad  j=1,2,\dots, \ell~,\quad n = 0,1,\dots, k-1~,
\end{equation}
where $N = k\ell$. We finally require that the product of all eigenvalues is one, so that the determinant of $\hatg$ is one. This leads to
\begin{equation}
    \sum_{j=1} ^\ell \lambda_j = -\frac{\pi \ell (k-1)}{k}  \quad  \mod \quad \frac{2\pi}{k}~.
\end{equation}
For $k=N$, there is only one $\lambda$ (since $\ell = 1$) and the above requires that $\lambda = -\pi \frac{(N-1)}{N} \text{ mod } \frac{2\pi}{N}$. If $k<N$ there is a continuous family of solutions with the desired property labeled by the $\lambda_j$'s. 

Up to permutations, these matrices can be taken to consist of $\ell$-many $k \times k $ diagonal blocks,
\begin{equation} \label{eq:gen_clock} 
C= \begin{pmatrix}
e^{i\lambda_1} C_{k} & 0  & \cdots & 0 \\
0 & e^{i\lambda_2} C_{k} &  \cdots & 0 \\ 
\vdots & \vdots & \ddots & \vdots \\
0 & 0 & \cdots &   e^{i\lambda_\ell} C_{k}
\end{pmatrix}~,
\end{equation}
where
\begin{equation} 
C_k = \begin{pmatrix}
1 & 0  & \cdots & 0 \\
0 & e^{\frac{2\pi i}{k}}&  \cdots & 0 \\ 
\vdots & \vdots & \ddots & \vdots \\
0 & 0 & \cdots &   e^{\frac{2\pi i}{k}(k-1)}
\end{pmatrix}~.
\end{equation}
Given this $\hatg = C$, it is easy to find an $\hath$ with the property \eqref{congugation}, for instance, one can take the block diagonal matrix consisting of $k\times k$ permutations, 
\begin{equation} \label{eq:gen_shift} 
S= \begin{pmatrix}
 S_{k} & 0  & \cdots & 0 \\
0 & S_{k} &  \cdots & 0 \\ 
\vdots & \vdots & \ddots & \vdots \\
0 & 0 & \cdots &   S_{k}
\end{pmatrix}~,
\end{equation}
where
\begin{equation} 
S_k = e^{-\frac{2\pi i}{k}\frac{k-1}{2}} 
\begin{pmatrix} 0 & 1 & 0 & \cdots & 0 \\
0 & 0 & 1 & \cdots & 0 \\
\vdots & \vdots & \ddots & \ddots & \vdots \\
0 & 0 & 0 & 0&1 \\
1 & 0 & 0 & \cdots & 0
\end{pmatrix}\,.
\end{equation}
The usual clock and shift matrices correspond to $k=N$ with $\lambda = -\pi \frac{N-1}{N}$. 
The above matrices satisfy
\begin{equation}\label{eq:commutation} 
 S_k\, C_k =e^{\frac{2\pi i}{k}} \, C_k\, S_k\,.
\end{equation}
Using this relation it is straightforward to show that for any choices of $\lambda_j$, the matrix $C$ in \eqref{eq:gen_clock} obeys 
\begin{equation}
\label{SC-comutator}
    S\, C =e^{\frac{2\pi i}{k}} \, C\, S~.
\end{equation}
Hence, we conclude that the elements $\hatg \in \hatG$ that satisfy \eqref{congugation} are the ones that belong to the conjugacy classes of the elements in \eqref{eq:gen_clock}.
Notice that permutations of the matrices in \eqref{eq:gen_clock} all belong to the same conjugacy class. 

If we choose $\lambda_j = - \pi \frac{k-1}{k}$, in addition to \eqref{SC-comutator} the above matrices satisfy
\begin{equation} 
C^k = S^k = e^{i \pi (k-1)}\, \mathbbm{1}_{N\times N}\,.
\end{equation}
Taking the quotient by $\ZZ_k$, these generalized clock and shift matrices define a subgroup $\ZZ_k \times \ZZ_k  \subset SU(N)/\ZZ_k$. In this case, Equations~\eqref{eq:gen_clock} and \eqref{eq:gen_shift} define a lift of $\ZZ_k \times \ZZ_k$ to $SU(N)$. For instance, we can take the lift
\begin{equation}
(a,b) \to C^b \, S^a \,, \quad a,b = 0,\ldots,k-1\,. 
\end{equation}
Alternatively, we can view this lift as providing a projective representation of $\ZZ_k\times\ZZ_k$ with the 2-cocycle 
\begin{equation}\label{eq:clockshift_cocycle} 
\omega_{(a_1,b_1),(a_2,b_2)} = \exp\left[ \frac{2\pi i}{k}\left( a_1 b_2  + \frac{k(k-1)}{2}\left(\beta(a_1,a_2) + \beta(b_1,b_2) \right) \right) \right]\,,
\end{equation}
where 
\begin{equation}
\beta(a_1,a_2) \equiv \frac{[a_1]_k + [a_2]_k - [a_1+a_2]_k}{k}\,,\quad [a]_k \equiv a \text{ mod } k \,. 
\end{equation} 

For other Lie groups $\hatG$, one can find $SU(N)$ subgroups whose center lies in the center of $\hatG$, and then use the construction of this Appendix to find elements which comnute up to elements of the center.

%%%%%%%%%%%%%%%%%%%%%%%%%%%%%
\section{Holonomy Effective Potential}
\label{app:holonomy}
%%%%%%%%%%%%%%%%%%%%%%%%%%%%%

In this Appendix we consider a free massive scalar field on $S^1\times \IR^{d-1}$ coupled to a background abelian gauge field $A$ with holonomy $\theta = \oint_{S^1}A$. Since the gauge field is non-dynamical, the exact effective action for $\theta$ (i.e. the $\theta$-dependence of the vacuum energy) can be obtained by computing a 1-loop functional determinant. The computation is fairly standard and similar calculations appear in~\cite{AFFLECK1980461,PhysRevD.24.475,GPY,Komargodski:2017dmc}. Assuming the circle has has a perimeter $L$, the effective potential for $\theta$ is given by
\eq{
V_{\rm eff}(\theta)&= \frac{1}{LV_{d-1}} \log\,\det(p^2+M^2)\\
&=   \frac{1}{L} \int \frac{d^{d-1}p}{(2\pi)^{d-1}} \sum_{k\in \IZ}\log\left[p^2+M^2+\left(\frac{2\pi k}{L}+\frac{\theta}{L}\right)^2\right]\,,
}
We can  compute the effective potential by rewriting
\eq{
V_{\rm eff}(\theta)&= -\lim_{\epsilon\to 0^+}\frac{1}{L} \int_0^\infty \frac{dt}{t^{1-\epsilon}}\int \frac{d^{d-1}p}{(2\pi)^{d-1}} \sum_{k\in \IZ}e^{-t\left[p^2 + M^2\left(\frac{2\pi k}{L}+\frac{\theta}{L}\right)^2 \right]}~,
}
which we can compute by using Poisson resummation
\eq{
V_{\rm eff}(\theta)&=- \lim_{\epsilon\to 0^+} \frac{1}{2\sqrt{\pi}} \int_0^\infty \frac{dt}{t^{3/2-\epsilon}}\int  \frac{d^{d-1}p}{(2\pi)^{d-1}}~e^{-t(p^2+M^2)} \sum_{n\in \IZ}e^{-\frac{n^2L^2}{4t} + in \theta}\\
&=- \lim_{\epsilon\to 0^+} \frac{1}{\sqrt{\pi}} \int_0^\infty \frac{dt}{t^{3/2-\epsilon}}\int  \frac{d^{d-1}p}{(2\pi)^{d-1}} ~e^{-t(p^2+M^2)}\left[ \frac{1}{2} + \sum_{n=1}^\infty e^{-\frac{n^2L^2}{4t}}\cos(n \theta)\right]\,,
}
Now we can subtract out the infinite $\theta$-independent contribution. Performing the momentum integral, we obtain
\eq{
V_{\rm eff}(\theta)&= -\frac{1}{2^{d-1}\pi^{\frac{d}{2}}}   \sum_{n=1}^\infty \cos(n \theta) \int_0^\infty \frac{dt}{t^{d/2+1}}\, e^{-tM^2-\frac{n^2L^2}{4t}}\,.
}
In the limit of large $ML \gg 1$ we can use the saddle point approximation to evaluate the proper time integral. This leads to 
\eq{
V_{\rm eff}(\theta)&= -\frac{2}{L^d}\left(\frac{ML}{2\pi}\right)^{\frac{d-1}{2}}  \sum_{n=1}^\infty \frac{1}{n^{\frac{d+1}{2}}} \cos(n \theta) \, e^{-n ML}\,. 
}
We use this result in Section~\ref{sec:ZNdeformations} where $\theta$ is the holonomy of a flat (but dynamical) $\ZZ_N$ gauge field.

%%%%%%%%%%%%%%%%%%%%%%%%%%%%%%%%%%%
\section{Projective Representations of $O(2)$}
\label{app:O(2)}
%%%%%%%%%%%%%%%%%%%%%%%%%%%%%%%%%%%

In this Appendix  we will discuss the projective representations of $O(2)$. In general $O(N)$ has a double cover $Pin^\pm(N)$ that fits into a commutative diagram with $Spin(N)$ and $SO(N)$:
\eq{
\xymatrix{Pin^\pm(N)\ar[r]\ar[d]&Spin(N)\ar[d]\\
O(N)\ar[r]&SO(N)}
}
The group $Pin^\pm(N)$ is given by the space of unit elements of the Clifford algebra $C\ell_1(\IR^N,\pm q)$ on $\IR^N$ defined with respect to the quadratic form $\pm q$ where $q$ is the standard Euclidean inner product. The group $Spin(N)$ can be realized as the subgroup of all even unit elements.  

The representation theory of $O(2)$ is well known, there are two one-dimensional representations -- the trivial and sign representations -- and an infinite number of real $2d$ representations generated by $r_{n\theta},y$ where $n\in \IZ$ and 
\eq{
r_\theta=\left(\begin{array}{cc}
e^{i \theta}&0\\0&e^{-i \theta}
\end{array}\right)\quad, \quad y=\left(\begin{array}{cc}
0&1\\1&0
\end{array}\right)
\label{O2matrixgenerators}
~.}
For the special case of $N=2$, $Pin^+(2)\cong O(2)$.  
Under the map $Pin^+(2)\to O(2)$, the $2d$ minimal $O(2)$ representation pulls back to the charge 2 $2d$ representation of $Pin^+(2)$. 
This means that the  minimal $2d$ representation of $Pin^+(2)$ is analogous to the `charge $\half$' representation of $O(2)$. Unlike the case of $U(1)$, this is a projective representation of $O(2)$ since there is no  choice of counterterms that both cancels the projective phase and is consistent with the $\IZ_2$ outer-automorphism of $U(1)$ in the decomposition $O(2)=U(1)\rtimes \IZ_2$. 

On the other hand $Pin^-(2)\not\cong O(2)$. The $2d$ representations of $Pin^-(2)$ are generated by $s_\theta,x$ for $\theta\in U(1)$ which satisfy 
\eq{
s_{2\pi}=1\quad, \quad 
x^2=s_\pi\quad, \quad x s_\theta x^{-1}=s_{-\theta}~. 
}
This has an infinite number of 
2-complex dimensional representation labeled by $n\in \IZ$ where $s_\theta,x$ are given by 
\eq{
s_\theta=\left(\begin{array}{cc}
e^{i n\theta}&0\\0&e^{-i n\theta}
\end{array}\right)\quad, \quad x=\left(\begin{array}{cc}
0&-1\\1&0
\end{array}\right)
~.\label{Pin-matrixgenerators}} 

\noindent These representations are projective representations of $O(2)$ for $n$ odd coming from the projection $\pi(s_\theta)=r_{\theta}$ and $\pi(y)=x$. One can show there does not exist a choice of counterterms that cancels the corresponding projective phase.

%%%%%%%%%%%%%%%%%%%%%%%%%%%%%%%
\section{Fusion Constraint for $\IZ_N$ Lines}
\label{app:fusion}
%%%%%%%%%%%%%%%%%%%%%%%%%%%%%%%

Let us consider a theory with topological lines $\scrL$ that generate a $K=\IZ_N^{(d-2)}$ symmetry on which the $G^\0$ symmetry is fractionalized.  
As described in Section~\ref{sec:no1formsymfrac}, the $G^\0$ junctions can act on the $K$ lines of charge $a=0,1,\ldots,N-1$ by the phase 
\begin{equation}
\omega_{g,h;a} :  G \times G\times K \to U(1)\,. 
\end{equation}
At the same time, the $a,b\in K$ lines can form point-like junctions which can in turn act on the codimension-1 $g\in G^\0$ surfaces. This action is captured by a phase 
\begin{equation}
\nu_{g;a,b} : G \times K \times K \to U(1)~.
\end{equation}
Associativity of these two actions implies
\begin{equation}
    (\delta_G \omega)_{g,h,k;a} = 1\,, \ (\delta_K\nu)_{g;a,b,c} = 1\,,
\end{equation}
where $\delta_G$ and $\delta_K$ are the group cohomology differentials with respect to $G,K$ respectively. Hence, if we fix $a \in K$, $\omega_a \in Z^2(BG,U(1))$ and if we fix $g \in G$, $\nu_g \in Z^2(BK,U(1))$. Consistency of the $G$ junctions in the presence of $K$ junctions further implies the relation
\begin{equation} \label{eq:omeganurelation}
(\delta_K \omega)_{g,h;a,b} = \frac{\omega_{g,h;a}\, \omega_{g,h;b}}{\omega_{g,h;ab}} = (\delta_G\nu)_{g,h;a,b} = \frac{\nu_{g;a,b}\, \nu_{h;a,b}}{\nu_{gh; a,b}} \,,
\end{equation}
A local counterterm assigns a phase $\alpha_{g;a}:G \times K \to U(1)$ to the junction of the line labeled by $a$ and the surface labeled by $g$. This changes both $\omega$ and $\nu$, 
\begin{equation}\label{eq:bigradedCT}
\omega \to \omega \, \delta_G\alpha\,, \quad \nu \to \nu \, \delta_K\alpha\,.
\end{equation}

\smallskip 
In the case $K = \ZZ_N$ we have $H^2(BK,U(1)) = 0$ so we can write $\nu_{g;a,b} = (\delta_K \beta)_{g;a,b}$ for some $\beta_{g;a}: G \times K \to U(1)$. Then from \eqref{eq:omeganurelation} we have
\begin{equation}
\delta_K \omega = \delta_K \delta_G \beta\,,
\end{equation}
which is solved by setting $\omega_{g,h;a} =\widetilde\omega_{g,h;a}\, (\delta_G\beta)_{g,h;a}$ where $(\delta_K\widetilde\omega)_{g,h;a,b} = 1$ and $(\delta_G \widetilde\omega)_{g,h,k;a} = 1$. Since for fixed $g,h$, $\widetilde\omega_{g,h;a}$ is a $K$ 1-cocycle, $\widetilde\omega_{g,h;a}$ can be regarded as an element of $Z^2(BG,Z^1(BK,U(1))$ --- it measures the failure of the trivialization of $\nu$ to trivialize $\omega$. 

Since $Z^1(BK,U(1))=H^1(BK,U(1))$, the non-trivial $\widetilde\omega$ classes are classified by 
\eq{
H^{2}(BG,H^1(BK,U(1))\,,
}
and when $K=\IZ_N$ we have
\eq{
H^2(BG,H^1(\IZ_N,U(1)))\cong H^2(BG,\IZ_N)\,.
}
Therefore, in this case the  anomalous phases $\omega$ are completely classified by elements of $H^2(BG,\IZ_N)$. 

More concretely, the fact that for fixed $g,h$ $\widetilde\omega_{g,h;a}$ is an element of $H^1(BK,U(1))$ is equivalent to saying that for each $g,h$, $\widetilde\omega_{g,h;a}$ is a one-dimensional representation of $K = \ZZ_N$. Hence, we can write a concrete expression
\begin{equation}
\widetilde\omega_{g,h;a} = \exp\left(\frac{2\pi ik}{N}\,a\,f_{g,h}\right)\,, \quad f_{g,h} : G \times G \to \ZZ_N\,. 
\end{equation}
where $k\in \IZ_N$ classifies an  element of $ H^1(BK,U(1))$ (it is the charge of the representation) and $f_{g,h}\in Z^2(BG,\IZ_N)$. 
% The $G$-closedness of $\tilde\omega$ implies that $(\delta f)_{g,h,k} = 0$ mod $p$. Hence $f \in Z^2(BG,\ZZ_p)$. 
In this construction, $f$ can be shifted by the counterterms which leave $\nu$ invariant ($\delta_K\alpha =1$),
\begin{equation}
\alpha_{g;a} = \exp\left(\frac{2\pi i}{N} \,a\,c_g \right)~,
\end{equation}
but transforms $f_{g,h} \to f_{g,h} + c_g + c_h - c_{gh}$. Hence $f$, and also $\widetilde\omega$, is characterized by $H^2(BG,\ZZ_N)$.

%%%%%%%%%%%%%%%%%%%%%%%%%%%%%%%%%%%%%%%%%%
\section{Symmetry Fractionalization and Screening in a \\ Lattice Gauge Theory}
%%%%%%%%%%%%%%%%%%%%%%%%%%%
\label{app:lattice} 
%%%%%%%%%%%%%%%%%%%%%%%%%%%%%%%%%%%%%%

Consider the non-abelian gauge-Higgs model first introduced in~\cite{fradkinshenker}. This is a $SU(N)$ gauge theory with Euclidean lattice action (and lattice spacing set to 1)
\begin{equation}
S = -\text{Re}\, \left[ \beta \sum_p \Tr (U_p) +\kappa \sum_{\ell} \Tr(\Phi^\dagger_x\, U_{(x,\mu)} \, \Phi_{x+\hat\mu}) \right]~.
\end{equation}
Here $\mu = 1,\ldots,d$ denote the $d$ Euclidean lattice directions, $U_{(x,\mu)}$ are $SU(N)$ group matrices on the links $\ell=(x,\mu)$ with endpoints $x,x+\hat\mu$, $U_p$ are the product of $U_\ell$ matrices around the plaquette boundary of a plaquette $p = (x,\mu\nu)$, and $\Phi_x \in SU(N)$ are group-valued Higgs fields on each site. We take the above trace in the fundamental representation. The parameter $\beta$ is interpreted as the inverse gauge coupling squared while the hopping parameter $\kappa$ is related to the mass of the Higgs fields. Gauge transformations act as
\begin{equation}
\Phi_x \to G_x\, \Phi_x\,, \quad U_{x,\mu} \to G_x \, U_{x,\mu}\, G_{x+\hat\mu}^\dagger\,, \quad G_x \in SU(N)\,.
\end{equation}
At first glance the theory appears to have a $SU(N)$ global symmetry that acts on $\Phi$ from the right, $\Phi_x \to \Phi_x V$ with $V \in SU(N)$. However, the $\ZZ_N$ center elements coincide with a gauge transformation, so the faithfully-acting global symmetry is $PSU(N)$. Of course it is more natural to couple the fields in the lattice action to background gauge fields for the lift $SU(N)$, by changing the hopping term to 
\begin{equation}
\sum_{\ell} \Tr(\Phi^\dagger_x\, U_{x,\mu} \, \Phi_{x+\hat\mu} \, \CU_{x,\mu}^\dagger ) \,,
\end{equation}
where $\CU_\ell \in SU(N)$ transforms under local background gauge transformations as $\CU_{x,\mu} \to V_{x}\, \CU_{x,\mu}\, V_{x+\hat\mu}^\dagger$. However, this coupling is incomplete --- it fails to capture the fact that the center elements where $\CU = e^{\frac{2\pi i }{N}} \mathbbm{1}$ are transparent to local operators. To remedy this, we also introduce a $\ZZ_N$-valued plaquette field $e^{\frac{2\pi i}{N}B_p}$ with $B_p \in \ZZ$, 
\begin{equation}
S[\{B_p\}, \{\CU_\ell\}] = -\text{Re}\, \left[ \beta \sum_p \Tr (U_p\, e^{-\frac{2\pi i}{N}B_p}) +\kappa \sum_{\ell} \Tr(\Phi^\dagger_x\, U_{x,\mu} \, \Phi_{x+\hat\mu} \, \CU_{x,\mu}^\dagger ) \right]\,.
\end{equation}
Now we have the additional invariance where we send $\CU_\ell \to e^{\frac{2\pi i}{N}K_\ell}\,\CU_\ell$ if we simultaneously transform the dynamical fields $U_\ell \to e^{\frac{2\pi i}{N}K_\ell} \,U_\ell$ and take $B_p \to B_p + (dK)_p$. Inserting a symmetry operator for the $PSU(N)$ global symmetry (lifted to $SU(N)$) is equivalent to turning on $\CU_\ell$ on a set of links which pierce a closed codimension-1 surface on the dual lattice. 

Below we consider two regimes where we can semi-quantitatively compute Wilson line expectation values.

\bigskip
%%%%%%%%%%%%%%%%%%%%%%%%%%%%%%%%
\noindent
\underline{\emph{The `hopping' regime: $\kappa \ll 1$}}
%%%%%%%%%%%%%%%%%%%%%%%%%%%%%%%%

\bigskip
\noindent
First we consider the Wilson loop expectation value in the hopping parameter expansion, valid for $\kappa \ll 1$. Specifically, we are interested in
\begin{equation}
    \langle W(\gamma) \rangle  = \prod_x \int d\Phi_x \prod_\ell \int dU_\ell \, \Tr( \prod_{\ell \in \gamma} U_\ell) \, e^{-S[\{B_p\},\{\CU_\ell\}]}\,,
\end{equation}
where we integrate with respect to the Haar measure of $SU(N)$ for both $U_\ell$ and $\Phi_x$. Let us also assume $\beta \ll 1$, so that we can perform a strong-coupling expansion. There are two types of contributions to the above expectation value. We can bring down plaquettes from the Wilson term in the action to tile the interior of the Wilson loop, yielding contributions which scale like $\beta^A$ where $A$ is the area of the bounding surface. There are also contributions where we bring down a sequence of links from the hopping term in the action. Such links must form a closed loop (otherwise the integration over $\Phi$'s causes it to vanish). Consider a generic contribution of the latter type, consisting of a sequence of $L$ links forming a closed loop. Integrating out the matter fields gives 
\begin{align}
&\cdots \left(\frac{\kappa}{2}\right)^L \prod_{i=1}^L \int d\Phi_i \, \Tr(\Phi_1^\dagger U_{12} \Phi_2 \CU_{12}^\dagger)\Tr(\Phi_2^\dagger U_{23} \Phi_3 \CU_{23}^\dagger) \cdots \Tr(\Phi_L^\dagger U_{L1} \Phi_1 \CU_{L1}^\dagger) \nonumber \cdots \\
= &\cdots \left(\frac{\kappa}{2}\right)^L\frac{1}{N^{L-1}} \int d\Phi_1 \, \Tr(\Phi_1^\dagger U_{12}U_{23}\cdots U_{L1} \Phi_1 \CU_{L1}^\dagger \cdots \CU_{23}^\dagger \CU_{12}^\dagger) \nonumber \cdots\\
= &\cdots \left(\frac{\kappa}{2N}\right)^L \Tr((\CU_{12}\CU_{23}\cdots \CU_{L1})^\dagger) \Tr(U_{12}U_{23}\cdots U_{L1}) \cdots\,. 
\end{align}
The Wilson loop expectation value is dominated (for large loops) by a contribution like the one above, where the hopping loop runs exactly opposite to the Wilson loop insertion (i.e. $U_{12}U_{23} \cdots U_{L1} = (\prod_{\ell\in \gamma}U_\ell)^\dagger$). Just keeping this leading contribution, we find
\begin{equation}
\langle W(\gamma) \rangle \approx \left(\frac{\kappa}{2N}\right)^L \Tr(\prod_{\ell\in \gamma}\CU_{\ell}) = e^{-M L}\, W_{\rm flavor}(\gamma)\, + \CO(\kappa,\beta)
\end{equation}
where we defined $M = \log(2N/\kappa)$. So we see that the dynamical Wilson line `flows' to the background Wilson line in the projective representation of $PSU(N)$. This is the lattice gauge theory analog of the quantum mechanics partition function in Eq.~\eqref{eq:bgd_wilson_line}.

\bigskip
%%%%%%%%%%%%%%%%%%%%%%%%%%%%%%%%
\noindent
\underline{\emph{The `Higgs' regime: $\kappa \gg 1$}}
%%%%%%%%%%%%%%%%%%%%%%%%%%%%%%%%

\bigskip
\noindent
Now we consider the opposite regime. When $\kappa\to\infty$ we get a constraint from the hopping term that
\begin{equation}
U_{x,\mu} = \Phi_x\, \CU_{x,\mu}\, \Phi_{x+\hat\mu}^\dagger\,. 
\end{equation}
Then the Wilson loop becomes 
\begin{equation}
\langle W(\gamma) \rangle \approx W_{\rm flavor}(\gamma)~,
\end{equation}
on the nose. We can easily write a local counterterm on the Wilson line which is a function $\kappa$ such that in the two limits considered above, we have $\langle W(\gamma) \, e^{\mu_{\rm c.t.}(\kappa)L} \rangle \approx W_{\rm flavor}(\gamma)$. Both the strong-coupling hopping regime and the Higgs regimes are trivially gapped phases which are continuously connected in this model~\cite{fradkinshenker}, and indeed we see that the Wilson line is well-approximated by the decoupled worldline quantum mechanics whose partition function yields the trace of the background Wilson line in the projective representation of $PSU(N)$. Indeed, let us consider the Wilson line expectation value in the presence of a $PSU(N)$ symmetry twist equal to the clock matrix, which is equivalent to setting $\prod_{\ell\in\gamma} \CU_\ell = C$ where $C$ is the clock matrix. In this case the flavor trace vanishes and we satisfy the selection rule in Eq.~\eqref{eq:selectionrule}.

\bibliographystyle{JHEP}
\bibliography{PQBib}

\providecommand{\href}[2]{#2}\begingroup\raggedright\begin{thebibliography}{100}

\bibitem{Wegner:1971app}
F.J.~Wegner, \emph{{Duality in Generalized Ising Models and Phase Transitions
  Without Local Order Parameters}},
  \href{https://doi.org/10.1063/1.1665530}{\emph{J. Math. Phys.} {\bfseries 12}
  (1971) 2259}.

\bibitem{Wilson:1974sk}
K.G.~Wilson, \emph{{Confinement of Quarks}},
  \href{https://doi.org/10.1103/PhysRevD.10.2445}{\emph{Phys. Rev. D}
  {\bfseries 10} (1974) 2445}.

\bibitem{tHooft:1979rat}
G.~'t~Hooft, \emph{{Naturalness, chiral symmetry, and spontaneous chiral
  symmetry breaking}},
  \href{https://doi.org/10.1007/978-1-4684-7571-5_9}{\emph{NATO Sci. Ser. B}
  {\bfseries 59} (1980) 135}.

\bibitem{Kondo:1964nea}
J.~Kondo, \emph{{Resistance Minimum in Dilute Magnetic Alloys}},
  \href{https://doi.org/10.1143/PTP.32.37}{\emph{Prog. Theor. Phys.} {\bfseries
  32} (1964) 37}.

\bibitem{Gaiotto:2014kfa}
D.~Gaiotto, A.~Kapustin, N.~Seiberg and B.~Willett, \emph{{Generalized Global
  Symmetries}}, \href{https://doi.org/10.1007/JHEP02(2015)172}{\emph{JHEP}
  {\bfseries 02} (2015) 172} [\href{https://arxiv.org/abs/1412.5148}{{\ttfamily
  1412.5148}}].

\bibitem{Freed:2022qnc}
D.S.~Freed, G.W.~Moore and C.~Teleman, \emph{{Topological symmetry in quantum
  field theory}},  \href{https://arxiv.org/abs/2209.07471}{{\ttfamily
  2209.07471}}.

\bibitem{McGreevy:2022oyu}
J.~McGreevy, \emph{{Generalized Symmetries in Condensed Matter}},
  \href{https://doi.org/10.1146/annurev-conmatphys-040721-021029}{\emph{Ann.
  Rev. Condensed Matter Phys.} {\bfseries 14} (2023) 57}
  [\href{https://arxiv.org/abs/2204.03045}{{\ttfamily 2204.03045}}].

\bibitem{Schafer-Nameki:2023jdn}
S.~Schafer-Nameki, \emph{{ICTP lectures on (non-)invertible generalized
  symmetries}},
  \href{https://doi.org/10.1016/j.physrep.2024.01.007}{\emph{Phys. Rept.}
  {\bfseries 1063} (2024) 1}
  [\href{https://arxiv.org/abs/2305.18296}{{\ttfamily 2305.18296}}].

\bibitem{Shao:2023gho}
S.-H.~Shao, \emph{{What's Done Cannot Be Undone: TASI Lectures on
  Non-Invertible Symmetries}},
  \href{https://arxiv.org/abs/2308.00747}{{\ttfamily 2308.00747}}.

\bibitem{Bhardwaj:2023kri}
L.~Bhardwaj, L.E.~Bottini, L.~Fraser-Taliente, L.~Gladden, D.S.W.~Gould,
  A.~Platschorre et~al., \emph{{Lectures on generalized symmetries}},
  \href{https://doi.org/10.1016/j.physrep.2023.11.002}{\emph{Phys. Rept.}
  {\bfseries 1051} (2024) 1}
  [\href{https://arxiv.org/abs/2307.07547}{{\ttfamily 2307.07547}}].

\bibitem{Freed:2022iao}
D.S.~Freed, \emph{{Introduction to topological symmetry in QFT}}, {\emph{Proc.
  Symp. Pure Math.} {\bfseries 107} (2024) 93}
  [\href{https://arxiv.org/abs/2212.00195}{{\ttfamily 2212.00195}}].

\bibitem{Brennan:2023mmt}
T.D.~Brennan and S.~Hong, \emph{{Introduction to Generalized Global Symmetries
  in QFT and Particle Physics}},
  \href{https://arxiv.org/abs/2306.00912}{{\ttfamily 2306.00912}}.

\bibitem{Costa:2024wks}
D.~Costa et~al., \emph{{Simons Lectures on Categorical Symmetries}},  11, 2024
  [\href{https://arxiv.org/abs/2411.09082}{{\ttfamily 2411.09082}}].

\bibitem{lake2018}
E.~Lake, \emph{Higher-form symmetries and spontaneous symmetry breaking},
  2018.

\bibitem{Hofman_2019}
D.~Hofman and N.~Iqbal, \emph{Goldstone modes and photonization for higher form
  symmetries},
  \href{https://doi.org/10.21468/scipostphys.6.1.006}{\emph{SciPost Physics}
  {\bfseries 6} (2019) }.

\bibitem{Iqbal:2021rkn}
N.~Iqbal and J.~McGreevy, \emph{{Mean string field theory: Landau-Ginzburg
  theory for 1-form symmetries}},
  \href{https://doi.org/10.21468/SciPostPhys.13.5.114}{\emph{SciPost Phys.}
  {\bfseries 13} (2022) 114}
  [\href{https://arxiv.org/abs/2106.12610}{{\ttfamily 2106.12610}}].

\bibitem{Bhardwaj:2023fca}
L.~Bhardwaj, L.E.~Bottini, D.~Pajer and S.~Schafer-Nameki, \emph{{Categorical
  Landau Paradigm for Gapped Phases}},
  \href{https://doi.org/10.1103/PhysRevLett.133.161601}{\emph{Phys. Rev. Lett.}
  {\bfseries 133} (2024) 161601}
  [\href{https://arxiv.org/abs/2310.03786}{{\ttfamily 2310.03786}}].

\bibitem{Baez:2003yaq}
J.C.~Baez and A.D.~Lauda, \emph{{Higher-Dimensional Algebra V: 2-Groups}},
  \href{https://arxiv.org/abs/math/0307200}{{\ttfamily math/0307200}}.

\bibitem{Baez:2004in}
J.~Baez and U.~Schreiber, \emph{{Higher gauge theory: 2-connections on
  2-bundles}},  \href{https://arxiv.org/abs/hep-th/0412325}{{\ttfamily
  hep-th/0412325}}.

\bibitem{Cordova:2018cvg}
C.~C\'ordova, T.T.~Dumitrescu and K.~Intriligator, \emph{{Exploring 2-Group
  Global Symmetries}},
  \href{https://doi.org/10.1007/JHEP02(2019)184}{\emph{JHEP} {\bfseries 02}
  (2019) 184} [\href{https://arxiv.org/abs/1802.04790}{{\ttfamily
  1802.04790}}].

\bibitem{Benini:2018reh}
F.~Benini, C.~C\'ordova and P.-S.~Hsin, \emph{{On 2-Group Global Symmetries and
  their Anomalies}}, \href{https://doi.org/10.1007/JHEP03(2019)118}{\emph{JHEP}
  {\bfseries 03} (2019) 118}
  [\href{https://arxiv.org/abs/1803.09336}{{\ttfamily 1803.09336}}].

\bibitem{Sharpe:2015mja}
E.~Sharpe, \emph{{Notes on generalized global symmetries in QFT}},
  \href{https://doi.org/10.1002/prop.201500048}{\emph{Fortsch. Phys.}
  {\bfseries 63} (2015) 659}
  [\href{https://arxiv.org/abs/1508.04770}{{\ttfamily 1508.04770}}].

\bibitem{Gukov:2013zka}
S.~Gukov and A.~Kapustin, \emph{{Topological Quantum Field Theory, Nonlocal
  Operators, and Gapped Phases of Gauge Theories}},
  \href{https://arxiv.org/abs/1307.4793}{{\ttfamily 1307.4793}}.

\bibitem{Brennan:2020ehu}
T.D.~Brennan and C.~Cordova, \emph{{Axions, higher-groups, and emergent
  symmetry}}, \href{https://doi.org/10.1007/JHEP02(2022)145}{\emph{JHEP}
  {\bfseries 02} (2022) 145}
  [\href{https://arxiv.org/abs/2011.09600}{{\ttfamily 2011.09600}}].

\bibitem{Delmastro:2022pfo}
D.G.~Delmastro, J.~Gomis, P.-S.~Hsin and Z.~Komargodski, \emph{{Anomalies and
  symmetry fractionalization}},
  \href{https://doi.org/10.21468/SciPostPhys.15.3.079}{\emph{SciPost Phys.}
  {\bfseries 15} (2023) 079}
  [\href{https://arxiv.org/abs/2206.15118}{{\ttfamily 2206.15118}}].

\bibitem{Barkeshli:2014cna}
M.~Barkeshli, P.~Bonderson, M.~Cheng and Z.~Wang, \emph{{Symmetry
  Fractionalization, Defects, and Gauging of Topological Phases}},
  \href{https://doi.org/10.1103/PhysRevB.100.115147}{\emph{Phys. Rev. B}
  {\bfseries 100} (2019) 115147}
  [\href{https://arxiv.org/abs/1410.4540}{{\ttfamily 1410.4540}}].

\bibitem{Brennan:2022tyl}
T.D.~Brennan, C.~Cordova and T.T.~Dumitrescu, \emph{{Line Defect Quantum
  Numbers \& Anomalies}},  \href{https://arxiv.org/abs/2206.15401}{{\ttfamily
  2206.15401}}.

\bibitem{Wang_2014}
C.~Wang, A.C.~Potter and T.~Senthil, \emph{Classification of interacting
  electronic topological insulators in three dimensions},
  \href{https://doi.org/10.1126/science.1243326}{\emph{Science} {\bfseries 343}
  (2014) 629–631}.

\bibitem{Thorngren:2014pza}
R.~Thorngren, \emph{{Framed Wilson Operators, Fermionic Strings, and
  Gravitational Anomaly in 4d}},
  \href{https://doi.org/10.1007/JHEP02(2015)152}{\emph{JHEP} {\bfseries 02}
  (2015) 152} [\href{https://arxiv.org/abs/1404.4385}{{\ttfamily 1404.4385}}].

\bibitem{Kravec:2014aza}
S.M.~Kravec, J.~McGreevy and B.~Swingle, \emph{{All-fermion electrodynamics and
  fermion number anomaly inflow}},
  \href{https://doi.org/10.1103/PhysRevD.92.085024}{\emph{Phys. Rev. D}
  {\bfseries 92} (2015) 085024}
  [\href{https://arxiv.org/abs/1409.8339}{{\ttfamily 1409.8339}}].

\bibitem{Zou:2017ppq}
L.~Zou, C.~Wang and T.~Senthil, \emph{{Symmetry enriched U(1) quantum spin
  liquids}}, \href{https://doi.org/10.1103/PhysRevB.97.195126}{\emph{Phys. Rev.
  B} {\bfseries 97} (2018) 195126}
  [\href{https://arxiv.org/abs/1710.00743}{{\ttfamily 1710.00743}}].

\bibitem{Wang:2018qoy}
J.~Wang, X.-G.~Wen and E.~Witten, \emph{{A New SU(2) Anomaly}},
  \href{https://doi.org/10.1063/1.5082852}{\emph{J. Math. Phys.} {\bfseries 60}
  (2019) 052301} [\href{https://arxiv.org/abs/1810.00844}{{\ttfamily
  1810.00844}}].

\bibitem{Hsin:2019fhf}
P.-S.~Hsin and A.~Turzillo, \emph{{Symmetry-enriched quantum spin liquids in (3
  + 1)$d$}}, \href{https://doi.org/10.1007/JHEP09(2020)022}{\emph{JHEP}
  {\bfseries 09} (2020) 022}
  [\href{https://arxiv.org/abs/1904.11550}{{\ttfamily 1904.11550}}].

\bibitem{Kan:2024fuu}
N.~Kan, K.~Kawabata and H.~Wada, \emph{{Symmetry fractionalization and duality
  defects in Maxwell theory}},
  \href{https://doi.org/10.1007/JHEP10(2024)238}{\emph{JHEP} {\bfseries 10}
  (2024) 238} [\href{https://arxiv.org/abs/2404.14481}{{\ttfamily
  2404.14481}}].

\bibitem{Jackiw:1975fn}
R.~Jackiw and C.~Rebbi, \emph{{Solitons with Fermion Number 1/2}},
  \href{https://doi.org/10.1103/PhysRevD.13.3398}{\emph{Phys. Rev. D}
  {\bfseries 13} (1976) 3398}.

\bibitem{Komargodski:2017dmc}
Z.~Komargodski, A.~Sharon, R.~Thorngren and X.~Zhou, \emph{{Comments on Abelian
  Higgs Models and Persistent Order}},
  \href{https://doi.org/10.21468/SciPostPhys.6.1.003}{\emph{SciPost Phys.}
  {\bfseries 6} (2019) 003} [\href{https://arxiv.org/abs/1705.04786}{{\ttfamily
  1705.04786}}].

\bibitem{Komargodski:2017smk}
Z.~Komargodski, T.~Sulejmanpasic and M.~\"Unsal, \emph{{Walls, anomalies, and
  deconfinement in quantum antiferromagnets}},
  \href{https://doi.org/10.1103/PhysRevB.97.054418}{\emph{Phys. Rev. B}
  {\bfseries 97} (2018) 054418}
  [\href{https://arxiv.org/abs/1706.05731}{{\ttfamily 1706.05731}}].

\bibitem{Benini:2017dus}
F.~Benini, P.-S.~Hsin and N.~Seiberg, \emph{{Comments on global symmetries,
  anomalies, and duality in (2 + 1)d}},
  \href{https://doi.org/10.1007/JHEP04(2017)135}{\emph{JHEP} {\bfseries 04}
  (2017) 135} [\href{https://arxiv.org/abs/1702.07035}{{\ttfamily
  1702.07035}}].

\bibitem{Cordova:2018acb}
C.~C\'ordova and T.T.~Dumitrescu, \emph{{Candidate phases for SU(2) adjoint
  QCD$_4$ with two flavors from $\mathcal{N}=2$ supersymmetric Yang-Mills
  theory}}, \href{https://doi.org/10.21468/SciPostPhys.16.5.139}{\emph{SciPost
  Phys.} {\bfseries 16} (2024) 139}
  [\href{https://arxiv.org/abs/1806.09592}{{\ttfamily 1806.09592}}].

\bibitem{Anber:2019nze}
M.M.~Anber and E.~Poppitz, \emph{{On the baryon-color-flavor (BCF) anomaly in
  vector-like theories}},
  \href{https://doi.org/10.1007/JHEP11(2019)063}{\emph{JHEP} {\bfseries 11}
  (2019) 063} [\href{https://arxiv.org/abs/1909.09027}{{\ttfamily
  1909.09027}}].

\bibitem{Cordova:2019uob}
C.~C\'ordova, D.S.~Freed, H.T.~Lam and N.~Seiberg, \emph{{Anomalies in the
  Space of Coupling Constants and Their Dynamical Applications II}},
  \href{https://doi.org/10.21468/SciPostPhys.8.1.002}{\emph{SciPost Phys.}
  {\bfseries 8} (2020) 002} [\href{https://arxiv.org/abs/1905.13361}{{\ttfamily
  1905.13361}}].

\bibitem{Anber:2020gig}
M.M.~Anber and E.~Poppitz, \emph{{Generalized \textquoteright{}t Hooft
  anomalies on non-spin manifolds}},
  \href{https://doi.org/10.1007/JHEP04(2020)097}{\emph{JHEP} {\bfseries 04}
  (2020) 097} [\href{https://arxiv.org/abs/2002.02037}{{\ttfamily
  2002.02037}}].

\bibitem{Anber:2021iip}
M.M.~Anber, S.~Hong and M.~Son, \emph{{New anomalies, TQFTs, and confinement in
  bosonic chiral gauge theories}},
  \href{https://doi.org/10.1007/JHEP02(2022)062}{\emph{JHEP} {\bfseries 02}
  (2022) 062} [\href{https://arxiv.org/abs/2109.03245}{{\ttfamily
  2109.03245}}].

\bibitem{Anber:2021lzb}
M.M.~Anber, \emph{{Condensates and anomaly cascade in vector-like theories}},
  \href{https://doi.org/10.1007/JHEP03(2021)191}{\emph{JHEP} {\bfseries 03}
  (2021) 191} [\href{https://arxiv.org/abs/2101.04132}{{\ttfamily
  2101.04132}}].

\bibitem{Apruzzi:2021vcu}
F.~Apruzzi, S.~Schafer-Nameki, L.~Bhardwaj and J.~Oh, \emph{{The Global Form of
  Flavor Symmetries and 2-Group Symmetries in 5d SCFTs}},
  \href{https://doi.org/10.21468/SciPostPhys.13.2.024}{\emph{SciPost Phys.}
  {\bfseries 13} (2022) 024}
  [\href{https://arxiv.org/abs/2105.08724}{{\ttfamily 2105.08724}}].

\bibitem{Heckman:2022suy}
J.J.~Heckman, C.~Lawrie, L.~Lin, H.Y.~Zhang and G.~Zoccarato, \emph{{6D SCFTs,
  center-flavor symmetries, and Stiefel-Whitney compactifications}},
  \href{https://doi.org/10.1103/PhysRevD.106.066003}{\emph{Phys. Rev. D}
  {\bfseries 106} (2022) 066003}
  [\href{https://arxiv.org/abs/2205.03411}{{\ttfamily 2205.03411}}].

\bibitem{Lohitsiri:2022jyz}
N.~Lohitsiri and T.~Sulejmanpasic, \emph{{Comments on QCD$_{3}$ and anomalies
  with fundamental and adjoint matter}},
  \href{https://doi.org/10.1007/JHEP10(2022)081}{\emph{JHEP} {\bfseries 10}
  (2022) 081} [\href{https://arxiv.org/abs/2205.07825}{{\ttfamily
  2205.07825}}].

\bibitem{Bhardwaj:2023zix}
L.~Bhardwaj, M.~Bullimore, A.E.V.~Ferrari and S.~Schafer-Nameki,
  \emph{{Generalized Symmetries and Anomalies of 3d N=4 SCFTs}},
  \href{https://doi.org/10.21468/SciPostPhys.16.3.080}{\emph{SciPost Phys.}
  {\bfseries 16} (2024) 080}
  [\href{https://arxiv.org/abs/2301.02249}{{\ttfamily 2301.02249}}].

\bibitem{Brennan:2023tae}
T.D.~Brennan, \emph{{A new solution to the Callan Rubakov effect}},
  \href{https://doi.org/10.1007/JHEP11(2024)170}{\emph{JHEP} {\bfseries 11}
  (2024) 170} [\href{https://arxiv.org/abs/2309.00680}{{\ttfamily
  2309.00680}}].

\bibitem{Anber:2023urd}
M.M.~Anber, N.~Lohitsiri and T.~Sulejmanpasic, \emph{{Remarks on QCD$_{4}$ with
  fundamental and adjoint matter}},
  \href{https://doi.org/10.1007/JHEP12(2023)063}{\emph{JHEP} {\bfseries 12}
  (2023) 063} [\href{https://arxiv.org/abs/2306.01849}{{\ttfamily
  2306.01849}}].

\bibitem{Brennan:2023ynm}
T.D.~Brennan and A.~Sheckler, \emph{{Anomaly Enforced Gaplessness for
  Background Flux Anomalies and Symmetry Fractionalization}},
  \href{https://arxiv.org/abs/2311.00093}{{\ttfamily 2311.00093}}.

\bibitem{Brennan:2023vsa}
T.D.~Brennan and K.~Intriligator, \emph{{Anomalies of 4d Spin$_{G}$ theories}},
  \href{https://doi.org/10.1007/JHEP07(2024)157}{\emph{JHEP} {\bfseries 07}
  (2024) 157} [\href{https://arxiv.org/abs/2312.04756}{{\ttfamily
  2312.04756}}].

\bibitem{Brennan:2023kpo}
T.D.~Brennan, \emph{{Anomaly enforced gaplessness and symmetry
  fractionalization for Spin$_{G}$ symmetries}},
  \href{https://doi.org/10.1007/JHEP02(2024)065}{\emph{JHEP} {\bfseries 02}
  (2024) 065} [\href{https://arxiv.org/abs/2308.12999}{{\ttfamily
  2308.12999}}].

\bibitem{COHEN1983183}
E.~Cohen and C.~Gomez, \emph{A computation of $\text{Tr}(-1)^f$ in
  supersymmetric gauge theories with matter},
  \href{https://doi.org/https://doi.org/10.1016/0550-3213(83)90100-1}{\emph{Nuclear
  Physics B} {\bfseries 223} (1983) 183}.

\bibitem{Kouno:2015sja}
H.~Kouno, K.~Kashiwa, J.~Takahashi, T.~Misumi and M.~Yahiro,
  \emph{{Understanding QCD at high density from a Z$_3$-symmetric QCD-like
  theory}}, \href{https://doi.org/10.1103/PhysRevD.93.056009}{\emph{Phys. Rev.
  D} {\bfseries 93} (2016) 056009}
  [\href{https://arxiv.org/abs/1504.07585}{{\ttfamily 1504.07585}}].

\bibitem{Iritani:2015ara}
T.~Iritani, E.~Itou and T.~Misumi, \emph{{Lattice study on QCD-like theory with
  exact center symmetry}},
  \href{https://doi.org/10.1007/JHEP11(2015)159}{\emph{JHEP} {\bfseries 11}
  (2015) 159} [\href{https://arxiv.org/abs/1508.07132}{{\ttfamily
  1508.07132}}].

\bibitem{Misumi:2015hfa}
T.~Misumi, T.~Iritani and E.~Itou, \emph{{Finite-temperature phase transition
  of $N_{f}=3$ QCD with exact center symmetry}},
  \href{https://doi.org/10.22323/1.251.0152}{\emph{PoS} {\bfseries LATTICE2015}
  (2016) 152} [\href{https://arxiv.org/abs/1510.07227}{{\ttfamily
  1510.07227}}].

\bibitem{Tanizaki:2017qhf}
Y.~Tanizaki, T.~Misumi and N.~Sakai, \emph{{Circle compactification and 't
  Hooft anomaly}}, \href{https://doi.org/10.1007/JHEP12(2017)056}{\emph{JHEP}
  {\bfseries 12} (2017) 056}
  [\href{https://arxiv.org/abs/1710.08923}{{\ttfamily 1710.08923}}].

\bibitem{Tanizaki:2017mtm}
Y.~Tanizaki, Y.~Kikuchi, T.~Misumi and N.~Sakai, \emph{{Anomaly matching for
  the phase diagram of massless $\mathbb{Z}_N$-QCD}},
  \href{https://doi.org/10.1103/PhysRevD.97.054012}{\emph{Phys. Rev. D}
  {\bfseries 97} (2018) 054012}
  [\href{https://arxiv.org/abs/1711.10487}{{\ttfamily 1711.10487}}].

\bibitem{Shimizu:2017asf}
H.~Shimizu and K.~Yonekura, \emph{{Anomaly constraints on deconfinement and
  chiral phase transition}},
  \href{https://doi.org/10.1103/PhysRevD.97.105011}{\emph{Phys. Rev. D}
  {\bfseries 97} (2018) 105011}
  [\href{https://arxiv.org/abs/1706.06104}{{\ttfamily 1706.06104}}].

\bibitem{Cherman:2017tey}
A.~Cherman, S.~Sen, M.~Unsal, M.L.~Wagman and L.G.~Yaffe, \emph{{Order
  parameters and color-flavor center symmetry in QCD}},
  \href{https://doi.org/10.1103/PhysRevLett.119.222001}{\emph{Phys. Rev. Lett.}
  {\bfseries 119} (2017) 222001}
  [\href{https://arxiv.org/abs/1706.05385}{{\ttfamily 1706.05385}}].

\bibitem{Yonekura:2019vyz}
K.~Yonekura, \emph{{Anomaly matching in QCD thermal phase transition}},
  \href{https://doi.org/10.1007/JHEP05(2019)062}{\emph{JHEP} {\bfseries 05}
  (2019) 062} [\href{https://arxiv.org/abs/1901.08188}{{\ttfamily
  1901.08188}}].

\bibitem{Kanazawa:2019tnf}
T.~Kanazawa and M.~\"Unsal, \emph{{Quantum distillation in QCD}},
  \href{https://doi.org/10.1103/PhysRevD.102.034013}{\emph{Phys. Rev. D}
  {\bfseries 102} (2020) 034013}
  [\href{https://arxiv.org/abs/1909.05222}{{\ttfamily 1909.05222}}].

\bibitem{Fujimori:2020zka}
T.~Fujimori, E.~Itou, T.~Misumi, M.~Nitta and N.~Sakai, \emph{{Lattice
  ${\mathbb C} P^{N-1}$ model with ${\mathbb Z}_{N}$ twisted boundary
  condition: bions, adiabatic continuity and pseudo-entropy}},
  \href{https://doi.org/10.1007/JHEP08(2020)011}{\emph{JHEP} {\bfseries 08}
  (2020) 011} [\href{https://arxiv.org/abs/2006.05106}{{\ttfamily
  2006.05106}}].

\bibitem{Nguyen:2022lie}
M.~Nguyen, Y.~Tanizaki and M.~\"Unsal, \emph{{Winding \ensuremath{\theta} and
  destructive interference of instantons}},
  \href{https://doi.org/10.1007/JHEP09(2023)033}{\emph{JHEP} {\bfseries 09}
  (2023) 033} [\href{https://arxiv.org/abs/2207.03008}{{\ttfamily
  2207.03008}}].

\bibitem{Nardoni:2024sos}
E.~Nardoni, M.~Sacchi, O.~Sela, G.~Zafrir and Y.~Zheng, \emph{{Dimensionally
  reducing generalized symmetries from (3+1)-dimensions}},
  \href{https://doi.org/10.1007/JHEP07(2024)110}{\emph{JHEP} {\bfseries 07}
  (2024) 110} [\href{https://arxiv.org/abs/2403.15995}{{\ttfamily
  2403.15995}}].

\bibitem{Tanizaki:2022plm}
Y.~Tanizaki and M.~\"Unsal, \emph{{Semiclassics with \textquoteright{}t Hooft
  flux background for QCD with 2-index quarks}},
  \href{https://doi.org/10.1007/JHEP08(2022)038}{\emph{JHEP} {\bfseries 08}
  (2022) 038} [\href{https://arxiv.org/abs/2205.11339}{{\ttfamily
  2205.11339}}].

\bibitem{Tanizaki:2022ngt}
Y.~Tanizaki and M.~\"Unsal, \emph{{Center vortex and confinement in
  Yang\textendash{}Mills theory and QCD with anomaly-preserving
  compactifications}}, \href{https://doi.org/10.1093/ptep/ptac042}{\emph{PTEP}
  {\bfseries 2022} (2022) 04A108}
  [\href{https://arxiv.org/abs/2201.06166}{{\ttfamily 2201.06166}}].

\bibitem{Hayashi:2023wwi}
Y.~Hayashi, Y.~Tanizaki and H.~Watanabe, \emph{{Semiclassical analysis of the
  bifundamental QCD on~$\mathbb{R}^2\times T^2$ with \textquoteright{}t Hooft
  flux}}, \href{https://doi.org/10.1007/JHEP10(2023)146}{\emph{JHEP} {\bfseries
  10} (2023) 146} [\href{https://arxiv.org/abs/2307.13954}{{\ttfamily
  2307.13954}}].

\bibitem{Hayashi:2024gxv}
Y.~Hayashi, Y.~Tanizaki and H.~Watanabe, \emph{{Non-supersymmetric duality
  cascade of QCD(BF) via semiclassics on
  \ensuremath{\mathbb{R}}$^{2}$\texttimes{} T$^{2}$ with the
  baryon-\textquoteright{}t Hooft flux}},
  \href{https://doi.org/10.1007/JHEP07(2024)033}{\emph{JHEP} {\bfseries 07}
  (2024) 033} [\href{https://arxiv.org/abs/2404.16803}{{\ttfamily
  2404.16803}}].

\bibitem{Hayashi:2024qkm}
Y.~Hayashi and Y.~Tanizaki, \emph{{Semiclassics for the QCD vacuum structure
  through T$^{2}$-compactification with the baryon-\textquoteright{}t Hooft
  flux}}, \href{https://doi.org/10.1007/JHEP08(2024)001}{\emph{JHEP} {\bfseries
  08} (2024) 001} [\href{https://arxiv.org/abs/2402.04320}{{\ttfamily
  2402.04320}}].

\bibitem{Sachdev:2003yk}
S.~Sachdev and M.~Vojta, \emph{{Quantum impurity in an antiferromagnet:
  Nonlinear sigma model theory}},
  \href{https://doi.org/10.1103/PhysRevB.68.064419}{\emph{Phys. Rev. B}
  {\bfseries 68} (2003) 064419}
  [\href{https://arxiv.org/abs/cond-mat/0303001}{{\ttfamily
  cond-mat/0303001}}].

\bibitem{Florens_2006}
S.~Florens, L.~Fritz and M.~Vojta, \emph{Kondo effect in bosonic spin liquids},
  \href{https://doi.org/10.1103/physrevlett.96.036601}{\emph{Physical Review
  Letters} {\bfseries 96} (2006) }.

\bibitem{Liu_2021}
S.~Liu, H.~Shapourian, A.~Vishwanath and M.A.~Metlitski, \emph{Magnetic
  impurities at quantum critical points: Large- <mml:math
  xmlns:mml="http://www.w3.org/1998/math/mathml"><mml:mi>n</mml:mi></mml:math>
  expansion and connections to symmetry-protected topological states},
  \href{https://doi.org/10.1103/physrevb.104.104201}{\emph{Physical Review B}
  {\bfseries 104} (2021) }.

\bibitem{Cuomo:2022xgw}
G.~Cuomo, Z.~Komargodski, M.~Mezei and A.~Raviv-Moshe, \emph{{Spin impurities,
  Wilson lines and semiclassics}},
  \href{https://doi.org/10.1007/JHEP06(2022)112}{\emph{JHEP} {\bfseries 06}
  (2022) 112} [\href{https://arxiv.org/abs/2202.00040}{{\ttfamily
  2202.00040}}].

\bibitem{Hsin:2024aqb}
P.-S.~Hsin, R.~Kobayashi and C.~Zhang, \emph{{Fractionalization of coset
  non-invertible symmetry and exotic Hall conductance}},
  \href{https://doi.org/10.21468/SciPostPhys.17.3.095}{\emph{SciPost Phys.}
  {\bfseries 17} (2024) 095}
  [\href{https://arxiv.org/abs/2405.20401}{{\ttfamily 2405.20401}}].

\bibitem{Wang:2015fmi}
C.~Wang and T.~Senthil, \emph{{Time-Reversal Symmetric $U(1)$ Quantum Spin
  Liquids}}, \href{https://doi.org/10.1103/PhysRevX.6.011034}{\emph{Phys. Rev.
  X} {\bfseries 6} (2016) 011034}
  [\href{https://arxiv.org/abs/1505.03520}{{\ttfamily 1505.03520}}].

\bibitem{Hsin:2019gvb}
P.-S.~Hsin and S.-H.~Shao, \emph{{Lorentz Symmetry Fractionalization and
  Dualities in (2+1)d}},
  \href{https://doi.org/10.21468/SciPostPhys.8.2.018}{\emph{SciPost Phys.}
  {\bfseries 8} (2020) 018} [\href{https://arxiv.org/abs/1909.07383}{{\ttfamily
  1909.07383}}].

\bibitem{Ning:2019ffr}
S.-Q.~Ning, L.~Zou and M.~Cheng, \emph{{Fractionalization and Anomalies in
  Symmetry-Enriched U(1) Gauge Theories}},
  \href{https://doi.org/10.1103/PhysRevResearch.2.043043}{\emph{Phys. Rev.
  Res.} {\bfseries 2} (2020) 043043}
  [\href{https://arxiv.org/abs/1905.03276}{{\ttfamily 1905.03276}}].

\bibitem{Bulmash:2021hmb}
D.~Bulmash and M.~Barkeshli, \emph{{Fermionic symmetry fractionalization in
  (2+1) dimensions}},
  \href{https://doi.org/10.1103/PhysRevB.105.125114}{\emph{Phys. Rev. B}
  {\bfseries 105} (2022) 125114}
  [\href{https://arxiv.org/abs/2109.10913}{{\ttfamily 2109.10913}}].

\bibitem{Aasen:2021vva}
D.~Aasen, P.~Bonderson and C.~Knapp, \emph{{Characterization and Classification
  of Fermionic Symmetry Enriched Topological Phases}},
  \href{https://arxiv.org/abs/2109.10911}{{\ttfamily 2109.10911}}.

\bibitem{Ang:2019txy}
J.P.~Ang, K.~Roumpedakis and S.~Seifnashri, \emph{{Line Operators of Gauge
  Theories on Non-Spin Manifolds}},
  \href{https://doi.org/10.1007/JHEP04(2020)087}{\emph{JHEP} {\bfseries 04}
  (2020) 087} [\href{https://arxiv.org/abs/1911.00589}{{\ttfamily
  1911.00589}}].

\bibitem{metlitski2015}
M.A.~Metlitski, \emph{$s$-duality of $u(1)$ gauge theory with $\theta =\pi$ on
  non-orientable manifolds: Applications to topological insulators and
  superconductors},  2015.

\bibitem{Geiko:2022qjy}
R.~Geiko and G.W.~Moore, \emph{{When Does a Three-Dimensional
  Chern\textendash{}Simons\textendash{}Witten Theory Have a Time Reversal
  Symmetry?}}, \href{https://doi.org/10.1007/s00023-023-01303-3}{\emph{Annales
  Henri Poincare} {\bfseries 25} (2024) 673}
  [\href{https://arxiv.org/abs/2209.04519}{{\ttfamily 2209.04519}}].

\bibitem{Delmastro:2019vnj}
D.~Delmastro and J.~Gomis, \emph{{Symmetries of Abelian Chern-Simons Theories
  and Arithmetic}}, \href{https://doi.org/10.1007/JHEP03(2021)006}{\emph{JHEP}
  {\bfseries 03} (2021) 006}
  [\href{https://arxiv.org/abs/1904.12884}{{\ttfamily 1904.12884}}].

\bibitem{Gagliano:2025gwr}
F.~Gagliano, A.~Grigoletto and K.~Ohmori, \emph{{Higher Representations and
  Quark Confinement}},  \href{https://arxiv.org/abs/2501.09069}{{\ttfamily
  2501.09069}}.

\bibitem{transmute}
N.~Seiberg and S.~Seifnashri, \emph{{Symmetry Transmutation and Anomaly
  Matching}}, {\emph{To appear\!} }.

\bibitem{Yu:2020twi}
M.~Yu, \emph{{Symmetries and anomalies of (1+1)d theories: 2-groups and
  symmetry fractionalization}},
  \href{https://doi.org/10.1007/JHEP08(2021)061}{\emph{JHEP} {\bfseries 08}
  (2021) 061} [\href{https://arxiv.org/abs/2010.01136}{{\ttfamily
  2010.01136}}].

\bibitem{Bartsch:2023pzl}
T.~Bartsch, M.~Bullimore and A.~Grigoletto, \emph{{Higher representations for
  extended operators}},  \href{https://arxiv.org/abs/2304.03789}{{\ttfamily
  2304.03789}}.

\bibitem{Bhardwaj:2023wzd}
L.~Bhardwaj and S.~Schafer-Nameki, \emph{{Generalized charges, part I:
  Invertible symmetries and higher representations}},
  \href{https://doi.org/10.21468/SciPostPhys.16.4.093}{\emph{SciPost Phys.}
  {\bfseries 16} (2024) 093}
  [\href{https://arxiv.org/abs/2304.02660}{{\ttfamily 2304.02660}}].

\bibitem{Senthil:1999czm}
T.~Senthil and M.P.A.~Fisher, \emph{{Z\_2 Gauge Theory of Electron
  Fractionalization in Strongly Correlated Systems}},
  \href{https://doi.org/10.1103/PhysRevB.62.7850}{\emph{Phys. Rev. B}
  {\bfseries 62} (2000) 7850}
  [\href{https://arxiv.org/abs/cond-mat/9910224}{{\ttfamily
  cond-mat/9910224}}].

\bibitem{Chen:2014wse}
X.~Chen, F.J.~Burnell, A.~Vishwanath and L.~Fidkowski, \emph{{Anomalous
  Symmetry Fractionalization and Surface Topological Order}},
  \href{https://doi.org/10.1103/PhysRevX.5.041013}{\emph{Phys. Rev. X}
  {\bfseries 5} (2015) 041013}
  [\href{https://arxiv.org/abs/1403.6491}{{\ttfamily 1403.6491}}].

\bibitem{Teo:2015xla}
J.C.Y.~Teo, T.L.~Hughes and E.~Fradkin, \emph{{Theory of Twist Liquids: Gauging
  an Anyonic Symmetry}},
  \href{https://doi.org/10.1016/j.aop.2015.05.012}{\emph{Annals Phys.}
  {\bfseries 360} (2015) 349}
  [\href{https://arxiv.org/abs/1503.06812}{{\ttfamily 1503.06812}}].

\bibitem{Fidkowski:2016svr}
L.~Fidkowski, N.H.~Lindner and N.~Tarantino, \emph{{Symmetry fractionalization
  and twist defects}},
  \href{https://doi.org/10.1088/1367-2630/18/3/035006}{\emph{New J. Phys.}
  {\bfseries 18} (2016) 035006}.

\bibitem{Chen:2016fxq}
X.~Chen, \emph{{Symmetry fractionalization in two dimensional topological
  phases}}, \href{https://doi.org/10.1016/j.revip.2017.02.002}{\emph{Rev.
  Phys.} {\bfseries 2} (2017) 3}
  [\href{https://arxiv.org/abs/1606.07569}{{\ttfamily 1606.07569}}].

\bibitem{Barkeshli:2019vtb}
M.~Barkeshli and M.~Cheng, \emph{{Relative Anomalies in (2+1)D Symmetry
  Enriched Topological States}},
  \href{https://doi.org/10.21468/SciPostPhys.8.2.028}{\emph{SciPost Phys.}
  {\bfseries 8} (2020) 028} [\href{https://arxiv.org/abs/1906.10691}{{\ttfamily
  1906.10691}}].

\bibitem{Antinucci:2024izg}
A.~Antinucci, C.~Copetti, G.~Galati and G.~Rizi, \emph{{Topological Constraints
  on Defect Dynamics}},  \href{https://arxiv.org/abs/2412.18652}{{\ttfamily
  2412.18652}}.

\bibitem{Jensen:2017eof}
K.~Jensen, E.~Shaverin and A.~Yarom, \emph{{\textquoteright{}t Hooft anomalies
  and boundaries}}, \href{https://doi.org/10.1007/JHEP01(2018)085}{\emph{JHEP}
  {\bfseries 01} (2018) 085}
  [\href{https://arxiv.org/abs/1710.07299}{{\ttfamily 1710.07299}}].

\bibitem{Thorngren:2020yht}
R.~Thorngren and Y.~Wang, \emph{{Anomalous symmetries end at the boundary}},
  \href{https://doi.org/10.1007/JHEP09(2021)017}{\emph{JHEP} {\bfseries 09}
  (2021) 017} [\href{https://arxiv.org/abs/2012.15861}{{\ttfamily
  2012.15861}}].

\bibitem{Rudelius:2020orz}
T.~Rudelius and S.-H.~Shao, \emph{{Topological Operators and Completeness of
  Spectrum in Discrete Gauge Theories}},
  \href{https://doi.org/10.1007/JHEP12(2020)172}{\emph{JHEP} {\bfseries 12}
  (2020) 172} [\href{https://arxiv.org/abs/2006.10052}{{\ttfamily
  2006.10052}}].

\bibitem{Tachikawa:2017gyf}
Y.~Tachikawa, \emph{{On gauging finite subgroups}},
  \href{https://doi.org/10.21468/SciPostPhys.8.1.015}{\emph{SciPost Phys.}
  {\bfseries 8} (2020) 015} [\href{https://arxiv.org/abs/1712.09542}{{\ttfamily
  1712.09542}}].

\bibitem{Bhardwaj:2022dyt}
L.~Bhardwaj, M.~Bullimore, A.E.V.~Ferrari and S.~Schafer-Nameki,
  \emph{{Anomalies of Generalized Symmetries from Solitonic Defects}},
  \href{https://doi.org/10.21468/SciPostPhys.16.3.087}{\emph{SciPost Phys.}
  {\bfseries 16} (2024) 087}
  [\href{https://arxiv.org/abs/2205.15330}{{\ttfamily 2205.15330}}].

\bibitem{Gaiotto:2010be}
D.~Gaiotto, G.W.~Moore and A.~Neitzke, \emph{{Framed BPS States}},
  \href{https://doi.org/10.4310/ATMP.2013.v17.n2.a1}{\emph{Adv. Theor. Math.
  Phys.} {\bfseries 17} (2013) 241}
  [\href{https://arxiv.org/abs/1006.0146}{{\ttfamily 1006.0146}}].

\bibitem{Nguyen:2024ikq}
M.~Nguyen, T.~Sulejmanpasic and M.~\"Unsal, \emph{{Phases of theories with
  $\mathbb{Z}_N$ 1-form symmetry and the roles of center vortices and magnetic
  monopoles}},  \href{https://arxiv.org/abs/2401.04800}{{\ttfamily
  2401.04800}}.

\bibitem{Robbins:2022wlr}
D.G.~Robbins, E.~Sharpe and T.~Vandermeulen, \emph{{Decomposition,
  trivially-acting symmetries, and topological operators}},
  \href{https://doi.org/10.1103/PhysRevD.107.085017}{\emph{Phys. Rev. D}
  {\bfseries 107} (2023) 085017}
  [\href{https://arxiv.org/abs/2211.14332}{{\ttfamily 2211.14332}}].

\bibitem{Alvarez:1984es}
O.~Alvarez, \emph{{Topological Quantization and Cohomology}},
  \href{https://doi.org/10.1007/BF01212452}{\emph{Commun. Math. Phys.}
  {\bfseries 100} (1985) 279}.

\bibitem{Freed:2006yc}
D.S.~Freed, G.W.~Moore and G.~Segal, \emph{{Heisenberg Groups and
  Noncommutative Fluxes}},
  \href{https://doi.org/10.1016/j.aop.2006.07.014}{\emph{Annals Phys.}
  {\bfseries 322} (2007) 236}
  [\href{https://arxiv.org/abs/hep-th/0605200}{{\ttfamily hep-th/0605200}}].

\bibitem{Freed:2006ya}
D.S.~Freed, G.W.~Moore and G.~Segal, \emph{{The Uncertainty of Fluxes}},
  \href{https://doi.org/10.1007/s00220-006-0181-3}{\emph{Commun. Math. Phys.}
  {\bfseries 271} (2007) 247}
  [\href{https://arxiv.org/abs/hep-th/0605198}{{\ttfamily hep-th/0605198}}].

\bibitem{Kapustin:2014gua}
A.~Kapustin and N.~Seiberg, \emph{{Coupling a QFT to a TQFT and Duality}},
  \href{https://doi.org/10.1007/JHEP04(2014)001}{\emph{JHEP} {\bfseries 04}
  (2014) 001} [\href{https://arxiv.org/abs/1401.0740}{{\ttfamily 1401.0740}}].

\bibitem{MooreDiffCoh}
G.~Moore, \emph{{A Minicourse on Generalized Abelian Gauge Theory, Self-Dual
  Theories, and Differential Cohomology}}, .

\bibitem{PhysRev.80.440}
R.P.~Feynman, \emph{Mathematical formulation of the quantum theory of
  electromagnetic interaction},
  \href{https://doi.org/10.1103/PhysRev.80.440}{\emph{Phys. Rev.} {\bfseries
  80} (1950) 440}.

\bibitem{PhysRev.84.108}
R.P.~Feynman, \emph{An operator calculus having applications in quantum
  electrodynamics}, \href{https://doi.org/10.1103/PhysRev.84.108}{\emph{Phys.
  Rev.} {\bfseries 84} (1951) 108}.

\bibitem{PhysRev.82.664}
J.~Schwinger, \emph{On gauge invariance and vacuum polarization},
  \href{https://doi.org/10.1103/PhysRev.82.664}{\emph{Phys. Rev.} {\bfseries
  82} (1951) 664}.

\bibitem{PhysRevLett.66.1669}
Z.~Bern and D.A.~Kosower, \emph{Efficient calculation of one-loop qcd
  amplitudes}, \href{https://doi.org/10.1103/PhysRevLett.66.1669}{\emph{Phys.
  Rev. Lett.} {\bfseries 66} (1991) 1669}.

\bibitem{BERN1992451}
Z.~Bern and D.A.~Kosower, \emph{The computation of loop amplitudes in gauge
  theories},
  \href{https://doi.org/https://doi.org/10.1016/0550-3213(92)90134-W}{\emph{Nuclear
  Physics B} {\bfseries 379} (1992) 451}.

\bibitem{STRASSLER1992145}
M.J.~Strassler, \emph{Field theory without feynman diagrams: One-loop effective
  actions},
  \href{https://doi.org/https://doi.org/10.1016/0550-3213(92)90098-V}{\emph{Nuclear
  Physics B} {\bfseries 385} (1992) 145}.

\bibitem{Gaiotto:2017yup}
D.~Gaiotto, A.~Kapustin, Z.~Komargodski and N.~Seiberg, \emph{{Theta, Time
  Reversal, and Temperature}},
  \href{https://doi.org/10.1007/JHEP05(2017)091}{\emph{JHEP} {\bfseries 05}
  (2017) 091} [\href{https://arxiv.org/abs/1703.00501}{{\ttfamily
  1703.00501}}].

\bibitem{Fujimori:2019skd}
T.~Fujimori, E.~Itou, T.~Misumi, M.~Nitta and N.~Sakai,
  \emph{{Confinement-deconfinement crossover in the lattice $\mathbb{C}P^{N-1}$
  model}}, \href{https://doi.org/10.1103/PhysRevD.100.094506}{\emph{Phys. Rev.
  D} {\bfseries 100} (2019) 094506}
  [\href{https://arxiv.org/abs/1907.06925}{{\ttfamily 1907.06925}}].

\bibitem{Misumi:2019upg}
T.~Misumi, T.~Fujimori, E.~Itou, M.~Nitta and N.~Sakai, \emph{{Lattice study on
  the twisted ${\mathbb C} P^{N-1}$ models on ${\mathbb R} \times S^1$}},
  \href{https://doi.org/10.22323/1.363.0015}{\emph{PoS} {\bfseries LATTICE2019}
  (2019) 015} [\href{https://arxiv.org/abs/1911.07398}{{\ttfamily
  1911.07398}}].

\bibitem{Cherman:2020hbe}
A.~Cherman, T.~Jacobson, S.~Sen and L.G.~Yaffe, \emph{{Higgs-confinement phase
  transitions with fundamental representation matter}},
  \href{https://doi.org/10.1103/PhysRevD.102.105021}{\emph{Phys. Rev. D}
  {\bfseries 102} (2020) 105021}
  [\href{https://arxiv.org/abs/2007.08539}{{\ttfamily 2007.08539}}].

\bibitem{Cherman:2024exo}
A.~Cherman, T.~Jacobson, S.~Sen and L.G.~Yaffe, \emph{{Line operators, vortex
  statistics, and Higgs versus confinement dynamics}},
  \href{https://doi.org/10.1007/JHEP06(2024)200}{\emph{JHEP} {\bfseries 06}
  (2024) 200} [\href{https://arxiv.org/abs/2401.17489}{{\ttfamily
  2401.17489}}].

\bibitem{Aharony:2013hda}
O.~Aharony, N.~Seiberg and Y.~Tachikawa, \emph{{Reading between the lines of
  four-dimensional gauge theories}},
  \href{https://doi.org/10.1007/JHEP08(2013)115}{\emph{JHEP} {\bfseries 08}
  (2013) 115} [\href{https://arxiv.org/abs/1305.0318}{{\ttfamily 1305.0318}}].

\bibitem{Witten:2000nv}
E.~Witten, \emph{{Supersymmetric index in four-dimensional gauge theories}},
  \href{https://doi.org/10.4310/ATMP.2001.v5.n5.a1}{\emph{Adv. Theor. Math.
  Phys.} {\bfseries 5} (2002) 841}
  [\href{https://arxiv.org/abs/hep-th/0006010}{{\ttfamily hep-th/0006010}}].

\bibitem{Witten:1983tx}
E.~Witten, \emph{{Current Algebra, Baryons, and Quark Confinement}},
  \href{https://doi.org/10.1016/0550-3213(83)90064-0}{\emph{Nucl. Phys. B}
  {\bfseries 223} (1983) 433}.

\bibitem{Aharony:2023amq}
O.~Aharony, G.~Cuomo, Z.~Komargodski, M.~Mezei and A.~Raviv-Moshe,
  \emph{{Phases of Wilson lines: conformality and screening}},
  \href{https://doi.org/10.1007/JHEP12(2023)183}{\emph{JHEP} {\bfseries 12}
  (2023) 183} [\href{https://arxiv.org/abs/2310.00045}{{\ttfamily
  2310.00045}}].

\bibitem{Cordova:2022rer}
C.~Cordova, K.~Ohmori and T.~Rudelius, \emph{{Generalized symmetry breaking
  scales and weak gravity conjectures}},
  \href{https://doi.org/10.1007/JHEP11(2022)154}{\emph{JHEP} {\bfseries 11}
  (2022) 154} [\href{https://arxiv.org/abs/2202.05866}{{\ttfamily
  2202.05866}}].

\bibitem{Cherman:2023xok}
A.~Cherman and T.~Jacobson, \emph{{Emergent 1-form symmetries}},
  \href{https://doi.org/10.1103/PhysRevD.109.125013}{\emph{Phys. Rev. D}
  {\bfseries 109} (2024) 125013}
  [\href{https://arxiv.org/abs/2304.13751}{{\ttfamily 2304.13751}}].

\bibitem{Hsin:2020nts}
P.-S.~Hsin and H.T.~Lam, \emph{{Discrete theta angles, symmetries and
  anomalies}},
  \href{https://doi.org/10.21468/SciPostPhys.10.2.032}{\emph{SciPost Phys.}
  {\bfseries 10} (2021) 032}
  [\href{https://arxiv.org/abs/2007.05915}{{\ttfamily 2007.05915}}].

\bibitem{Bhardwaj:2021wif}
L.~Bhardwaj, \emph{{2-Group symmetries in class S}},
  \href{https://doi.org/10.21468/SciPostPhys.12.5.152}{\emph{SciPost Phys.}
  {\bfseries 12} (2022) 152}
  [\href{https://arxiv.org/abs/2107.06816}{{\ttfamily 2107.06816}}].

\bibitem{Freed:2017rlk}
D.S.~Freed, Z.~Komargodski and N.~Seiberg, \emph{{The Sum Over Topological
  Sectors and $\theta$ in the 2+1-Dimensional $\mathbb{C}\mathbb{P}^1$
  $\sigma$-Model}},
  \href{https://doi.org/10.1007/s00220-018-3093-0}{\emph{Commun. Math. Phys.}
  {\bfseries 362} (2018) 167}
  [\href{https://arxiv.org/abs/1707.05448}{{\ttfamily 1707.05448}}].

\bibitem{Pisarski:1986gr}
R.D.~Pisarski, \emph{{Magnetic Monopoles in Topologically Massive Gauge
  Theories}}, \href{https://doi.org/10.1103/PhysRevD.34.3851}{\emph{Phys. Rev.
  D} {\bfseries 34} (1986) 3851}.

\bibitem{Affleck:1989qf}
I.~Affleck, J.A.~Harvey, L.~Palla and G.W.~Semenoff, \emph{{The {Chern-Simons}
  Term Versus the Monopole}},
  \href{https://doi.org/10.1016/0550-3213(89)90220-4}{\emph{Nucl. Phys. B}
  {\bfseries 328} (1989) 575}.

\bibitem{Seiberg:2024yig}
N.~Seiberg, \emph{{Anomalous Continuous Translations}},
  \href{https://arxiv.org/abs/2412.14434}{{\ttfamily 2412.14434}}.

\bibitem{RABINOVICI1984523}
E.~Rabinovici, A.~Schwimmer and S.~Yankielowicz, \emph{Quantization in the
  presence of wess-zumino terms},
  \href{https://doi.org/https://doi.org/10.1016/0550-3213(84)90609-6}{\emph{Nuclear
  Physics B} {\bfseries 248} (1984) 523}.

\bibitem{Tong:2014yla}
D.~Tong and K.~Wong, \emph{{Monopoles and Wilson Lines}},
  \href{https://doi.org/10.1007/JHEP06(2014)048}{\emph{JHEP} {\bfseries 06}
  (2014) 048} [\href{https://arxiv.org/abs/1401.6167}{{\ttfamily 1401.6167}}].

\bibitem{Moore:2015szp}
G.W.~Moore, A.B.~Royston and D.~Van~den Bleeken, \emph{{Semiclassical framed
  BPS states}}, \href{https://doi.org/10.1007/JHEP07(2016)071}{\emph{JHEP}
  {\bfseries 07} (2016) 071}
  [\href{https://arxiv.org/abs/1512.08924}{{\ttfamily 1512.08924}}].

\bibitem{Brennan:2016znk}
T.D.~Brennan and G.W.~Moore, \emph{{A note on the semiclassical formulation of
  BPS states in four-dimensional $N=$ 2 theories}},
  \href{https://doi.org/10.1093/ptep/ptw159}{\emph{PTEP} {\bfseries 2016}
  (2016) 12C110} [\href{https://arxiv.org/abs/1610.00697}{{\ttfamily
  1610.00697}}].

\bibitem{Brennan:2018ura}
T.D.~Brennan, G.W.~Moore and A.B.~Royston, \emph{{Wall Crossing from Dirac
  Zeromodes}}, \href{https://doi.org/10.1007/JHEP09(2018)038}{\emph{JHEP}
  {\bfseries 09} (2018) 038}
  [\href{https://arxiv.org/abs/1805.08783}{{\ttfamily 1805.08783}}].

\bibitem{Chuang:2013wt}
W.-y.~Chuang, D.-E.~Diaconescu, J.~Manschot, G.W.~Moore and Y.~Soibelman,
  \emph{{Geometric engineering of (framed) BPS states}},
  \href{https://doi.org/10.4310/ATMP.2014.v18.n5.a3}{\emph{Adv. Theor. Math.
  Phys.} {\bfseries 18} (2014) 1063}
  [\href{https://arxiv.org/abs/1301.3065}{{\ttfamily 1301.3065}}].

\bibitem{Cordova:2013bza}
C.~C\'ordova and A.~Neitzke, \emph{{Line Defects, Tropicalization, and
  Multi-Centered Quiver Quantum Mechanics}},
  \href{https://doi.org/10.1007/JHEP09(2014)099}{\emph{JHEP} {\bfseries 09}
  (2014) 099} [\href{https://arxiv.org/abs/1308.6829}{{\ttfamily 1308.6829}}].

\bibitem{Brennan:2018rcn}
T.D.~Brennan, A.~Dey and G.W.~Moore, \emph{{\textquoteright{}t Hooft defects
  and wall crossing in SQM}},
  \href{https://doi.org/10.1007/JHEP10(2019)173}{\emph{JHEP} {\bfseries 10}
  (2019) 173} [\href{https://arxiv.org/abs/1810.07191}{{\ttfamily
  1810.07191}}].

\bibitem{Lee:2011ph}
S.~Lee and P.~Yi, \emph{{Framed BPS States, Moduli Dynamics, and
  Wall-Crossing}}, \href{https://doi.org/10.1007/JHEP04(2011)098}{\emph{JHEP}
  {\bfseries 04} (2011) 098} [\href{https://arxiv.org/abs/1102.1729}{{\ttfamily
  1102.1729}}].

\bibitem{Cordova:2016uwk}
C.~Cordova, D.~Gaiotto and S.-H.~Shao, \emph{{Infrared Computations of Defect
  Schur Indices}}, \href{https://doi.org/10.1007/JHEP11(2016)106}{\emph{JHEP}
  {\bfseries 11} (2016) 106}
  [\href{https://arxiv.org/abs/1606.08429}{{\ttfamily 1606.08429}}].

\bibitem{Brennan:2019hzm}
T.D.~Brennan and G.W.~Moore, \emph{{Index-Like Theorems from Line Defect
  Vevs}}, \href{https://doi.org/10.1007/JHEP09(2019)073}{\emph{JHEP} {\bfseries
  09} (2019) 073} [\href{https://arxiv.org/abs/1903.08172}{{\ttfamily
  1903.08172}}].

\bibitem{Vafa:1983tf}
C.~Vafa and E.~Witten, \emph{{Restrictions on Symmetry Breaking in Vector-Like
  Gauge Theories}},
  \href{https://doi.org/10.1016/0550-3213(84)90230-X}{\emph{Nucl. Phys. B}
  {\bfseries 234} (1984) 173}.

\bibitem{Tong:2021phe}
D.~Tong, \emph{{Comments on symmetric mass generation in 2d and 4d}},
  \href{https://doi.org/10.1007/JHEP07(2022)001}{\emph{JHEP} {\bfseries 07}
  (2022) 001} [\href{https://arxiv.org/abs/2104.03997}{{\ttfamily
  2104.03997}}].

\bibitem{Bars:1981se}
I.~Bars and S.~Yankielowicz, \emph{{Composite Quarks and Leptons as Solutions
  of Anomaly Constraints}},
  \href{https://doi.org/10.1016/0370-2693(81)90664-X}{\emph{Phys. Lett. B}
  {\bfseries 101} (1981) 159}.

\bibitem{Wang:2022ucy}
J.~Wang and Y.-Z.~You, \emph{{Symmetric Mass Generation}},
  \href{https://doi.org/10.3390/sym14071475}{\emph{Symmetry} {\bfseries 14}
  (2022) 1475} [\href{https://arxiv.org/abs/2204.14271}{{\ttfamily
  2204.14271}}].

\bibitem{Shirman:2023hhk}
Y.~Shirman, S.~Shukla and M.~Waterbury, \emph{{Chirality changing RG flows:
  dynamics and models}},
  \href{https://doi.org/10.1007/JHEP06(2023)168}{\emph{JHEP} {\bfseries 06}
  (2023) 168} [\href{https://arxiv.org/abs/2303.08847}{{\ttfamily
  2303.08847}}].

\bibitem{Creutz:1996xc}
M.~Creutz, M.~Tytgat, C.~Rebbi and S.-S.~Xue, \emph{{Lattice formulation of the
  standard model}},
  \href{https://doi.org/10.1016/S0370-2693(97)00463-2}{\emph{Phys. Lett. B}
  {\bfseries 402} (1997) 341}
  [\href{https://arxiv.org/abs/hep-lat/9612017}{{\ttfamily hep-lat/9612017}}].

\bibitem{Poppitz:2010at}
E.~Poppitz and Y.~Shang, \emph{{Chiral Lattice Gauge Theories Via
  Mirror-Fermion Decoupling: A Mission (im)Possible?}},
  \href{https://doi.org/10.1142/S0217751X10049852}{\emph{Int. J. Mod. Phys. A}
  {\bfseries 25} (2010) 2761}
  [\href{https://arxiv.org/abs/1003.5896}{{\ttfamily 1003.5896}}].

\bibitem{Wen:2013ppa}
X.-G.~Wen, \emph{{A lattice non-perturbative definition of an SO(10) chiral
  gauge theory and its induced standard model}},
  \href{https://doi.org/10.1088/0256-307X/30/11/111101}{\emph{Chin. Phys.
  Lett.} {\bfseries 30} (2013) 111101}
  [\href{https://arxiv.org/abs/1305.1045}{{\ttfamily 1305.1045}}].

\bibitem{DeMarco:2017gcb}
M.~DeMarco and X.-G.~Wen, \emph{{A Novel Non-Perturbative Lattice
  Regularization of an Anomaly-Free $1 + 1d$ Chiral $SU(2)$ Gauge Theory}},
  \href{https://arxiv.org/abs/1706.04648}{{\ttfamily 1706.04648}}.

\bibitem{Kikukawa:2017ngf}
Y.~Kikukawa, \emph{{On the gauge invariant path-integral measure for the
  overlap Weyl fermions in $\underline{16}$ of SO(10)}},
  \href{https://doi.org/10.1093/ptep/ptz115}{\emph{PTEP} {\bfseries 2019}
  (2019) 113B03} [\href{https://arxiv.org/abs/1710.11618}{{\ttfamily
  1710.11618}}].

\bibitem{Wang:2018ugf}
J.~Wang and X.-G.~Wen, \emph{{A Solution to the 1+1D Gauged Chiral Fermion
  Problem}}, \href{https://doi.org/10.1103/PhysRevD.99.111501}{\emph{Phys. Rev.
  D} {\bfseries 99} (2018) 111501}
  [\href{https://arxiv.org/abs/1807.05998}{{\ttfamily 1807.05998}}].

\bibitem{Butt:2018nkn}
N.~Butt, S.~Catterall and D.~Schaich, \emph{{$SO(4)$ invariant Higgs-Yukawa
  model with reduced staggered fermions}},
  \href{https://doi.org/10.1103/PhysRevD.98.114514}{\emph{Phys. Rev. D}
  {\bfseries 98} (2018) 114514}
  [\href{https://arxiv.org/abs/1810.06117}{{\ttfamily 1810.06117}}].

\bibitem{Haldane:1995xgi}
F.D.M.~Haldane, \emph{{Stability of Chiral Luttinger Liquids and Abelian
  Quantum Hall States.}},
  \href{https://doi.org/10.1103/PhysRevLett.74.2090}{\emph{Phys. Rev. Lett.}
  {\bfseries 74} (1995) 2090}
  [\href{https://arxiv.org/abs/cond-mat/9501007}{{\ttfamily
  cond-mat/9501007}}].

\bibitem{Chen:2022cyw}
S.~Chen and Y.~Tanizaki, \emph{{Solitonic Symmetry beyond Homotopy:
  Invertibility from Bordism and Noninvertibility from Topological Quantum
  Field Theory}},
  \href{https://doi.org/10.1103/PhysRevLett.131.011602}{\emph{Phys. Rev. Lett.}
  {\bfseries 131} (2023) 011602}
  [\href{https://arxiv.org/abs/2210.13780}{{\ttfamily 2210.13780}}].

\bibitem{Chen:2023czk}
S.~Chen and Y.~Tanizaki, \emph{{Solitonic symmetry as non-invertible symmetry:
  cohomology theories with TQFT coefficients}},
  \href{https://arxiv.org/abs/2307.00939}{{\ttfamily 2307.00939}}.

\bibitem{Pace:2023kyi}
S.D.~Pace, \emph{{Emergent generalized symmetries in ordered phases and
  applications to quantum disordering}},
  \href{https://doi.org/10.21468/SciPostPhys.17.3.080}{\emph{SciPost Phys.}
  {\bfseries 17} (2024) 080}
  [\href{https://arxiv.org/abs/2308.05730}{{\ttfamily 2308.05730}}].

\bibitem{Pace:2023mdo}
S.D.~Pace, C.~Zhu, A.~Beaudry and X.-G.~Wen, \emph{{Generalized symmetries in
  singularity-free nonlinear \ensuremath{\sigma} models and their disordered
  phases}}, \href{https://doi.org/10.1103/PhysRevB.110.195149}{\emph{Phys. Rev.
  B} {\bfseries 110} (2024) 195149}
  [\href{https://arxiv.org/abs/2310.08554}{{\ttfamily 2310.08554}}].

\bibitem{Hsin:2025ria}
P.-S.~Hsin, R.~Kobayashi and C.~Zhang, \emph{{Anomalies of Coset Non-Invertible
  Symmetries}},  \href{https://arxiv.org/abs/2503.00105}{{\ttfamily
  2503.00105}}.

\bibitem{Sheckler:2025rlk}
A.~Sheckler, \emph{{Mixed Anomalies of Magnetic Symmetries}},
  \href{https://arxiv.org/abs/2503.08789}{{\ttfamily 2503.08789}}.

\bibitem{DHoker:2024vii}
E.~D'Hoker, T.T.~Dumitrescu, E.~Gerchkovitz and E.~Nardoni, \emph{{Cascading
  from $\mathscr{N}=2$ Supersymmetric Yang-Mills Theory to Confinement and
  Chiral Symmetry Breaking in Adjoint QCD}},
  \href{https://arxiv.org/abs/2412.20547}{{\ttfamily 2412.20547}}.

\bibitem{Lee:2020ojw}
Y.~Lee, K.~Ohmori and Y.~Tachikawa, \emph{{Revisiting Wess-Zumino-Witten
  terms}}, \href{https://doi.org/10.21468/SciPostPhys.10.3.061}{\emph{SciPost
  Phys.} {\bfseries 10} (2021) 061}
  [\href{https://arxiv.org/abs/2009.00033}{{\ttfamily 2009.00033}}].

\bibitem{AFFLECK1980461}
I.~Affleck, \emph{The role of instantons in scale-invariant gauge theories},
  \href{https://doi.org/https://doi.org/10.1016/0550-3213(80)90350-8}{\emph{Nuclear
  Physics B} {\bfseries 162} (1980) 461}.

\bibitem{PhysRevD.24.475}
N.~Weiss, \emph{Effective potential for the order parameter of gauge theories
  at finite temperature},
  \href{https://doi.org/10.1103/PhysRevD.24.475}{\emph{Phys. Rev. D} {\bfseries
  24} (1981) 475}.

\bibitem{GPY}
D.J.~Gross, R.D.~Pisarski and L.G.~Yaffe, \emph{Qcd and instantons at finite
  temperature}, \href{https://doi.org/10.1103/RevModPhys.53.43}{\emph{Rev. Mod.
  Phys.} {\bfseries 53} (1981) 43}.

\bibitem{fradkinshenker}
E.~Fradkin and S.H.~Shenker, \emph{Phase diagrams of lattice gauge theories
  with higgs fields},
  \href{https://doi.org/10.1103/PhysRevD.19.3682}{\emph{Phys. Rev. D}
  {\bfseries 19} (1979) 3682}.

\end{thebibliography}\endgroup
\end{document}